\documentclass[12pt,letter]{article}
\usepackage{graphicx,subfigure}
\usepackage{epsfig}
\usepackage{epstopdf}
\usepackage{amsmath}
\usepackage{amsfonts}
\usepackage{amssymb}
\newcommand{\rom}[1]{\mathrm{#1}}

\newcommand{\beq}{\begin{equation}}
\newcommand{\eeq}{\end{equation}}
\newcommand{\be}{\begin{equation}}
\newcommand{\ee}{\end{equation}}
\newcommand{\beqa}{\begin{eqnarray}}
\newcommand{\eeqa}{\end{eqnarray}}
\newcommand{\beqar}{\begin{eqnarray*}}
\newcommand{\eeqar}{\end{eqnarray*}}
\newcommand{\bea}{\begin{eqnarray}}
\newcommand{\eea}{\end{eqnarray}}

\newcommand{\bc}{\bar{c}}
\newcommand{\pa}{\partial}
\newcommand{\ka}{\kappa}
\newcommand{\bz}{\bar{z}}

\newcommand{\reef}[1]{(\ref{#1})}

\newcommand{\Tr}{\textrm{Tr}}
\newcommand{\diag}{\textrm{diag}}

\newcommand{\lsim}{\mathrel{\raisebox{-.6ex}{$\stackrel{\textstyle<}{\sim}$}}}

\numberwithin{equation}{section}

\begin{document}

\setlength{\unitlength}{1mm}

\begin{titlepage}

\begin{flushright}
MIT-CTP-3796
\end{flushright}
\vspace{1cm}

\begin{center}
{\bf \Large Black Saturn}
\end{center}

\vspace{7mm}

\begin{center}
Henriette Elvang$^{a}$ and  Pau Figueras$^{b}$

\vspace{.5cm}
{\small {\textit{$^{a}$Center for Theoretical Physics, }}\\
{\small \textit{Massachusetts
Institute of Technology, Cambridge, MA 02139, USA}}}\\
\vspace{2mm}
{\small \textit{$^{b}$Departament de F{\'\i}sica Fonamental}}\\
{\small \textit{Universitat de Barcelona, Diagonal 647, E-08028,
    Barcelona, Spain}}

\vspace*{0.5cm}
{\small {\tt elvang@lns.mit.edu, pfigueras@ffn.ub.es}}
\end{center}

\vspace{1cm}

\begin{abstract}

Using the inverse scattering method we construct an exact stationary
asymptotically 
flat 4+1-dimensional vacuum solution describing ``black 
saturn'': a spherical black hole surrounded by a black ring. Angular
momentum keeps the configuration in equilibrium.
Black saturn reveals a number of interesting gravitational
phenomena: 
(1) The balanced solution exhibits 2-fold continuous non-uniqueness
for fixed mass and angular momentum; 
(2) Remarkably, the 4+1d Schwarzschild black hole is not unique, since
the black 
ring and black hole of black saturn can counter-rotate to give zero
total angular momentum at infinity, while maintaining balance; 
(3) The system cleanly demonstrates rotational frame-dragging when a
black hole with vanishing Komar angular momentum is rotating as
the black ring drags the surrounding spacetime. 
Possible generalizations include multiple rings of
saturn as well as doubly spinning black saturn configurations.   

\end{abstract}

\end{titlepage}

\tableofcontents

\newpage

\section{Introduction}

Multi-black hole spacetimes play an interesting role in
black hole physics.
A central question is how to keep a configuration of multiple black
holes in equilibrium.  
Two Schwarzschild black holes attract
each other and cannot be in equilibrium without external forces to
hold them in place.
The simplest way
to achieve a stationary balanced configuration is by adding enough
electric charge to each black hole, so that the electromagnetic
repulsion exactly cancels the gravitational attraction. In 3+1
dimensions, the resulting
solution, and its generalization to multiple black holes, is of course
the well-known extremal multi-Reissner Nordstrom black hole solution
\cite{multiRN}.   
 
For asymptotically flat vacuum solutions, rotation seems to be the
only candidate for keeping black holes apart. However, for the
3+1-dimensional axisymmetric double Kerr solution \cite{KN}, the
spin-spin interaction \cite{wald} is not sufficiently strong to balance
the gravitational attraction of black holes with regular horizons
\cite{CH,WDCH,MR,Stephani:2003tm}.    
Hence multi-Kerr black hole spacetimes are not in equilibrium, but suffer
from singular struts which provide the pressure to keep the black
holes apart \cite{WDCH}.    

We present here a 4+1-dimensional stationary vacuum solution for
which angular momentum does provide sufficient force to keep two
black objects apart. The possibility of balanced, regular multi-black
hole vacuum spacetimes can be motivated as follows. 
The five-dimensional vacuum Einstein's equations admit black ring
solutions \cite{ER} which have horizons of topology $S^2 \times S^1$. 
Rotation prevents the black ring from collapsing.
Very thin black rings are kept in equilibrium by a Newtonian
force balance between a string-like tension and a centrifugal force
arising from the rotation \cite{EEV} (see also \cite{Hovdebo}). With
this Newtonian balance in mind, it is 
natural to ask if rotation provides a sufficiently strong force to also
keep a black ring  
in equilibrium in an ``external potential''. This could for instance
be in the gravitational field of a Myers-Perry black hole
\cite{MPBH} at the center of the black ring. Our solution realizes
this possibility: a black ring balanced by rotation around a concentric
spherical black hole in an asymptotically flat spacetime. 
We call this balanced configuration
\emph{black saturn}.

It should be emphasized that the black hole and the black ring
generally have strong gravitational backreactions, so that 
only for very thin black
rings with large $S^1$ radius does the motivation of a black ring in
an external potential apply. 
On the other hand, the gravitational
interactions between the two objects 
give rise to interesting phenomena, such as frame-dragging, which
we examine in detail.
We summarize here a selection of physical properties of black saturn:

\begin{itemize}
\item[-] {\sl Continuous non-uniqueness}: The total mass $M$ and
  angular momentum $J$ measured at infinity can be distributed
  continuously between the two black objects in the balanced saturn
  configuration. Thus the solution exhibits 2-fold continuous
  non-uniqueness. 
  An additional discrete non-uniqueness exists in regimes that admit
  both thin and fat black rings. 
\item[-] {\sl Counter rotation}:
  The black ring and the $S^3$ black hole have
  independent rotation parameters, and they can be co-rotating as well
  as counter-rotating while maintaining balance. (We define co- and
  counter-rotation in terms of the relative sign of the angular
  velocities.)
\item[-] {\sl Non-uniqueness of the 4+1d Schwarzschild black hole}:
  Strikingly, the black ring and $S^3$ black hole can be
  counter-rotating to give zero total ADM angular momentum at
  infinity. 
  This means that the 4+1-dimensional
  Schwarzschild-Tangherlini black hole is not the only asymptotically
  flat black hole solution with $J=0$ at infinity; in fact the $J=0$
  black saturn configurations are 2-fold continuously non-unique.
  
  The existence of the $J=0$ black saturn solutions does not
  contradict the uniqueness theorem \cite{Gibbons:2002bh} that the
  Schwarzschild black hole is the only static
  asymptotically flat vacuum black hole solution; the reason is simply
  that black saturn, while being stationary, is non-static. 
 
  We also conclude that the slowly spinning Myers-Perry black hole is not
  unique; allowing for non-connected horizons one can
  get around the perturbative results of \cite{kodama}. 

\item[-] {\sl Rotational frame-dragging}: The gravitational
  interaction between the black ring and the $S^3$ black hole
  manifests itself in form of rotational frame-dragging. This is most
  cleanly illustrated when the intrinsic angular momentum (measured
  by the Komar integral) of the $S^3$ black hole is set to zero,
  $J_\rom{Komar}^\rom{BH} = 0$. The angular velocity
  $\Omega^\rom{BH}$, however, is not zero but follows the behavior of
  the angular velocity $\Omega^\rom{BR}$ of the black ring. We interpret
  this as frame-dragging: the rotating black ring 
  drags the spacetime around with it, and in effect the black hole
  rotates too, despite having no intrinsic spin,
  $J_\rom{Komar}^\rom{BH} =  0$. 
  It is exciting to have access to rotational frame-dragging in an
  exact solution.  
\item[-] {\sl Countering frame-dragging}: Counter-rotation makes it
  possible to tune the intrinsic rotation $J_\rom{Komar}^\rom{BH}$ of
  the $S^3$ black hole, so that it ``cancels" the effect of 
  dragging caused by the surrounding black ring. This gives a solution for
  which the angular velocity of the black hole vanishes:
  $\Omega^\rom{BH}=0$ while $J_\rom{Komar}^\rom{BH} \ne 0$.

\end{itemize} 

We have found no black saturn configurations
($J=0$ or $J$ nonzero) for which the total horizon area of the $S^3$
black hole and black ring exceeds the area $a_\rom{H}^\rom{Schw}$ of 
the static 4+1-dimensional Schwarzschild black hole of the same ADM
mass,\footnote{This observation leads to the
general expectation that for fixed mass the entropy of the
$d$-dimensional Schwarzschild black hole serves as an upper bound on
the total entropy in any stationary $d$-dimensional asymptotically
flat balanced black hole vacuum spacetime.}
 however there are saturn configurations with total area 
arbitrarily close to $a_\rom{H}^\rom{Schw}$ for any value of $J$.  
The resulting phase diagram of 4+1-dimensional black holes is
discussed in more detail in \cite{EEF2}, where black saturn
thermodynamics is also studied. 

It is worth noting that for 4+1-dimensional asymptotically flat black
hole spacetimes the continuous non-uniqueness will go much 
further than the 2-fold continuous non-uniqueness of the simple black
saturn system presented here. 
An obvious generalization of our solutions includes multiple
rings of saturn. As argued above, the total mass and angular momentum
can be distributed continuously between the $n$ black objects in such a
spacetime, subject to balance conditions, and the result is
$2(n-1)$-fold continuous non-uniqueness. Including the
second angular momentum gives doubly spinning multiple black
saturns with $3(n-1)$-fold continuous non-uniqueness, also for
the $J_1=J_2=0$ configurations. 
If, as anticipated, the total area is
bounded by $a_\rom{H}^\rom{Schw}$ for given total mass, each
component of an $n$-black hole system will necessarily have smaller
area as $n$ increases. 

Supersymmetric black hole solutions with one or more concentric
balanced black rings around a rotating $S^3$ black hole were
constructed by Gauntlett and Gutowski \cite{GG}. Being supersymmetric,
the solutions are extremally charged and saturate the BPS bound of 
4+1-dimensional supergravity with $U(1)$ vector
multiplets. For the supersymmetric solutions, it is not possible to
observe dragging effects or counter-rotation, as we do for our
non-supersymmetric vacuum solutions, because the supersymmetric
solutions have vanishing angular velocities. 
The first order nature of the supersymmetry conditions
\cite{GGHPR,GMR} makes the construction of multi-black hole solutions
a fairly straightforward superposition of harmonic functions. For
non-supersymmetric black holes we do not have this luxury, and instead we
have to solve the full second order Einstein's equations. 

The black saturn solution is found using the inverse scattering
method. This solution generating method was first adapted to
Einstein's equations by Belinsky and Zakharov \cite{BZ1,BZ2}, and has
been used extensively to generate four-dimensional vacuum solutions
(see for instance \cite{BV} and references therein). 
Recently, the inverse scattering method, and closely related solution
generating techniques, have been applied to generate five-dimensional
rotating black hole vacuum solutions. The Myers-Perry black hole with
two independent rotation parameters was constructed by a smart
implementation of the inverse scattering method by Pomeransky
\cite{Pom}. Also, the unbalanced black ring with rotation on 
the $S^2$ was constructed \cite{Mishima:2005id,Tom1}; this solution
was constructed independently in \cite{PF} without use of solution
generating techniques. The original balanced $S^1$ rotating
black ring \cite{ER} has also been constructed by these methods 
\cite{Iguchi:2006rd,Tom2}. Most recently, Pomeransky and Sen'kov
have succeeded in constructing a doubly-spinning black ring
solution \cite{Pom2} using the inverse scattering method (numerical
results were also obtained recently \cite{kudoh}). 

\vspace{3mm}
We briefly review relevant aspects of the inverse scattering method in
section \ref{s:IMS}, where we also provide details of the construction
of the black saturn solution. Section \ref{s:analysis} contains an
analysis of the solution, including computations of the physical
parameters and the balance condition. 
The physics of the black saturn system is studied in section
\ref{s:physics}. 
Open questions are discussed in section \ref{s:discussion}.


\section{Construction of the solution}
\label{s:IMS}

We review in section \ref{s:review} the inverse scattering method 
with focus on the Belinsky-Zakharov (BZ) $n$-soliton transformations
\cite{BZ1,BZ2} (a detailed review can be found in the book
\cite{BV}). In section  \ref{s:BZing} we discuss the seed solution and
generate the black saturn solution by soliton 
transformations. The final result for the
metric is presented in section \ref{s:saturnsol}.

\subsection{The inverse scattering method}
\label{s:review}

The inverse scattering method can used as a solution generating method
for stationary axisymmetric spacetimes. These are
$D$-dimensional spacetimes with $D-2$ commuting Killing vector fields,
one of which is time.
The method allows construction of new solutions from known ones by
means of purely algebraic manipulations.  

We write the $D$-dimensional stationary axisymmetric
spacetime as
\begin{equation}
  ds^2=G_{ab}\,dx^a dx^b+e^{2\nu}\left(d\rho^2+dz^2\right)\;,
\label{eqn:stmetric} 
\end{equation}
where $a,b=1,\dots,D-2$ and
all compoments of the metric are functions of $\rho$ and $z$
only: $G_{ab}=G_{ab}(\rho,z)$ and $\nu=\nu(\rho,z)$. Without loss of
generality the 
coordinates can be chosen such that
\begin{equation}
  \det G= - \rho^2\, .
  \label{eqn:detG}
\end{equation}
Then Einstein's equations separate into two groups, one for the
$(D-2) \times (D-2)$ matrix $G$,
\begin{equation}
  \partial_\rho U+\partial_z V=0\;, 
  \label{eqn:eqG}
\end{equation}
where
\bea
 \label{UV}
 U=\rho\, (\partial_\rho G)\, G^{-1} \, , \hspace{5mm}
 V=\rho\, (\partial_zG)\, G^{-1} \, ,
\eea
and the other for the metric factor $e^{2\nu}$, 
\begin{equation}
  \label{eqn:eqnu}
  \partial_\rho\nu=\frac{1}{2}
  \left[-\frac{1}{\rho}+\frac{1}{4\rho}\Tr(U^2-V^2)\right]\,,
  \hspace{5mm}
  \partial_z\nu=\frac{1}{4\rho}\Tr(UV)\, .
\end{equation}
The equations \eqref{eqn:eqnu} for $\nu$ satisfy the integrability
condition $\partial_\rho\partial_z \nu=\partial_z\partial_\rho\nu$ 
as a consequence of \eqref{eqn:eqG}. Hence, once a solution
$G_{ij}(\rho,z)$ to \eqref{eqn:eqG} is found, one can determine 
$\nu(\rho,z)$ by direct integration. 

The matrix equations \reef{eqn:detG} and \eqref{eqn:eqG} form a
completely integrable system, meaning that one can find a set of
spectral equations (a ``Lax  pair'' or ``L-A pair'') whose
compatibility conditions 
are exactly \reef{eqn:detG} and \eqref{eqn:eqG}. 
The spectral equations for \reef{eqn:detG} and \eqref{eqn:eqG} are 
\begin{eqnarray}
  D_1\Psi=\frac{\rho V-\lambda U}{\lambda^2+\rho^2}\Psi\;, 
  && D_2\Psi=\frac{\rho U+\lambda V}{\lambda^2+\rho^2}\Psi\;,
  \label{eqn:LAeqs}
\end{eqnarray}
with commuting differential operators $D_1$ and $D_2$ given by
\begin{eqnarray}
  D_1=\partial_z-\frac{2\lambda^2}{\lambda^2+\rho^2}\partial_\lambda\;,&& 
  D_2=\partial_\rho+\frac{2\lambda\rho}{\lambda^2+\rho^2}\partial_\lambda\;,
\end{eqnarray}
The complex spectral parameter $\lambda$ is 
independent of $\rho$ and $z$, and
the generating function $\Psi(\lambda,\rho,z)$ is a $(D-2)\times
(D-2)$ matrix such that $\Psi(0,\rho,z)=G(\rho,z)$. 

The linearity of \reef{eqn:LAeqs} allows algebraic construction of new
solutions from known solutions based on the ``dressing method''. Given
a known ``seed'' solution $G_0$, one constructs the corresponding matrices
$U_0$ and $V_0$ in \reef{UV}, and determines a generating matrix
$\psi_0$ which solves \reef{eqn:LAeqs} with $U_0$ and $V_0$. Then one
seeks a new solution 
of the form
\begin{equation}
 \Psi=\chi\, \Psi_0\, ,
 \label{eqn:dress}
\end{equation}
where $\chi=\chi(\lambda,\rho,z)$ is the dressing matrix.
Inserting \reef{eqn:dress} into \eqref{eqn:LAeqs} now gives a set of
equations for $\chi$. The matrix  $\chi$ is further constrained by
requiring that the new metric $G=\Psi(\lambda=0,\rho,z)$ is real and
symmetric.  

We are here interested in so-called ``$n$-soliton'' dressing matrices,
which are characterized by having $n$ simple poles in the complex
$\lambda$-plane, and we further restrict to cases where the poles are
located on the real axis; this determines the location of the poles to
be \cite{BZ1,BZ2,BV} 
\begin{equation}
  \label{mutilde}
   \tilde\mu_k=\pm\sqrt{\rho^2+(z-a_k)^2}-(z-a_k)\;, 
\end{equation}
where $a_k$ are $n$ real constants. We refer to the ``$+$" pole as
a soliton and denote it by $\mu_k$, while the ``$-$" pole is an
anti-soliton denoted by $\bar{\mu}_k$. Note $\mu_k \bar{\mu}_k = -\rho^2$.

In addition to the $n$ real constants  $a_k$, an $n$-soliton
transformation is determined by $n$ arbitrary constant real
$(D-2)$-component vectors $m_{0}^{(k)}$, which we shall refer to as
the BZ vectors. The components of these vectors will be called BZ
parameters. In our applications, the BZ vectors control the
addition of angular momentum to a static seed solution. 

Given a seed solution $G_0$, the $n$-soliton transformation yields a
new solution $G$ with components
\begin{equation}
  G_{ab}=(G_0)_{ab}-
  \sum_{k,l=1}^{n}\frac{ 
  (G_0)_{ac}\, m_c^{(k)}\,  (\Gamma^{-1})_{kl}\;  m_d^{(l)}\, (G_0)_{db}}
                       {\tilde\mu_k\tilde\mu_l}\; .  
\label{eqn:unG}
\end{equation}
(Repeated spacetime indices $a,b,c,d=1,\dots,D-2$ are summed.)
The components of the vectors $m^{(k)}$ are
\begin{equation}
  m_a^{(k)}=m_{0b}^{(k)}\left[\Psi_0^{-1}(\tilde\mu_k,\rho,z)\right]_{ba}\, ,
\end{equation}
where $\Psi_0$ is the generating matrix which solves \reef{eqn:LAeqs}
with $U_0$ and $V_0$ determined by $G_0$ as in \reef{UV}, and
$m_{0b}^{(k)}$ are the BZ parameters.

The symmetric matrix $\Gamma$ is defined as
\begin{equation}
  \label{gamma}
  \Gamma_{kl}=\frac{m_a^{(k)}\, (G_0)_{ab}\, m_b^{(l)}}
   {\rho^2+\tilde\mu_k \tilde\mu_l}\, ,
\end{equation}
and the inverse $\Gamma^{-1}$ of $\Gamma$ appears in \reef{eqn:unG}.

The new matrix $G$ of \reef{eqn:unG} does not obey \eqref{eqn:detG};
instead, an $n$-soliton transformation gives 
\begin{equation}
\det
G=(-1)^n\rho^{2n}\left(\prod_{k=1}^n{\tilde\mu_k}^{\;-2}\right)\det
G_0\;, 
\label{renG}
\end{equation}
with $\det G_0=-\rho^2$. One can deal with this problem and
obtain a physical solution $G^{(\textrm{ph})}$ 
such that $\det G^{(\textrm{ph})}=-\rho^2$,
by multiplying $G$ by a suitable factor of $\rho$ and 
$\tilde\mu_k$'s. 
In four spacetime dimensions, this method of uniform renormalization
works well and allows one to construct for instance
(multi)Kerr-NUT solutions from just flat Minkowski space. In
higher dimensions, however, uniform renormalization typically leads to
nakedly singular solutions.

One way around this
problem is to restrict the soliton transformation to a 
$2\times 2$ block of the seed solution and perform uniform
renormalization on this block. 
This has been applied to reproduce
black ring solutions with a single angular momentum
\cite{Tom1,Tom2}. 
The drawback of this method
is clearly that it can only produce solutions with rotation in at most
a single plane. This would be sufficient for our purposes here,
but we prefer 
to present the solution generating method in a more general setting so
as to facilitate generalization of our black saturn solution to
include angular momentum in two independent planes. We therefore 
follow the strategy of \cite{Pom} which is applicable in any spacetime
dimension and does not suffer from the above-mentioned limitations. 

The idea is to note that the factor multiplying $\det G_0$ in
\reef{renG} is independent of the BZ 
vectors $m_0^{(k)}$. Start with a diagonal seed solution $(G_0,e^{2\nu_0})$ and
remove first solitons with trivial BZ parameters (so as to not
introduce any off-diagonal components in the matrix $G$). Then add
back the \emph{same} solitons but now with general BZ parameters. 
The resulting solution $G$ satisfies $\det G = -\rho^2$ by
construction. Moreover, the  
metric factor $e^{2\nu}$ of the full solution can easily be
obtained from the seed $G_0$ as \cite{Pom}
\begin{equation}
  e^{2\nu}=e^{2\nu_0}~\frac{\det \Gamma_{kl}}{\det\Gamma_{kl}^{(0)}}\;,
  \label{eqn:cf}
\end{equation}
where $\Gamma^{(0)}$ and $\Gamma$ are constructed as in \reef{gamma} 
using $G_0$ and $G$, respectively.

We now turn from the general discussion to the construction
of the black saturn solution.

\subsection{Seed and soliton transformation for black saturn}
\label{s:BZing}

{}For the analysis of axisymmetric solutions we make use of
the results of \cite{ER2,Har}. We refer to these
papers for general discussions of higher-dimensional Weyl solutions
and the analysis of the corresponding rod configurations.

\begin{figure}[t!]
\begin{picture}(0,0)(0,0)
\put(-3,29){$t$}
\put(-3,15){$\phi$}
\put(-3,0){$\psi$}
\put(22,-5){$a_1$}
\put(37,-5){$a_5$}
\put(51,-5){$a_4$}
\put(66,-5){$a_3$}
\put(79,-5){$a_2$}
\end{picture}
\centering{\includegraphics[width=4in]{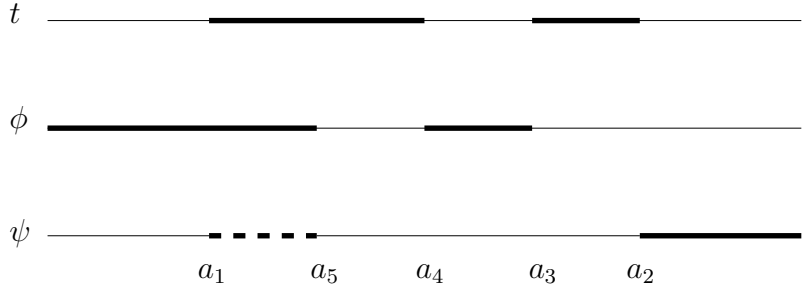}}
\bigskip
\caption{Sources for the seed metric $G_0$. The solid rods have
  positive density and the dashed rod has negative density. The rods
  are located at the $z$-axis with $\rho=0$ and add up to an infinite
  rod with uniform density such that $\det G_0 = -\rho^2$. The
labeling of the rod endpoints is a
little untraditional, but is simply motivated by the fact that we are
going to use the inverse scattering method to add solitons at $z=a_1$,
  $a_2$ and $a_3$.} 
\label{fig:blacksaturn}
\end{figure}

The rod configuration for the seed of black saturn is shown in figure
\ref{fig:blacksaturn}. The thick solid black lines correspond to rod
sources of uniform density $+1/2$, whereas the dashed line
segment corresponds to a rod source of uniform negative density
$-1/2$. The rods in the $t$ direction correspond to black
hole horizons. Note that for $a_1=a_5$ the negative rod is eliminated
and the solution describes a static black ring around an $S^3$ black
hole. This is an unbalanced configuration with a conically singular
membrane keeping the black ring and the $S^3$ black hole apart.
The negative density rod is included in
order to facilitate adding angular momentum to the black ring. 

Using the techniques of \cite{ER2} we construct the full
4+1-dimensional vacuum 
solution corresponding to the rod configuration in figure
\ref{fig:blacksaturn}. We find
\bea
   \label{seed1}
   G_0 = \diag\left\{ 
   - \frac{\mu_1\, \mu_3}{\mu_2\,  \mu_4},\,
   \frac{\rho^2\,  \mu_4}{\mu_5 \, \mu_3},\,
   \frac{\mu_5\,  \mu_2}{\mu_1}
   \right\} \, ,
   ~~~~~
   \det{G_0}=-\rho^2 \, .
\eea
The first term in $G_0$ corresponds to
the $tt$-component, the second to the $\phi\phi$-component and the
third to the $\psi\psi$-component. The $\mu_i$ are ``solitons'' as introduced in \reef{mutilde}, i.e.~
\bea
  \label{themuis}
  \mu_i = \sqrt{\rho^2+(z-a_i)^2} - (z-a_i) \, ,
\eea
where the $a_i$ are the rod endpoints in figure \ref{fig:blacksaturn}.
The metric factor $e^{2\nu}$ of the seed can written 
\bea
   \label{Hexp2nuB}
   e^{2\nu} =  k^2\,
   \frac{\mu_2 \, \mu_5
        (\rho^2+\mu_1\,\mu_2)^2(\rho^2+\mu_1\,\mu_4)
        (\rho^2+\mu_1\,\mu_5)(\rho^2+\mu_2\mu_3)
        (\rho^2+\mu_3\,\mu_4)^2 (\rho^2+\mu_4\,\mu_5)}
     {\mu_1 (\rho^2+\mu_3\mu_5)(\rho^2+\mu_1\,\mu_3)
      (\rho^2+\mu_2\,\mu_4)(\rho^2+\mu_2\,\mu_5)
     \prod_{i=1}^{5}(\rho^2+\mu_i^2)}\, .
     ~~~        
\eea
The integration constant $k$ will be fixed in section \ref{s:asympt}
   for the full black saturn solution.

We assume the ordering 
\bea
  \label{order}
  a_1 \le a_5 \le a_4 \le a_3 \le a_2 \, 
\eea
of the rod endpoints.\footnote{If instead we had chosen the different ordering 
$a_5  \le a_1 \le a_4 \le a_3 \le a_2$, then there would have been no
negative density rod, 
and the solution \reef{seed1} and \reef{Hexp2nuB} would describe two
$S^3$ black holes and two conical singularities, one for each of the
two finite rods in the angular directions. We will not use this
ordering, but always take the rod endpoints to satisfy \reef{order}.} 

The solution \reef{seed1} and \reef{Hexp2nuB}  
with the ordering \reef{order} is singular
and not in itself of physical interest. 
However, with a 1-soliton transformation we add an anti-soliton which
mixes the $t$ and $\psi$ directions in such a way that 
the negative density rod moves to the $t$-direction and cancels the
segment $[a_1,a_5]$ 
of the positive density rod.
It turns out that this leaves a naked singularity at $z=a_1$, but
choosing the BZ vector appropriately completely eliminates that
singularity (see section \ref{s:rods}).  
Taking $a_2=a_3$ in the seed solution, this 1-soliton transformation
gives the $S^1$ rotating black ring of \cite{ER}.   
We show this explicitly in appendix \ref{app:BR}.

Keeping $a_3 < a_2$, the above sketched 1-soliton transformation gives 
a rotating black ring around an $S^3$ black hole. This configuration
can be balanced and we study its physical properties in detail in
section \ref{s:cc2zero}. Including two more soliton transformations allow
us to give the $S^3$ black hole independent rotation in two planes. 
The steps of generating the black saturn solution by a 3-soliton
transformations are as follows: 

\begin{enumerate}
\item Perform the following three 1-soliton transformations on the
  seed solution \reef{seed1}: 
\begin{itemize}
  \item Remove an anti-soliton at $z=a_1$ with trivial BZ vector
  (1,0,0); this is equivalent to dividing $(G_0)_{tt}$ by
  $-\rho^2/\bar{\mu}_1^2 =-\mu_1^2/\rho^2$. 
  \item Remove a soliton at $z=a_2$ with trivial BZ vector (1,0,0);
    this is equivalent to dividing $(G_0)_{tt}$ by
    $\left(-\rho^2/\mu_2^2\right)$. 
  \item Remove an anti-soliton at $z=a_3$ with trivial BZ vector (1,0,0);
    this is equivalent to dividing $(G_0)_{tt}$ by
    $-\rho^2/\bar{\mu}_3^2 =-\mu_3^2/\rho^2$.
\end{itemize}
The result is the metric matrix
\bea
   G_0'=\diag\left\{\frac{\rho^2 \mu_2}{\mu_1\mu_3\mu_4},
   \frac{\rho^2\mu_4}{\mu_3\mu_5},     
   \frac{\mu_2\mu_5}{\mu_1}\right\}\, .
\eea
\item Rescale $G_0'$ by a factor of $\frac{\mu_1\mu_3}{\rho^2 \mu_2}$
  to find 
\begin{equation}
  \tilde{G}_0=\frac{\mu_1\mu_3}{\rho^2 \mu_2}G_0'
  =\diag\left\{\frac{1}{\mu_4},
   \frac{\mu_1\mu_4}{\mu_2\mu_5},
   -\frac{\mu_3}{\bar{\mu}_5}\right\}\;,
\end{equation}
where $\bar\mu_5=-\rho^2/\mu_5$. This will be the seed for the next
soliton transformation. 
\item The generating matrix 
\bea
  \tilde{\Psi}_0(\lambda,\rho,z) 
  =\diag\left\{\frac{1}{(\mu_4-\lambda)},
    \frac{(\mu_1-\lambda)(\mu_4-\lambda)}
    {(\mu_2-\lambda)(\mu_5-\lambda)},
    -\frac{(\mu_3-\lambda)}{(\bar{\mu}_5-\lambda)}\right\}
\eea
solves \reef{eqn:LAeqs} with $\tilde{G}_0$.
Note $\tilde{\Psi}(0,\rho,z) = \tilde{G}_0$.
\item Perform now a 3-soliton transformation with $ \tilde{G}_0$ as
  seed: 
\begin{itemize}
\item Add an anti-soliton at $z=a_1$ (pole at $\lambda=\bar\mu_1$)
 with BZ vector  
 $m_0^{(1)} =(1,0,c_1)$, 
\item Add a soliton at  $z=a_2$ (pole at $\lambda=\mu_2$) with BZ vector 
 $m_0^{(2)} =(1,0,c_2)$, and
\item Add an anti-soliton at $z=a_3$ (pole at $\lambda=\bar\mu_3$)
 with BZ vector  
 $m_0^{(3)} =(1,b_3,0)$.
 \end{itemize}
Denote the resulting metric $\tilde{G}$. The constants $c_1$, $c_2$,
and $b_3$ are the BZ parameters of the transformation. 
\item Rescale $\tilde{G}$ to find
\bea
  G = \frac{\rho^2 \mu_2}{\mu_1\mu_3} \tilde{G} \, .
\eea
This is needed to undo the rescaling of step 2, so that $\det G =
-\rho^2$. 
\item Construct $e^{2\nu}$ using  \reef{eqn:cf}. Note that $\Gamma$
  was found in the process of constructing $G$ and that
  $\Gamma_0=\Gamma\big|_{c_1=c_2=b_3=0}$. 
The result $(G, e^{2\nu})$ is the solution we want.
\end{enumerate}

Some comments are in order. 
First, the rescaling in step 2 is simply a choice of convenience that
yields a simple form for the generating matrix $\tilde{\Psi}_0$.
Secondly, with $c_1=c_2=b_3=0$, 
the effect of the 3-soliton transformation in step 4 is 
simply to undo the transformation of step 1. Since \reef{renG} is
independent of the BZ parameters $c_1$, $c_2$ and $b_3$, we are
guaranteed to have $\det G = \det G_0 =- \rho^2$, after step 5 has
undone the rescaling of step 2. 
Finally, in step 4 we could have added the
(anti-)solitons with general BZ vectors $m_0^{(k)} =
(a^{(k)},b^{(k)},c^{(k)})$ for $k=1,2,3$. However, 
$b^{(k)} \ne 0$, $k=1,2$, or $c^{(3)} \ne 0$ lead to irremovable
singularities and we therefore set 
$b^{(1)} = b^{(2)} = c^{(3)} = 0$. Finally, the solution is invariant under
rescalings of the BZ vectors, $m_0^{(k)} \to \sigma_k \, m_0^{(k)}$
(no sum on $k$) for any nonzero $\sigma_k$, and we use the
scaling freedom to set $a^{(k)}=1$ without loss of generality. 

In this paper we focus entirely on the black saturn solution with
angular momentum only in a single plane, so we set $b_3=0$ in the
following. The more general solution with $b_3 \ne 0$ remains to be
analyzed.  

\subsubsection*{2-soliton transformation}
With $b_3=0$ the transformation described above is essentially a
2-soliton transformation. In fact, the saturn solution with $b_3=0$
can be produced by a 2-soliton transformation in much the same way as
above. The resulting metric takes a slightly different form, but can
be shown, using the explicit form of the $\mu_i$'s in \reef{themuis},
to be identical to the metric resulting from the 3-soliton
transformation after a constant rescaling of the BZ parameters $c_1$
and $c_2$.


\subsection{Saturn solution}
\label{s:saturnsol}
The black saturn solution constructed by the above 3-soliton
 transformation with $b_3=0$ can be written\footnote{After  
 performing the BZ transformation, we shift 
 $t$ as $t \to t-q\, \psi$ in order to ensure asymptotic flatness. At
 this point $\psi$ is not assumed  
  to be periodic so the shift does not effect the 
  global structure of the solution. The periodicities of
  $\psi$ and $\phi$ will be fixed in section \ref{s:analysis}. We have
  also reversed the sense of rotation by taking $\psi \to -\psi$.}  
\bea
  \label{SaturnMetric}
  ds^2 =
  -\frac{H_y}{H_x} \Big[dt + \Big(\frac{\omega_\psi}{H_y}+q\Big) \,
  d\psi \Big]^2 
  + H_x \bigg\{ k^2 \, P \Big( d\rho^2 + dz^2 \Big) 
       + \frac{G_y}{H_y} \, d\psi^2 + \frac{G_x}{H_x}\, d\phi^2 \bigg\} \, .
\eea
For convenience we have chosen to write $e^{2\nu} = k^2\, H_x\, P$.
Here $k$ is the integration constant for the metric factor
  $e^{2\nu_0}$ given in \reef{Hexp2nuB}, and $G_{x,y}$, $H_{x,y}$, and
  $P$ are functions of $\rho$ and $z$ which will be given below. The
  constant $q$ is included in order to ensure asymptotic flatness (we
  determine the value of $q$ in the analysis of section
  \ref{s:asympt}).  
 
With $b_3=0$, our soliton transformations leave the $\phi\phi$-part
of the metric 
invariant, so from the static seed \reef{seed1} we have 
\bea
  G_x = (G_0)_{\phi\phi}= \frac{\rho^2\mu_4}{\mu_3\, \mu_5} \, .
\eea
The metric \reef{SaturnMetric} involves the functions
\bea
  P =  (\mu_3\, \mu_4+ \rho^2)^2
      (\mu_1\, \mu_5+ \rho^2)
      (\mu_4\, \mu_5+ \rho^2) \, ,
      \label{eqn:defP}
\eea
and
\bea
   H_x &=& F^{-1} \, 
   \bigg[ M_0 + c_1^2 \, M_1 + c_2^2\,  M_2 
   +  c_1\, c_2\, M_3 + c_1^2 c_2^2\, M_4 \bigg] \, , \label{eqn:defHx}\\[2mm]
    H_y &=& F^{-1} \, 
   \frac{\mu_3}{\mu_4}\, 
   \bigg[ M_0 \frac{\mu_1}{\mu_2} 
   - c_1^2 \, M_1 \frac{\rho^2}{\mu_1\,\mu_2} 
   - c_2^2\,  M_2 \frac{\mu_1\,\mu_2}{\rho^2}
   +  c_1\, c_2\, M_3 
   + c_1^2 c_2^2\, M_4 \frac{\mu_2}{\mu_1} \bigg] \, ,
\eea
where
\bea
  M_0 &=& \mu_2\, \mu_5^2 (\mu_1-\mu_3)^2 (\mu_2-\mu_4)^2
   (\rho^2+\mu_1\,\mu_2)^2(\rho^2+\mu_1\,\mu_4)^2
   (\rho^2+\mu_2\,\mu_3)^2 \, , \\[2mm]
  M_1 &=& \mu_1^2 \, \mu_2 \, \mu_3\, \mu_4 \, \mu_5 \, \rho^2\,
  (\mu_1-\mu_2)^2 (\mu_2-\mu_4)^2(\mu_1-\mu_5)^2
  (\rho^2+\mu_2\,\mu_3)^2  \, , \\[2mm]
  M_2 &=& \mu_2 \, \mu_3\, \mu_4 \, \mu_5 \, \rho^2\,
  (\mu_1-\mu_2)^2 (\mu_1-\mu_3)^2
  (\rho^2+\mu_1\,\mu_4)^2(\rho^2+\mu_2\, \mu_5)^2  \, ,\\[2mm]
  M_3 &=& 2 \mu_1 \mu_2 \, \mu_3\, \mu_4 \, \mu_5 \,
  (\mu_1-\mu_3) (\mu_1-\mu_5)(\mu_2-\mu_4)
  (\rho^2+\mu_1^2)(\rho^2+\mu_2^2) \nonumber \\[1mm]
  &&\hspace{5cm} \times
  (\rho^2+\mu_1\,\mu_4)(\rho^2+\mu_2\, \mu_3)
  (\rho^2+\mu_2\, \mu_5)  \, ,\\[2mm]
  M_4 &=& \mu_1^2 \, \mu_2\, \mu_3^2 \, \mu_4^2 \,
  (\mu_1-\mu_5)^2
  (\rho^2+\mu_1\,\mu_2)^2(\rho^2+\mu_2\, \mu_5)^2  \, ,
\eea
and
\bea
  F &=& \mu_1\, \mu_5\,  (\mu_1-\mu_3)^2(\mu_2-\mu_4)^2
  (\rho^2+\mu_1\,\mu_3)
  (\rho^2+\mu_2\,\mu_3)
  (\rho^2+\mu_1\,\mu_4) \nonumber\\
  && \hspace{3cm} \times
  (\rho^2+\mu_2\,\mu_4)
  (\rho^2+\mu_2\,\mu_5)
  (\rho^2+\mu_3\,\mu_5)
  \prod_{i=1}^5 (\rho^2+\mu_i^2) \, .
\eea
Finally we have
\bea
  G_y = \frac{\mu_3\, \mu_5}{\mu_4} \, ,
\eea
and the off-diagonal part of the metric is given by
\bea
  \omega_\psi
  &=&
  2 \frac{
     c_1\, R_1\, \sqrt{M_0 M_1}
    -c_2\, R_2\, \sqrt{M_0 M_2}
    +c_1^2\,c_2\, R_2\, \sqrt{M_1 M_4}
    -c_1\,c_2^2\, R_1\, \sqrt{M_2 M_4}
  }
  {F \sqrt{G_x}} \, .
\eea
Here $R_i = \sqrt{\rho^2 + (z-a_i)^2}$.

Setting $c_1 = c_2 = 0$ gives $\omega_\psi=0$
  and $G_y H_x/H_y = \mu_2\, \mu_5/\mu_1= (G_0)_{\psi\psi}$. The 
  full solution can be seen to simply reduce to the seed solution
  \reef{seed1} and \reef{Hexp2nuB} in this limit.  

Taking $c_1=0$ and then setting $a_1=a_5=a_4$ we obtain the singly
spinning Myers-Perry black hole, which was constructed similarly in
\cite{Pom}. For details, see appendix \ref{app:MP}.  
Taking instead $c_2=0$ and then setting $a_2=a_3$ we obtain the $S^1$
spinning black ring of \cite{ER}. Appendix \ref{app:BR} presents the
explicit coordinate transformation from Weyl coordinates $(\rho,z)$ to
ring coordinates $(x,y)$. The black ring was obtained in
\cite{Iguchi:2006rd} and \cite{Tom2} with a different transformation
which involved two solitons and started with a different seed
metric. The 1-soliton transformation used here appears to be
simpler.\footnote{We thank Roberto Emparan for sharing with us the
  idea of obtaining the $S^1$-spinning black ring by a 1-soliton
  transformation.} 

It is useful to note  that the only effect of changing the
signs of both BZ parameters $c_1$ and $c_2$, taking $(c_1,c_2) \to
(-c_1,-c_2)$, is a change of sense of the overall direction of
rotation, i.e.~the only effect is $G_{t\psi} \to -G_{t\psi}$. 

The metric \reef{SaturnMetric} is sufficiently complicated that it
is difficult to check algebraically that the Einstein vacuum equations
are solved. We have resorted to numerical methods in order to  
check the vanishing of all components of the Ricci tensor.

Next we present an analysis of the main properties of the black
saturn solution.


\section{Analysis}
\label{s:analysis}

We introduce a convenient parameterization of the
solution, and then analyze the rod structure.
The BZ parameter $c_1$ will be fixed in order to eliminate the
singularity left-over from the negative density rod of the seed
solution. Next it is shown that the solution is asymptotically
flat. Regularity is analyzed and the 
balance condition obtained by elimination of a
conical singularity. We analyze the horizon structure, and compute a
number of physical quantities for the solution: the ADM mass and
angular momentum, as well as angular velocities, temperatures and
horizon areas of the two black holes. We compute the Komar integrals
for mass and angular momentum and obtain a Smarr relation.  
We study various limits of the solution, and we comment on the
analysis of closed timelike curves (of which we find none).

\subsection{Parameterization}
\label{sec:parametrization}

The seed solution \reef{seed1}-\reef{Hexp2nuB} contains five
dimensionfull parameters, namely the rod endpoints $a_i$,
$i=1,\dots,5$.
Since the whole rod configuration can be shifted along the
$z$-axis without changing the solution,
the description in terms of the $a_i$'s is redundant; in addition to
the ordering \reef{order} and the directions of the rods as given in
figure \ref{fig:blacksaturn} we only need the lengths of the rods.
It is useful to also take out the overall scale of the solution so
that the seed solution is described in terms of three dimensionless
parameters and an overall scale. 

We choose the overall scale $L$ to be\footnote{The coordinates $\rho$
  and $z$, and hence the rod endpoints $a_i$, have dimensions
  (length)$^2$.} 
\bea
  L^2 = a_2-a_1 \, ,
\eea
and we introduce three dimensionless parameters $\ka_i$ as
\bea
  \kappa_{i} = \frac{a_{i+2} - a_1}{L^2}
  \, ,~~~~~~\rom{for}~~i=1,2,3 \, .
\eea
As a consequence of the ordering \reef{order}, the $\ka_i$'s satisfy
\bea
  \label{order2}
  0 \le \ka_3 \le \ka_2< \ka_1 \le 1\, .
\eea
(We exclude $\ka_2 = \ka_1$ for the balanced solution for reasons
which will be apparent in section \ref{s:balance}.) 
We shift and scale the $z$ coordinate accordingly: set
\bea
  z = L^2 \bar{z} +a_1 \, .
\eea
Then $\bar{z}$ is dimensionless. As we shall see in the following, the
  black ring horizon is located at $\rho=0$ for 
$\bar{z} \in [\ka_3, \ka_2]$, 
and the $S^3$ black hole horizon at  
$\rho=0$ for $\bar{z} \in [\ka_1, 1]$. 

The new parameterization effectively corresponds to taking
\bea
  a_1 \to 0\, , ~~~~
  a_5 \to \ka_3 \, , ~~~~
  a_4 \to \ka_2 \, , ~~~~
  a_3 \to \ka_1 \, , ~~~~
  a_2 \to 1 \, ,
\eea
while carefully keeping track of the scale $L$.

\vspace{3mm}

The soliton transformations introduce the two dimensionfull BZ
parameters, $c_1$ and $c_2$.  
It is convenient to redefine the BZ parameter $c_2$ by
introducing the dimensionless parameter $\bc_2$ as
\bea
  \label{bc2}
  \bc_2 = \frac{c_2}{c_1(1-\ka_2)} \, .
\eea
With this parameterization many expressions for the physical
parameters simplify.


\subsection{Rod structure}
\label{s:rods}

The rod structure at $\rho = 0$ is illustrated in figure
\ref{fig:saturnrod}. Harmark \cite{Har} introduces the ``direction''
of a given rod as the zero eigenvalue eigenvector of the metric matrix
$G$ at $\rho=0$. The direction of each rod is indicated in figure
\ref{fig:saturnrod}(b). To summarize:

\begin{figure}[t]
\begin{picture}(0,0)(0,0)
\put(5,-5){\footnotesize $t$}
\put(5,-15){\footnotesize $\phi$}
\put(5,-25){\footnotesize $\psi$}
\put(86,-5){\footnotesize $t$}
\put(86,-15){\footnotesize $\phi$}
\put(86,-25){\footnotesize $\psi$}
\put(23.3,-29){\footnotesize$0$}
\put(32.6,-29){\footnotesize $\ka_3$}
\put(42.5,-29){\footnotesize $\ka_2$}
\put(52.8,-29){\footnotesize $\ka_1$}
\put(63,-29){\footnotesize$1$}
\put(104,-29){\footnotesize$0$}
\put(114,-29){\footnotesize $\ka_3$}
\put(124,-29){\footnotesize $\ka_2$}
\put(132.5,-29){\footnotesize $\ka_1$}
\put(143.2,-29){\footnotesize $1$}
\put(113,-2.5){\scriptsize$(1,0,\Omega_\psi^\rom{BR})$}
\put(133.5,-2.5){\scriptsize$(1,0,\Omega_\psi^\rom{BH})$}
\put(99,-13){\scriptsize$(0,1,0)$}
\put(124.5,-13){\scriptsize$(0,1,0)$}
\put(148,-23){\scriptsize$(0,0,1)$}
\put(33,-36){\footnotesize Figure \ref{fig:saturnrod}(a)}
\put(116,-36){\footnotesize Figure \ref{fig:saturnrod}(b)}
\end{picture}
  \begin{center}
      \includegraphics[width=7cm]{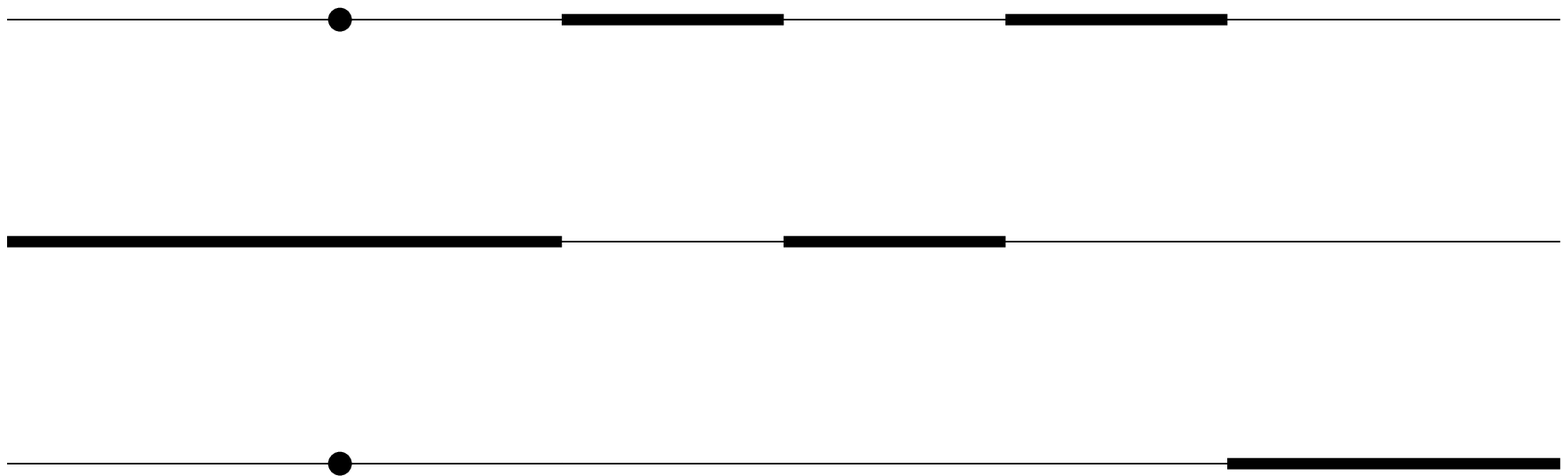}
      \hspace{8mm}
      \includegraphics[width=7cm]{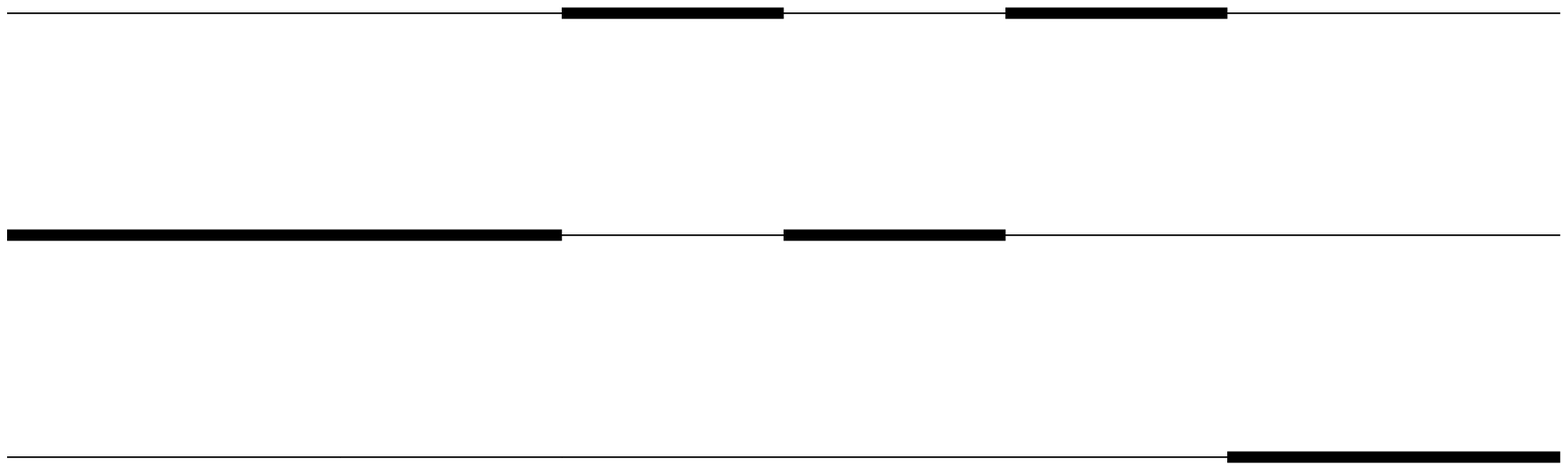}
\vspace{1cm}
    \caption{\small 
Rod structure of the black saturn solution. Note that the rods are
placed on the $\bar z$-axis, see section \ref{s:rods} for the
definition of $\bar{z}$. 
The dots in figure \ref{fig:saturnrod}(a) denote singularities
at $\bar{z}=0$, which are removed by the fixing $c_1$ according to
\reef{eqn:c1} (figure \ref{fig:saturnrod}(b)). This choice makes the
$\rho=0$ metric smooth across $\bar{z}=0$. Figure
\ref{fig:saturnrod}(b) also shows the directions of the rods.}
    \label{fig:saturnrod}
  \end{center}
\end{figure}

\begin{itemize}
\item The semi-infinite rod $\bar{z} \in ]-\infty,\ka_3]$ and the
  finite rod $[\ka_2,\ka_1]$ have directions $(0,1,0)$, i.e.~they are
  sources for the $\phi\phi$-part of the metric. 
\item The semi-infinite rod $[1,\infty[$ has direction $(0,0,1)$,
  i.e.~it is sourcing the $\psi\psi$-part of the metric. 
\item The finite rod $[\ka_3,\ka_2]$ corresponds to the location of
  the black ring horizon. It has direction
  $(1,0,\Omega_\psi^\rom{BR})$. The finite rod $[\ka_1,1]$ corresponds
  to the location of the $S^3$ black hole horizon. It has direction
  $(1,0,\Omega_\psi^\rom{BH})$. 
The angular velocities $\Omega_\psi^\rom{BR}$ and
  $\Omega_\psi^\rom{BH}$ will be given in section \ref{s:horizons}. 
\end{itemize}

Note that the negative density rod of the seed solution figure
\ref{fig:blacksaturn} is no longer present. 
The soliton transformation which added the anti-soliton at
$z=a_1$ has made the $+1/2$ and $-1/2$ density rods in the $t$ and
$\psi$ direction cancel. However, the cancellation of the rods 
left a singularity at $z=a_1$, i.e.~$\bar{z}=0$. This shows up as a
$(z-a_1)^{-1} \sim \bar{z}^{-1}$ divergence in $G_{tt}$ and
$G_{\psi\psi}$, as indicated by dots in figure  
\ref{fig:saturnrod}(a). Luckily, the singularities are removed
completely by setting 
\bea
  |c_1| = L \sqrt{\frac{2 \ka_1 \ka_2}{\ka_3}} \, .
  \label{eqn:c1}
\eea
With $c_1$ fixed according to \reef{eqn:c1}, the metric at $\rho=0$ is
completely smooth across $\bar{z}=0$. This means that we have
succesfully removed the negative density rod at $\bar{z}\in [0,\ka_3]$
($z\in [a_1,a_5])$, and there is no longer any significance to the
point $\bar{z} = 0$  ($z=a_1$) in the metric, as illustrated in figure
\ref{fig:saturnrod}(b). 
  
The condition \reef{eqn:c1} will be imposed throughout the rest of the
paper. 
Since $(c_1,c_2) \to (-c_1,-c_2)$ just changes the 
overall direction of rotation, we choose $c_1 > 0$ without loss of
generality.


\subsection{Asymptotics}
\label{s:asympt}
We introduce asymptotic coordinates $(r,\theta)$
\bea
  \rho = \frac{1}{2} r^2 \sin{2\theta} \, ,
  ~~~~
  z =  \frac{1}{2} r^2 \cos{2\theta} \, ,
\eea
such that
\bea
  d\rho^2+dz^2 =  r^2\, (dr^2 +r^2 d\theta^2) \, .
\eea
The asymptotic limit is $r^2 = 2 \sqrt{\rho^2 + z^2} \to \infty$.
Requiring that $G_{t\psi} \to 0$ when $r \to \infty$ determines the
constant $q$ in the metric \reef{SaturnMetric} to be 
\bea
  q 
  = L\, \sqrt{\frac{2 \ka_1 \ka_2}{\ka_3}}\, \frac{\bc_2}{1+\ka_2\,\bc_2} \, .
\eea
We have used the definition \reef{bc2} of $\bc_2$ and imposed
\reef{eqn:c1} for $c_1$. 
 
To leading order, the asymptotic metric is
\bea
  e^{2\nu}
  ~=~ k^2\Big[ 1+\ka_2\,\bc_2 \Big]^2 \, \frac{1}{r^2} + \dots\, ,
\eea
which motivates us to choose the constant $k$ to be
\bea
  \label{thek}
  k ~=~\Big| 1+\ka_2\,\bc_2 \Big|^{-1} \, .
\eea
We shall assume\footnote{The solution with  $\bc_2 = - \ka_2^{-1}$
  is nakedly singular. See sections \ref{s:balance} and \ref{s:negeps}
  for further comments.} 
 that $\bc_2 \ne - \ka_2^{-1}$. The asymptotic metric then takes the form
\bea
  ds^2 = -dt^2 + dr^2 + r^2\, d\theta^2
  +  r^2 \sin^2\theta\, d\psi^2
  +  r^2 \cos^2\theta\, d\phi^2 \, .
\eea
Below we show that the angles $\psi$ and $\phi$ have periodicities
\bea
  \Delta\psi = \Delta\phi = 2\pi \, ,
\eea  
so that the solution indeed is asymptotically flat.


\subsection{Regularity and balance}
\label{s:balance}

In order to avoid a conical singularity at the location of a 
rod, the period $\Delta\eta$ of a spacelike coordinate $\eta(=\psi,
\phi)$ must be fixed as 
\begin{equation}
  \label{Deltas}
  \Delta\eta=2\pi \lim_{\rho\to 0}\sqrt{\frac{\rho^2
  e^{2\nu}}{g_{\eta\eta}}}\;. 
\end{equation}
Requiring regularity on the rod $\bar{z}\in [1,\,\infty]$ fixes the
period of $\psi$ to be  
$\Delta\psi=2 \pi$, and regularity  on the rod $\bar{z}\in
[-\,\infty,\ka_3]$ determines 
 $\Delta\phi=2 \pi$. 
 We have used \reef{eqn:c1} and \reef{thek}.\footnote{If we had not
   imposed the condition \reef{eqn:c1}, which removes the singularity
   at $\bar{z}=0$, then \reef{Deltas} would have given $\Delta\phi
   =\frac{\pi}{L}\sqrt{\frac{2\ka_3}{\ka_1 \ka_2}} \, c_1$ for
   $\bar{z}\in [0,\ka_3]$. Requiring $\Delta\phi=2 \pi$ is precisely
   the condition \reef{eqn:c1}.} 
 According to the discussion in the previous section this ensures
 asymptotic flatness of the solution. 

Next we consider regularity as $\rho \to 0$ for the finite rod  
$\bar{z}\in[\ka_2,\, \ka_1]$. Eq.~\reef{Deltas} gives  
\begin{equation}
   \label{deltaphi3}
   \Delta \phi=2\pi\,
   \frac{\ka_1-\ka_2}
   {\big| 1+\ka_2\,\bc_2 \big|\sqrt{\ka_1
   (1-\ka_2)(1-\ka_3)(\ka_1-\ka_3)}}\, . 
\end{equation}
When no constraints other than \reef{eqn:c1} are imposed, the metric
has a conical singular membrane in the plane of the ring, 
extending from the inner $S^1$ radius of the black ring to the horizon
of the $S^3$ black hole.   

We can avoid this conical singularity and balance the solution by
requiring the right hand side of \reef{deltaphi3} to be equal to
2$\pi$. 
Solving for $\bc_2$ this gives us the balancing, or equilibrium,
condition for black saturn, i.e.  
\bea
  \nonumber
  \bc_2 = \frac{1}{\ka_2}
  \left[
   \epsilon
   \frac{\ka_1-\ka_2}
    { \sqrt{\ka_1 (1-\ka_2)(1-\ka_3)(\ka_1-\ka_3)} } - 1
  \right] \, , 
  \hspace{3mm}   
  \rom{with}
  \hspace{3mm}   
  \left\{
  \begin{array}{ccc}
  \epsilon = +1 & 
  ~\rom{when}~ &\bc_2 > -\ka_2^{-1}  \\ 
  \epsilon = -1 &~\rom{when}~ & \bc_2 < -\ka_2^{-1}
  \end{array} 
   \right. 
  \, . \\
  \label{balance}
\eea
The solution with $\bc_2 = -\ka_2^{-1}$ is nakedly singular. Thus
the choice of sign $\epsilon$ divides the balanced black saturn solutions into two separate sectors. 
The limit of removing the $S^3$ black hole to leave just the balanced black
ring requires setting $\bc_2=0$ and according to \reef{balance} this
is only possible for $\epsilon=+1$. We are going to study the
$\epsilon=+1$ solutions in detail in section \ref{s:physics}, but will
also discuss some properties of the $\epsilon=-1$ solutions (see sections
\ref{s:komar} and \ref{s:negeps}).



\subsection{Horizons}
\label{s:horizons}

The rod analysis of section \ref{s:rods} showed that the two horizon
rods had directions $(1,0,\Omega_\psi^i)$, $i=\rom{BR},\rom{BH}$, for
the black ring and the $S^3$ black hole. Equivalently, the Killing
vectors $\xi = \pa_t+\Omega^i_{\psi} \pa_{\psi}$ are null on the
respective horizons. The angular velocities $\Omega^i_{\psi}$ are 
\bea
 \Omega^\rom{BH}_\psi
  &=& 
  \frac{1}{L} \, \big[ 1+\ka_2\,\bc_2 \big]\,
  \sqrt{\frac{\ka_2 \ka_3}{2 \ka_1}} \, 
    \frac{\ka_3 (1-\ka_1) - \ka_1 (1-\ka_2) (1-\ka_3) \bc_2}
     {\ka_3(1-\ka_1) + \ka_1 \ka_2 (1-\ka_2) (1-\ka_3) \bc_2^2}  \, , \\[2mm]
   \Omega^\rom{BR}_\psi
  &=& 
  \frac{1}{L} \, \big[ 1+\ka_2\,\bc_2 \big]\,
  \sqrt{\frac{\ka_1 \ka_3}{2 \ka_2}} \, 
    \frac{\ka_3  - \ka_2 (1-\ka_3) \bc_2}
     {\ka_3 -  \ka_3 (\ka_1 -\ka_2) \bc_2 + \ka_1 \ka_2 (1-\ka_3) \bc_2^2}  \, .
\eea 
The black ring and the $S^3$ black hole generally have different
angular velocities. 

\subsubsection*{Myers-Perry black hole horizon geometry}
 
One black hole horizon is located at $\rho=0$ for $\ka_1 \le
\bar{z} \le 1$ and the metric on a spatial cross-section of the
horizon can be written  
\bea
  ds_\rom{BH}^2 = 
  \frac{2L^2 (\bz-\ka_1)(\bz-\ka_3)}{(\bz-\ka_2)}\, d\phi^2
  +  L^2 s_\rom{BH}^2 \, g(\bz) (1-\bz)\, d\psi^2
  + \frac{L^2\, (\bz-\ka_2)\, d\bz^2}
         {(1-\bz)(\bz-\ka_1)(\bz-\ka_3)g(\bz)} \, ,~~~
\eea
where the constant $s_\rom{BH}$ is 
\bea
  s_\rom{BH}=
  \frac{\ka_3(1-\ka_1) 
  +\ka_1 \ka_2 (1-\ka_2) (1-\ka_3) \bc_2^2}
    {\ka_3 \sqrt{(1-\ka_1) (1-\ka_2) (1-\ka_3)} 
      \,\big[ 1+ \ka_2 \bc_2 \big]^2} \, ,
\eea
and the function $g(\bz)$ is
\bea
  \nonumber
  g(\bz) &=& 2 \ka_1 \ka_3 (1-\ka_1) (1-\ka_2) (1-\ka_3)  (\bz-\ka_2) \\[2mm]
    &&\nonumber \times
    \big[ 1+ \ka_2 \bc_2 \big]^2 
    \bigg[
      (1-\ka_1)^2 \ka_3
      \Big[ \ka_1 (\bz - \ka_2)
         - \ka_3 \Big( \ka_1- \ka_2(1-\bz)^2 - \ka_1 \ka_2(2-\bz)\Big)
      \Big]  \\[2mm]
    &&\nonumber \hspace{1.3cm}
    + 2 \ka_1 \ka_2 \ka_3   (1-\ka_1) (1-\ka_2) (1-\ka_3)  
      (1-\bz) (\bz - \ka_1) \, \bc_2 \\[2mm]
    && \hspace{1.3cm}
    + \ka_1^2 \ka_2 (1-\ka_2)^2 (1-\ka_3)^2 \, \bz\, (\bz - \ka_1)\, \bc_2^2    
    \bigg]^{-1} \, .
\eea
Note that $s_\rom{BH} \ge 0$.
One can check that $g(\bz)$ is positive for $\ka_1 \le \bz \le 1$, so
for $s_\rom{BH} > 0$, the horizon is topologically an
$S^3$. Metrically the $S^3$ is distorted by rotation, as is the case
for a Myers-Perry black hole, and here the horizon is further
deformed by the presence of the black ring.

\subsubsection*{Black ring horizon geometry}

The black ring horizon is located at $\rho=0$ for $\ka_3 \le \bz \le
\ka_2$. The metric of a spatial cross section of the horizon can be
written 
\bea
  ds_\rom{BR}^2 = \frac{2L^2 (\ka_2-\bz)(\bz-\ka_3)}{(\ka_1-\bz)}\, d\phi^2
  +  L^2 s_\rom{BR}^2\, f(\bz) (\ka_1-\bz)\, d\psi^2
  + \frac{L^2d\bz^2}{ (\ka_2-\bz)(\bz-\ka_3)f(\bz)} \, ,
\eea
where the constant $s_\rom{BR}$ is 
\bea
  s_\rom{BR}= \sqrt{\frac{\ka_2-\ka_3}{\ka_1 (\ka_1-\ka_3) (1-\ka_3)}}
  \,\frac{\big( \ka_3- \ka_3(\ka_1-\ka_2) \bc_2 +\ka_1 \ka_2 (1-\ka_3)
    \bc_2^2 \big)}
    {\ka_3\left[ 1+ \ka_2 \bc_2 \right]^2} \, ,
\eea
and the function $f(\bz)$ is
\bea
  f(\bz) &=& 2\ka_1\ka_3(\ka_1-\ka_3)(1-\ka_3)(1-\bz)\nonumber\\
    &\;&\times
    \big[ 1+ \ka_2 \bc_2 \big]^2 (\ka_2-\ka_3)^{-1}\bigg[
    \ka_3\Big[\ka_2(\ka_1-\bz)+\ka_3\Big(\ka_2\big(1-\ka_1(2-\bz)\big)-\ka_1(1-\bz)^2\Big)\Big]\nonumber\\
    &\;&\hspace{3.5cm}+2\ka_1\ka_2\ka_3(1-\ka_3)(1-\bz)(\ka_2-\bz)\bc_2\nonumber\\
    &\;&\hspace{3.5cm}+\ka_1\ka_2^2(1-\ka_3)^2\bz(\ka_2-\bz)\bc_2^2
    \bigg]^{-1}\;. 
\eea
It follows from \reef{order2} that $s_\rom{BR} \ge 0$, and it can be checked 
that $f(\bz)$ is positive for $\ka_3 \le \bz \le \ka_2$.  
The coordinate $\psi$ parametrizes a circle whose radius depends on $\bar{z}$. The coordinates $(\bz,\phi)$ parameterize a deformed 
two-sphere. The topology of the horizon is therefore $S^1 \times
S^2$. As is the case for the black ring \cite{ER,EEV}, the
metric of the horizon is not a direct product of the $S^1 \times S^2$ 
(contrary to the supersymmetric case \cite{EEMR1,EEMR2,Bena,GG}). For black
saturn the black ring 
horizon is further distorted by the presence of the $S^3$ black hole.  

\subsubsection*{Horizon areas}
It is straightforward to compute the horizon areas. We find
\bea
\label{BHarea}
\mathcal{A}_\rom{BH}
  &=& 4 L^3 \pi^2 
  \sqrt{\frac{2(1-\ka_1)^3}{(1-\ka_2)(1-\ka_3)}}~
  \frac{1 + 
      \frac{\ka_1 \ka_2 (1-\ka_2) (1-\ka_3)}{\ka_3(1-\ka_1)} \,\bc_2^2}
             { \big( 1+\ka_2\, \bc_2 \big)^{2}}
         \, , \\[2mm]
\label{BRarea}
\mathcal{A}_\rom{BR}
  &=& 4 L^3 \pi^2  
  \sqrt{\frac{2 \ka_2 (\ka_2-\ka_3)^3}{\ka_1 (\ka_1-\ka_3)(1-\ka_3)}}~
  \frac{1- (\ka_1-\ka_2) \bc_2 
   + \frac{\ka_1 \ka_2 (1-\ka_3)}{\ka_3}\,  \bc_2^2  }
       { \big( 1+\ka_2\, \bc_2 \big)^{2}} \, .
\eea
Note that for all real $\bc_2$ and $0 < \ka_3 < \ka_2 < \ka_1 < 1$,
  the expressions for the horizon areas are real and positive, hence 
  well-defined. In particular, there are no signs of closed timelike
  curves. 


\subsubsection*{Temperatures}
We compute the temperatures using \cite{Har} and find
\begin{eqnarray}
T_{\textrm{H}}^{\textrm{BH}}&=&
\frac{1}{2 L\, \pi} 
\sqrt{\frac{(1-\ka_2)(1-\ka_3)}{2(1-\ka_1)}}\,
  \frac{ \big( 1+\ka_2\, \bc_2 \big)^{2}}
     {1 + \frac{\ka_1 \ka_2 (1-\ka_2) (1-\ka_3)}{\ka_3(1-\ka_1)} \,\bc_2^2} \, ,   
        \\[4mm] \nonumber
T_{\textrm H}^{\textrm{BR}}&=&
\frac{1}{2 L \, \pi} 
\sqrt{\frac{\ka_1(1-\ka_3)(\ka_1-\ka_3)}{2\ka_2(\ka_2-\ka_3)}}\,
    \frac{\big( 1+\ka_2\, \bc_2 \big)^{2}}
      {1- (\ka_1-\ka_2) \bc_2 
   + \frac{\ka_1 \ka_2 (1-\ka_3)}{\ka_3}\,  \bc_2^2} \, .
\end{eqnarray}
The ordering \reef{order2} ensures that the temperatures are
non-negative.  

The expressions for the temperatures are complimentary to those for
the horizon areas \reef{BHarea}-\reef{BRarea}: with the entropy being
one quarter times the horizon area,  
$S=\mathcal{A}/(4G)$, we have two very simple expressions: 
\bea
  T_{\textrm{H}}^{\textrm{BH}} S^{\textrm{BH}} 
  = \frac{\pi}{2G} \,L^2  (1-\ka_1) \, , 
  \hspace{1cm}
  T_{\textrm{H}}^{\textrm{BR}} S^{\textrm{BR}} 
  = \frac{\pi}{2G} \,L^2  (\ka_2-\ka_3) \, .
\eea
The former vanishes in the limit  $\ka_1 \to1$ which gives an extremal
rotating $S^3$ black hole. The latter vanishes when $\ka_2=\ka_3$,
which we interpret as the limit where the black ring becomes singular,
as the $j=1$ limit of the fat black rings. 


\subsection{ADM mass and angular momentum}
The solution is asymptotically flat and it is straightforward to
compute the ADM mass $M$ and angular momentum $J$ using the asymptotic
coordinates introduced in section \ref{s:asympt}. We find
\begin{eqnarray}
 \label{ADMmass}
M=\frac{3\pi\, L^2}{4G}\,
  \frac{\ka_3(1-\ka_1+\ka_2)-2\ka_2\ka_3(\ka_1-\ka_2)\bc_2 
        +\ka_2\big[\ka_1-\ka_2\ka_3(1+\ka_1-\ka_2)\big]\bc_2^2}
       {\ka_3 \big[ 1+ \ka_2 \bc_2 \big]^2} \, 
\end{eqnarray}  
and
\begin{eqnarray}
\label{Jadm}
J&=& \frac{\pi\,L^3}{G}\frac{1}{\ka_3 \big[ 1+ \ka_2 \bc_2 \big]^3}
        \sqrt{\frac{\ka_2}{2\ka_1\ka_3}}        
        \, \bigg[
        \ka_3^2
        -\bar{c}_2\ka_3\Big[
                                (\ka_1-\ka_2)(1-\ka_1+\ka_3)+\ka_2(1-\ka_3)
                                \Big] \nonumber\\[2mm]
        &\;&\hspace{4.5cm}+\bar{c}_2^2\ka_2\ka_3\Big[
           (\ka_1-\ka_2)(\ka_1-\ka_3)+\ka_1(1+\ka_1-\ka_2-\ka_3) \Big] 
         \nonumber\\
        &\;&\hspace{4.5cm}-\bar{c}_2^3\ka_1\ka_2\Big[           
                \ka_1-\ka_2\ka_3(2+\ka_1-\ka_2-\ka_3)\Big] 
        \bigg]\, .
\end{eqnarray}
It is worth noting that for any $\bc_2 \in \mathbb{R}$ the ADM mass
\reef{ADMmass} is positive as a simple consequence of the ordering
\reef{order2}.


\subsection{Komar integrals}
\label{s:komar}
 
Komar integrals evaluated on the horizon of each black hole allow us
to compute a measure of the mass and angular momentum of the two
objects of the saturn system.

\subsubsection*{Komar masses}

In five spacetime dimensions, the Komar mass is given by 
\begin{equation}
M_\textrm{Komar}=\frac{3}{32\pi G}\int_S *d\xi \;, 
\label{eqn:komarm}
\end{equation}
where $\xi$ is the dual 1-form associated to the asymptotic time
translation Killing field $\partial_t$ and $S$ is the boundary of any
spacelike hypersurface. Eq.~\eqref{eqn:komarm} measures the mass
contained in $S$, so the mass of each black hole in
a multi-black hole spacetime is computed by taking $S$ to be at the
horizon $H_i$.
Instead, if we take $S$ to be the
$S^3$ at infinity, then \eqref{eqn:komarm} gives the total mass of the
system, which coincides with the ADM mass. In terms of the metric
components we have
\begin{eqnarray}
M_\textrm{Komar}^i=\frac{3}{32\pi G}\int_{H_i}dz \, d\phi\,
        d\psi~\frac{1}{\sqrt{-\textrm{det}~g}}~g_{zz}g_{\phi\phi} 
        \big[-g_{\psi\psi}\, \partial_\rho g_{tt}+g_{t\psi}\,
        \partial_\rho g_{t\psi}\big]\; ,
\end{eqnarray}
which for the saturn solution gives
\begin{eqnarray}
M_\rom{Komar}^\rom{BH} &=&
  \frac{3 \pi L^2}{4 G} \,
  \frac{\ka_3 (1-\ka_1)+ \ka_1 \ka_2 (1-\ka_2)(1-\ka_3)
    \, \bc_2^2}{\ka_3(1+\bc_2\, \ka_2)} \, ,\label{mKbh} \\[2mm]  
 M_\rom{Komar}^\rom{BR} &=&
  \frac{3 \pi L^2}{4 G} \,
  \frac{\ka_2 \big[ 1- (1-\ka_2)\, \bc_2\big]
   \big[  \ka_3 - \ka_3 (\ka_1- \ka_2)\,  \bc_2 
      + \ka_1 \ka_2 (1-\ka_3)\,  \bc_2^2\big]}
       {\ka_3(1+\bc_2\,  \ka_2)^2} \, .  \label{mKbr}  
\end{eqnarray}
Note that
\reef{mKbh}-\reef{mKbr} give
\begin{equation}
M_\textrm{ADM}=M_\textrm{Komar}^\textrm{BR}+M_\textrm{Komar}^\textrm{BH}\; ,
\end{equation}
so the Komar masses add up to the ADM mass \reef{ADMmass}, even in the
presence of the conical singularity. 
We discuss the sign of the Komar masses at the end of this subsection.

\subsubsection*{Komar angular momenta}
The angular momentum Komar integral is given by 
\begin{eqnarray}
J_\textrm{Komar}=\frac{1}{16\pi G}\int_S*d\zeta\;,  \label{eqn:komarj}
\end{eqnarray}
where $\zeta$ is the 1-form dual to the Killing field $\partial_\psi$,
and $S$ is the boundary of any spacelike hypersurface. Now
\eqref{eqn:komarj} measures the angular momentum contained within $S$,
and therefore, if we choose $S$ to be the horizons
$H_i$, we can compute the ``intrinsic'' angular momentum of each black
object. We have
\begin{eqnarray}
J_\textrm{Komar}^i=\frac{1}{16\pi G}\int_{H_i}dz \, d\phi \,
        d\psi~\frac{1}{\sqrt{-\textrm{det}~g}}~g_{zz}g_{\phi\phi} 
        \big[-g_{\psi\psi} \, \partial_\rho g_{t\psi}+g_{t\psi} \,
        \partial_\rho g_{\psi\psi}\big]\; ,
\end{eqnarray}
which gives
\begin{eqnarray}
\label{jKbh}
  J_\rom{Komar}^\rom{BH} &=&
  - \frac{\pi L^3}{G} \,
  \sqrt{\frac{\ka_1 \ka_2}{2 \ka_3}}\, 
  \frac{\bc_2 \big[ \ka_3 (1-\ka_1)+ \ka_1 \ka_2 (1-\ka_2)(1-\ka_3)
    \, \bc_2^2 \big]}
       {\ka_3(1+\bc_2\,  \ka_2)^2} \, , \\[2mm]
\label{jKbr}
  J_\rom{Komar}^\rom{BR} &=&
  \frac{\pi L^3}{G} \,
  \sqrt{\frac{\ka_2}{2 \ka_1 \ka_3}}\, \\ \nonumber
  && \times
  \frac{\big[ \ka_3 - \ka_2(\ka_1-\ka_3)\, \bc_2 
              + \ka_1 \ka_2 (1-\ka_2)\, \bc_2^2 \big]
        \big[ \ka_3 - \ka_3(\ka_1-\ka_2)\, \bc_2 
              + \ka_1 \ka_2 (1-\ka_3)\, \bc_2^2 \big]}
       {\ka_3(1+\bc_2\,  \ka_2)^3} \, .
\end{eqnarray}
The Komar angular momenta add to up to $J_\rom{ADM}$ given
in \reef{Jadm}, 
\begin{equation}
  J_\textrm{ADM}=J_\textrm{Komar}^\textrm{BR}+J_\textrm{Komar}^\textrm{BH}\; ,
\end{equation}
even without imposing the balance condition \reef{balance}. 

We shall refer to the Komar angular momentum as the ``intrinsic''
angular momentum of the black hole. Note that for $\bc_2=0$, the $S^3$
black hole carries no intrinsic spin
$J_\textrm{Komar}^\textrm{BH}=0$. This was expected since the soliton
transformation with $c_2=0$ did not add spin to the $S^3$ black hole  
directly.  

\subsubsection*{Smarr relations}

Black rings \cite{ER} and Myers-Perry black holes \cite{MPBH} satisfy
the same Smarr formula 
\bea
  \label{theSmarr}
  \frac{2}{3}M = T_\rom{H} S + J \, \Omega \, .  
\eea
Using the expressions of the Komar masses \reef{mKbh}-\reef{mKbr} and
the Komar angular momenta \reef{jKbh}-\reef{jKbr} we find that both
the black ring and the black hole separately obey this Smarr
relation:
\begin{eqnarray}
\label{SatSmarr}
\frac{2}{3}\, M_\textrm{Komar}^\textrm{BR} = 
        T_\textrm{H}^\textrm{BR}S^\textrm{BR} 
        +\Omega_\psi^\textrm{BR}J_\textrm{Komar}^\textrm{BR}\;,
\hspace{1cm}
\frac{2}{3}\, M_\textrm{Komar}^\textrm{BH} = 
        T_\textrm{H}^\textrm{BH}S^\textrm{BH}
	+\Omega_\psi^\textrm{BH}J_\textrm{Komar}^\textrm{BH}\;. 
\end{eqnarray}
These Smarr relations are mathematical identities which relates the
physical quantities measured at the horizon, and they can be derived
quite generally for multi-black hole vacuum spacetimes
\cite{EEF2}. The relations \reef{SatSmarr} hold without imposing the 
balance condition \reef{balance}.  

\subsubsection*{Sign of Komar masses}
We have already noted that the total ADM mass \reef{ADMmass} is always
positive. 
Positivity of $M^\rom{BH}_\rom{Komar}$ in \reef{mKbh} requires that
$\bc_2 > -\ka_2^{-1}$, and this selects the $\epsilon = +1$ case
of the balance condition in section \ref{s:balance}. Furthermore,
imposing the balance condition 
\reef{balance} with $\epsilon=+1$ implies that $\bc_2$ takes values
$-\ka_2^{-1} < \bc_2 < (1-\ka_2)^{-1}$, and thus both
$M^\rom{BH}_\rom{Komar}$ in \reef{mKbh} and $M^\rom{BR}_\rom{Komar}$
in \reef{mKbr} are positive. 

On the other hand, imposing the balance condition \reef{balance} with
$\epsilon=-1$ means that \mbox{$\bc_2<-\ka_2^{-1}$}, and --- as can be
seen from \reef{mKbh}-\reef{mKbr} --- this gives
$M^\rom{BH}_\rom{Komar} < 0$ while $M^\rom{BR}_\rom{Komar} > 0$.  

One might take as a criterium for establishing the physical relevance
of a multi-black 
hole system that each of the components in the system has positive
mass. Clearly, at large separations the Komar mass of each object
should agree with the positive ADM mass of the object, but that does
not imply that the Komar masses in tightly bound gravitational systems
need to be positive.\footnote{We thank Roberto Emparan for discussions
about this point, and also Harvey Reall for helpful comments.} How,
physically, can a solution with negative Komar mass occur? 

It follows from the Smarr relation \reef{SatSmarr} that the Komar mass
can be negative provided that the angular velocity $\Omega$ and Komar
angular momentum $J_\rom{Komar}$ have opposite signs \emph{and} that
$\Omega J_\rom{Komar}$ is sufficiently large and negative to overwhelm
the positive $T_\rom{H} S$-term. In black saturn, the physical
mechanism behind 
opposite signs of $\Omega^\rom{BH}$ and $J_\rom{Komar}^\rom{BH}$ is
rotational frame-dragging: the rotating black ring drags the $S^3$
black hole so that its horizon is spinning in the opposite direction
of its ``intrinsic'' angular momentum. We examine this effect in
detail in section \ref{s:cc2zero}. In that section we focus on
solutions with $\bc_2=0$, hence $\epsilon=+1$, and these have
$J_\rom{Komar}^\rom{BH}=0$; however, that analysis also serves to
illustrate the physics which lies behind having $\Omega^\rom{BH}
J_\rom{Komar}^\rom{BH}<0$.  

In conclusion, the solutions with $M_\rom{Komar}^\rom{BH}>0$
($\epsilon=+1$) and $M_\rom{Komar}^\rom{BH}<0$ ($\epsilon=-1$) appear
to be equally valid.  
We shall primarily focus on the $M_\rom{Komar}^\rom{BH}>0$ solutions
when we study the physics of black saturn in section \ref{s:physics},
but we comment briefly on the $M_\rom{Komar}^\rom{BH}<0$ solutions in
section \ref{s:negeps}. 

Finally, let us remark that single black hole spacetimes with
counter-rotation (in the sense of $ \Omega J<0$) and negative Komar
mass, but positive ADM mass, have been constructed numerically as
solutions of five-dimensional Einstein-Maxwell theory with a
Chern-Simons term \cite{kunz}. In that case, part of the energy and
angular momentum is carried by the electromagnetic fields making
counter-rotation and negative Komar mass possible.



\subsection{Closed timelike curves}

One might expect the plane of the ring to be a natural place
for closed timelike curves  (CTCs) to appear, and we have 
focused our analysis on this region. For the case $\bc_2=0$, we find
analytically that $G_{\psi\psi} > 0$ for $\rho=0$ and $z<\ka_3$ (the plane
outside the ring) and $\ka_2<z<\ka_1$ (the plane between the ring and
the black hole). So for $\bc_2=0$ there are no CTCs in the plane of
the ring (cf.~\cite{Cvetic:2005zi}). 

When $\bc_2 \ne 0$ the metric components are sufficiently complicated
that we resort to numerical checks. 
We have performed such checks
for examples where the $S^3$ black hole and the black ring are
counter-rotating as well as co-rotating. Among other examples we have
checked the counter-rotating cases with $J=0$; no CTCs were found. 

CTCs tend to appear when solutions are over-spinning, at least that
is the case for supersymmetric black holes
\cite{BMPV,EEMR1,Bena,EEMR2,GG}. Hence we have checked in
detail cases where the black hole and the ring are co-rotating and
fast spinning. One such example is studied in section
\ref{s:reachjz}. For this 1-parameter family of solutions the $S^3 $
black hole angular velocity covers a large range of co- and
counter-rotation; we have checked numerically 
for CTCs in the plane of the ring and found none. 
 
While we have found no signs of the appearance of closed
timelike curves in our analysis, we emphasize that our numerical
checks are not exhaustive. Rewriting the solution in ring coordinates
$(x,y)$ will probably be helpful for checking for CTCs.

\subsection{Limits}
\label{s:limits}

Black saturn combines a singly spinning Myers-Perry spherical black
hole with a black ring in a balanced configuration, and it is possible
to obtain either of these solutions as limits of the balanced black
saturn solution with $\epsilon=+1$.  
We describe here the appropriate limits, while details 
are relegated to the appendix. 

\subsubsection*{Myers-Perry black hole limit}
In the general solution, one can remove the black ring by first
setting the BZ parameter $c_1=0$, thus eliminating the black ring
spin, and then removing the black ring rod by taking $\kappa_2 = \kappa_3 =
0$. For the physical solution, where the singularity at $\bar{z}=0$
has been removed, $c_1$ is fixed by \reef{eqn:c1} and we have to take the
limit $\kappa_2, \kappa_3 \to 0$, in such a way that $c_1$ remains
finite. This can be accomplished by first taking $\ka_2 \to \ka_3$ and
then $\ka_3 \to 0$. We provide details of this limit in appendix
\ref{app:MP}.

\subsubsection*{Black ring limit}
The black ring \cite{ER} is obtained by simply removing the $S^3$
black hole from the saturn configuration. This is done by first
setting the angular momentum of the black hole to zero by taking
$\bc_2=0$, and then setting $\ka_1=1$, which removes the $S^3$ black
hole. We show in appendix \ref{app:BR} that the remaining solution is
exactly the black ring of \cite{ER} by rewriting 
the solution explicitly in ring coordinates $x,y$. The balance
condition \reef{balance} becomes the familiar equilibrium condition
for a single black ring.

\subsubsection*{No merger limit}
It would be interesting if one could use the black saturn system to study a
controlled merger of the $S^3$ black hole and the
black ring. Unfortunately, this is not possible. Based on the rod
configuration given in figure \ref{fig:saturnrod}(b), the merger
should correspond to merging the two horizon rods, $[\ka_3,\ka_2]$ of
the black ring and $[\ka_1,1]$ of the $S^3$ black hole. 
Thus the merger would correspond to taking $\ka_1 \to \ka_2$. Imposing
the balance condition \reef{balance}, $\ka_1 \to \ka_2$ implies $\bc_2
\to - \ka_2^{-1}$. The solution with $\bc_2 = - \ka_2^{-1}$ is nakedly
singular, and hence the suggested merger limit is singular. 
As a side remark, we point out that the 
singular nature of the merger limit is in fact very similar to why
two balanced Kaluza-Klein black holes held apart by a static
bubble-of-nothing cannot be merged by taking a similar limit
\cite{EH}.

\section{Physics of black saturn}
\label{s:physics}

We examine a selection of  interesting physical properties of
black saturn. 
In section \ref{s:nonuniq1} we establish
that black saturn has 2-fold continuous
non-uniqueness. 
Section \ref{s:MPBR} reviews basic properties of the
Myers-Perry black hole and the black ring; properties which will be 
helpful for understanding the physics of black saturn.

It is useful to clarify notions of rotation and intrinsic spin:
\begin{itemize}
 \item[$\cdot$] A black hole is rotating when its angular velocity
 $\Omega^i$ is nonzero. 
 \item[$\cdot$]  Co(counter)-rotation means $\Omega^\rom{BH}$ and
 $\Omega^\rom{BR}$ have the same (opposite) sign. 
  \item[$\cdot$]  We use the term \emph{intrinsic} angular momentum to
  refer to the angular momentum  $J_\rom{Komar}$ measured by the Komar
  integral evaluated at the horizon of the black hole. 
\end{itemize}
The two black objects in black saturn interact gravitationally,
and one effect is frame-dragging. This is cleanly illustrated for the
case where the $S^3$ black hole has vanishing intrinsic angular momentum,
$J_\rom{Komar}^\rom{BH}=0$, but is nonetheless rotating,
$\Omega^\rom{BH}\ne 0$.  
We found in section \ref{s:komar} that $J_\rom{Komar}^\rom{BH}=0$ for
$\bc_2=0$, so in 
section \ref{s:cc2zero} we study the $\bc_2=0$ subfamily of
black saturn configurations. 

The general black saturn configurations with $\bc_2 \ne 0$ are studied
in sections \ref{s:BHwspin} and \ref{s:nonuniq2}. For $\bc_2 \ne 0$ 
the $S^3$ black hole and the black ring have independent rotation 
parameters, and this makes it possible to have counter-rotating
solutions and configurations with vanishing total angular momentum,
$J=0$. Having $\bc_2 \ne 0$ is also necessary for realizing the full
2-fold continuous non-uniqueness.

Note that we are imposing the balance condition \reef{s:balance} with
$\epsilon=+1$ throughout this section, with the exception of
subsection \ref{s:negeps}.   


\subsection{Parameter counting and non-uniqueness}
\label{s:nonuniq1}

We begin by counting the parameters in the saturn solution.
The full solution has six para\-meters: $\ka_{1,2,3}$, satisfying 
$0\le \ka_3 \le \ka_2 < \ka_1 \le 1$, one scale $L$, and the two BZ
parameters $c_1$ and $c_2$.  
The parameter $c_1$ is fixed according to \reef{eqn:c1}
in order to avoid a naked singularity at $\bar{z}=0$.
We conveniently rescaled $c_2$ to introduce the dimensionless parameter 
$\bc_2 \propto c_2$ in \reef{bc2}.
So the unbalanced solution has four dimensionless parameters,
$\ka_{1,2,3}$ and $\bc_2$, and the scale $L$.
The balance condition \reef{balance} imposes a constraint between
$\bc_2$ and $\ka_{1,2,3}$, and in conclusion, the balanced black
saturn solution has three dimensionless parameters and one scale $L$.  

{}Fixing the ADM mass $M$ of the full system fixes the scale $L$, and
leaves three dimensionless parameters. Fixing further the only 
other conserved asymptotic quantity, namely the angular momentum $J$,
leaves two free dimensionless parameters. Thus black saturn has 
2-fold continuous non-uniqueness. We examine the non-uniqueness in
greater detail in the following sections. 

\subsubsection*{Fixed mass reduced parameters}
We introduce the fixed mass reduced parameters
\bea
  \label{redpar}
  \begin{aligned}
  &j^2 =\frac{27 \pi}{32 G}\frac{J^2}{M^3}\;,&
  \hspace{1cm}
  &a_{\textrm{H}}^i =\frac{3}{16}\sqrt{\frac{3}{\pi}}\frac{{\cal
    A}_i}{(G M)^{3/2}}\;, \\[2mm]
  &\omega_i = \sqrt{\frac{8}{3\pi}}~
               \Omega^{i}_{\psi}(GM)^{1/2}\;,&
  \hspace{1cm}
  &\,\tau_i\, = \sqrt{\frac{32 \pi}{3}}~T^{i}_{\textrm{H}}(G M)^{1/2}\;, 
  \end{aligned}
\eea
which allow us to compare physical properties of configurations with
the same ADM mass $M$.  
The script $i$ labels the quantity corresponding to the black ring
($i=$BR) or the $S^3$ black hole ($i$=BH). 
We will also use the total horizon area, 
\beq
   a_{\textrm{H}}^\rom{total}
    =a_{\textrm{H}}^\textrm{BR}+a_{\textrm{H}}^\textrm{BH}\, , 
\eeq
in order to study the ``phase diagram'' (total entropy vs.~$j^2$) of
the black saturn. Occasionally we simply use $a_\rom{H}$ for
$a_{\textrm{H}}^\rom{total}$.

The reduced temperature and angular velocity are normalized such that
$\tau_\rom{BH} = 1$ for the five-dimensional Schwarzschild black hole
($j=0$), and $\omega_\rom{BH} = 1$ for the maximally rotating (singular)
Myers-Perry black hole ($j=1$).\footnote{Our normalizations of $\tau_i$
  and $\omega_i$ differ from the conventions used in \cite{EEV}.}

\subsection{Myers-Perry black hole and black rings}
\label{s:MPBR}

\begin{figure}[t]
\begin{picture}(0,0)(0,0)
\put(8,-3){\scriptsize$a_\textrm{H}$}
\put(54,-33){\scriptsize$j^2$}
\put(24,-12){\scriptsize MP black hole}
\put(35,-22){\scriptsize thin black ring}
\put(17,-28){\scriptsize fat black ring}
\put(0.3,-5){\scriptsize$2\sqrt{2}$}
\put(4.3,-13){\scriptsize$2$}
\put(4.3,-23){\scriptsize$1$}
\put(20,-35.5){\scriptsize$0.5$}
\put(30.8,-36){\scriptsize$\frac{27}{32}$}
\put(36.4,-35.5){\scriptsize$1$}
\put(62,-3){\scriptsize$\omega_\textrm{H}$}
\put(105,-36){\scriptsize$j^2$}
\put(65,-7){\scriptsize MP black hole}
\put(93,-22){\scriptsize black ring}
\put(58.5,-5){\scriptsize$1$}
\put(57,-18.8){\scriptsize$0.5$}
\put(75,-35.5){\scriptsize$0.5$}
\put(85.7,-36){\scriptsize$\frac{27}{32}$}
\put(91.3,-35.5){\scriptsize$1$}
\put(116,-3){\scriptsize$\tau_\textrm{H}$}
\put(163,-33){\scriptsize$j^2$}
\put(120,-20){\scriptsize MP black hole}
\put(152,-11){\scriptsize black ring}
\put(114,-13){\scriptsize$1$}
\put(109.3,-33.5){\scriptsize $10^{-1}$}
\put(129.8,-35.5){\scriptsize$0.5$}
\put(140.3,-36){\scriptsize$\frac{27}{32}$}
\put(145.9,-35.5){\scriptsize$1$}
\put(19.9,-42){\footnotesize Figure \ref{fig:cc2zeroBR2}(a)}
\put(74.9,-42){\footnotesize Figure \ref{fig:cc2zeroBR2}(b)}
\put(130.9,-42){\footnotesize Figure \ref{fig:cc2zeroBR2}(c)}
\end{picture}
  \begin{center}
      \includegraphics[width=4.6cm]{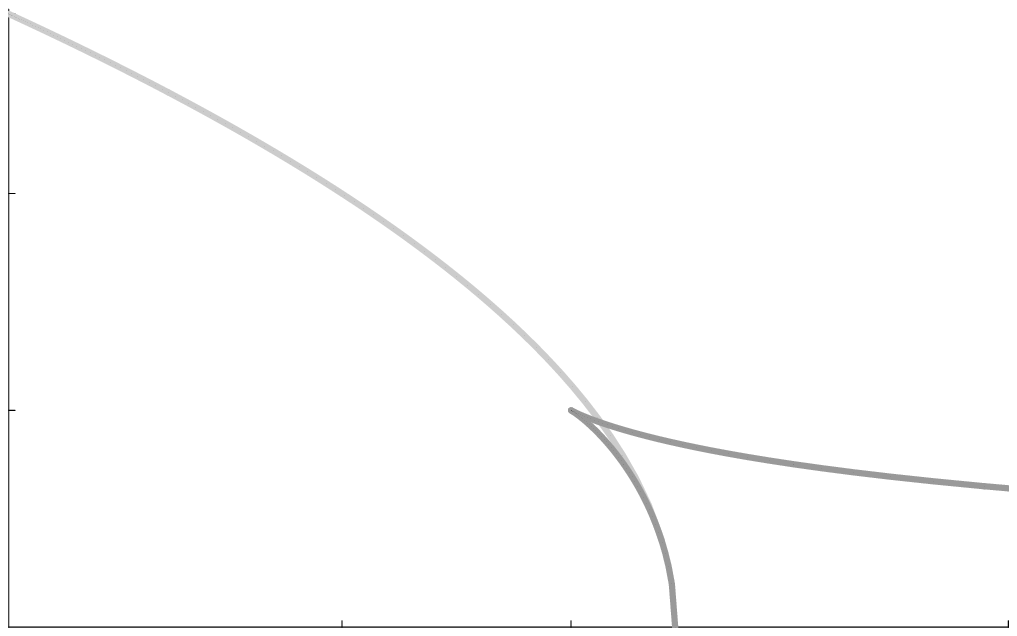}
      \hspace{6mm}
      \includegraphics[width=4.6cm]{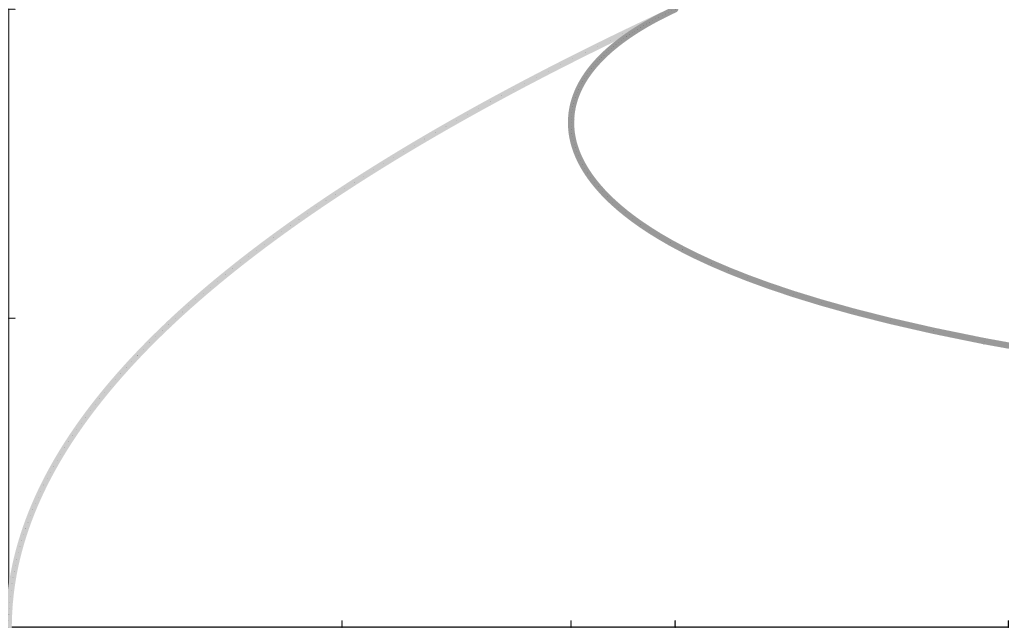}
      \hspace{6mm}
      \includegraphics[width=4.6cm]{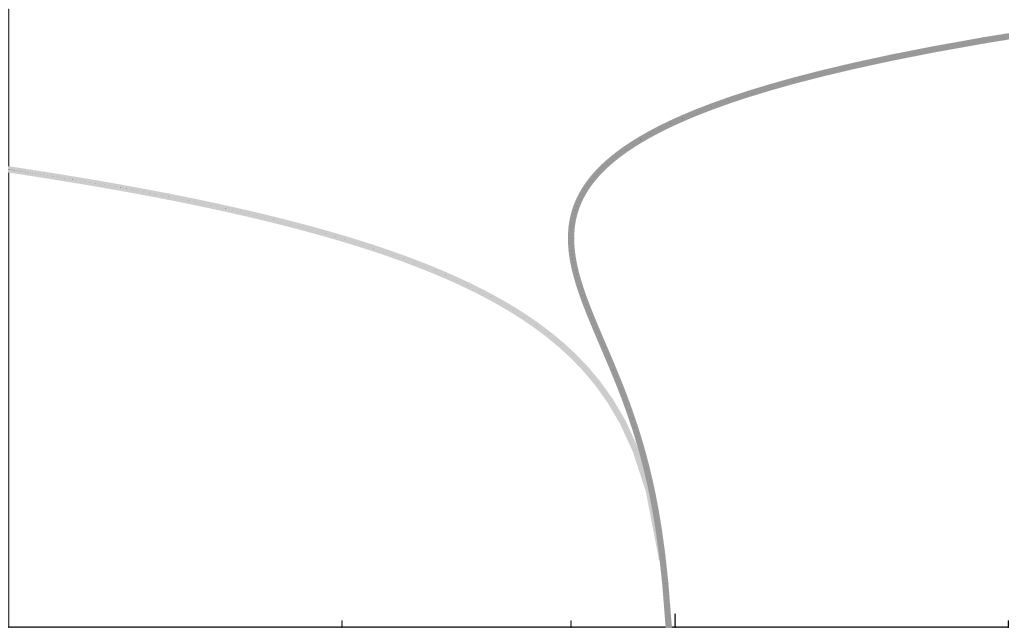}
    \vspace{1cm}
    \caption{\small Behavior of the reduced physical parameters for the 
    Myers-Perry black hole (light gray) and the black ring (dark
    gray). Note that we are using a logarithmic scale for the
    temperature.}
    \label{fig:cc2zeroBR2}
  \end{center}
\end{figure}

In preparation for studying the physical properties of black saturn,
we review the basic properties of the Myers-Perry black hole
\cite{MPBH} and the black ring \cite{ER} with a single angular
momentum.  
Figure \ref{fig:cc2zeroBR2} shows for fixed mass the behaviors of the
area, angular velocity and temperature of the Myers-Perry black hole
and the black ring as the reduced angular momentum
$j$ is varied.

{}For the Myers-Perry black hole the reduced angular velocity grows
linearly with the reduced angular momentum --- in fact 
our normalization is such that $\omega_\rom{MP} = j_\rom{MP}$. (In
order to better represent the near-$j=1$ behavior we choose here to
plot the physical properties vs.~$j^2$ rather than $j$, as it was done
in \cite{EEV}.) Increasing $j$, the area $a_\rom{H}^\rom{MP}$
decreases and the black hole gets colder ($\tau_\rom{MP}$ decreases).
As $j \to 1$, the $S^3$ horizon flattens out as a pancake in the plane
of rotation, and at   $j=1$ the solution becomes nakedly singular.

Black rings come in two types: thin and fat black rings. The
distinction is based on the ``phase diagram'' showing
$a_\rom{H}^\rom{BR}$ vs.~$j^2$ (see figure \ref{fig:cc2zeroBR2}(a)):
thin rings are those on the upper
branch, while the fat rings are those on the lower branch. 
As $j \to 1$, the $S^2$ of fat rings flatten out in
the plane of rotation, and the inner $S^1$ radius gets smaller while
the outer $S^1$ radius grows   
(the shape of black rings was studied in detail in \cite{EEV}). 
As $j$ increases, the fat
rings spin faster and become colder, much like the fast spinning
Myers-Perry black hole. As $j \to 1$, the fat rings approach the same
naked ring singularity of the $j=1$ Myers-Perry solution. 

A thin black ring has a nearly round $S^2$,
and the $S^1$ radius is larger than the $S^2$ radius. As
$j$ increases, thin black rings get hotter as the $S^2$ gets smaller
(and the ring thinner), and the angular velocity decreases. 
We shall see that many ``phases'' of black saturn also have
versions of the ``thin" and the ``fat" black ring branches.


\subsection{Configurations with $J^\rom{BH}_\rom{Komar}=0$}
\label{s:cc2zero}

Throughout this section we study the subfamily of black saturn with
$\bc_2=0$. It was shown in section \ref{s:komar} that for $\bc_2=0$
 the intrinsic angular momentum of the
$S^3$ black hole vanishes, $J_\rom{Komar}^\rom{BH} = 0$. 

When $\bc_2=0$ it is simple to solve the balance condition 
\reef{balance} 
for $\ka_3$: there are two solutions, but only one of
them satisfies the constraints 
$0<\ka_3(\ka_1,\ka_2)<\ka_2<\ka_1<1$. 
In order to illustrate the 
physics of the solution, we choose to further fix a physical quantity, 
so that we are left with a 1-parameter family of solutions. The extra
physical parameter to be fixed will be either the reduced area of the
black ring  $a_\rom{H}^\rom{BR}$ (section \ref{s:fixedaBR}) or 
the $S^3$ black hole  $a_\rom{H}^\rom{BH}$ (section \ref{s:fixedaBH}). 
Alternatively, we fix in section \ref{s:test} the Komar mass of the
black hole and test the gravitational interaction between the $S^3$
black hole and the black ring. 


\subsubsection{Fixed area black ring}
\label{s:fixedaBR}

\begin{figure}[p]
\begin{picture}(0,0)(0,0)
\put(13,90){\scriptsize$a_\textrm{H}$}
\put(82,48){\scriptsize$j^2$}
\put(97,90){\scriptsize$a_\textrm{H}$}
\put(165,48){\scriptsize$j^2$}
\put(12,29){\scriptsize$\omega_\textrm{H}$}
\put(82,-13){\scriptsize$j^2$}
\put(96,29){\scriptsize$\omega_\textrm{H}$}
\put(164,-13){\scriptsize$j^2$}
\put(12,-40){\scriptsize$\log\tau_\textrm{H}$}
\put(82,-83){\scriptsize$j^2$}
\put(97,-40){\scriptsize$\log\tau_\textrm{H}$}
\put(165,-82){\scriptsize$j^2$}
%
\put(9,39){\footnotesize Figure \ref{fig:cc2zeroBR}(a): Total $a_\textrm{H}$ for
        $a_\textrm{H}^\textrm{BR}=0.05$ (solid) and}
\put(9,34){\footnotesize  $a_\textrm{H}^\textrm{BR}=0.1$ (dotted).}
\put(104,39){\footnotesize Figure \ref{fig:cc2zeroBR}(b): Total $a_\textrm{H}$ for $a_\textrm{H}^\textrm{BR}=0.8$.}
\put(8,-22){\footnotesize Figure \ref{fig:cc2zeroBR}(c): $\omega_\textrm{BR}$ (upper curve) and
        $\omega_\textrm{BH}$ (lower }
\put(8,-27){\footnotesize  curve) for $a_\textrm{H}^\textrm{BR}=0.05$ (solid) and $a_\textrm{H}^\textrm{BR}=0.1$ }
\put(8,-32){\footnotesize (dotted).}
\put(92,-22){\footnotesize Figure \ref{fig:cc2zeroBR}(d): $\omega_\textrm{BR}$ (upper curve) and
        $\omega_\textrm{BH}$ (lower}
\put(92,-27){\footnotesize curve) for $a_\textrm{H}^\textrm{BR}=0.8$.}
\put(10,-92){\footnotesize Figure \ref{fig:cc2zeroBR}(e): $\tau_\textrm{BR}$ (lower curve) and
        $\tau_\textrm{BH}$ (upper}
\put(10,-97){\footnotesize curve) for
        $a_\textrm{H}^\textrm{BR}=0.05$ (solid) and
        $a_\textrm{H}^\textrm{BR}=0.1$ }
\put(10,-102){\footnotesize (dotted). }
\put(94,-93){\footnotesize Figure \ref{fig:cc2zeroBR}(f): $\tau_\textrm{BR}$ (lower curve) and
        $\tau_\textrm{BH}$ (upper}
\put(94,-98){\footnotesize curve) for
        $a_\textrm{H}^\textrm{BR}=0.8$.}
\end{picture}
\centerline{
       \begin{tabular}{cc}
        \includegraphics[width=7.2cm]{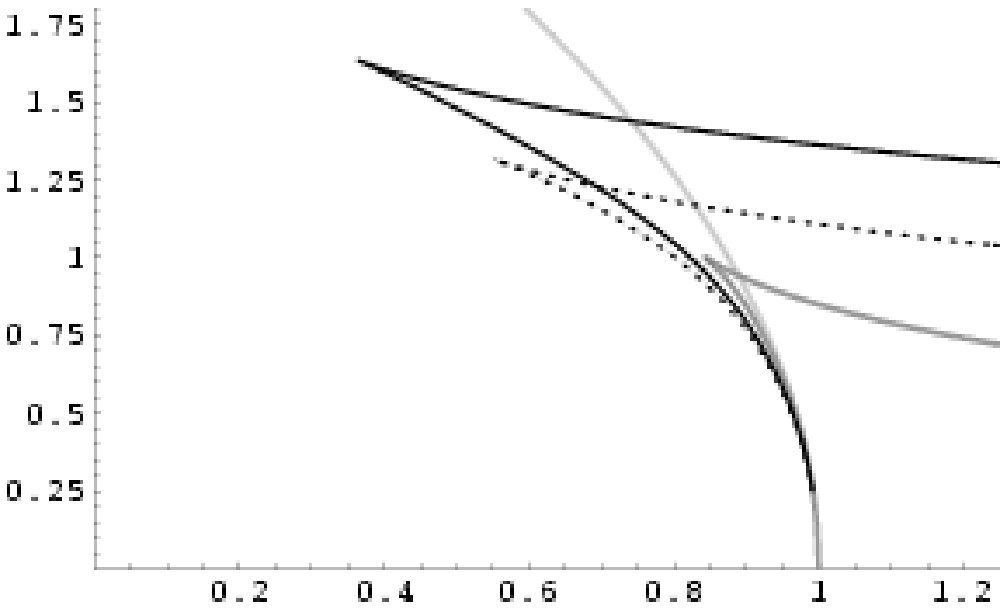}
        &\hspace{6mm}
        \includegraphics[width=7.2cm]{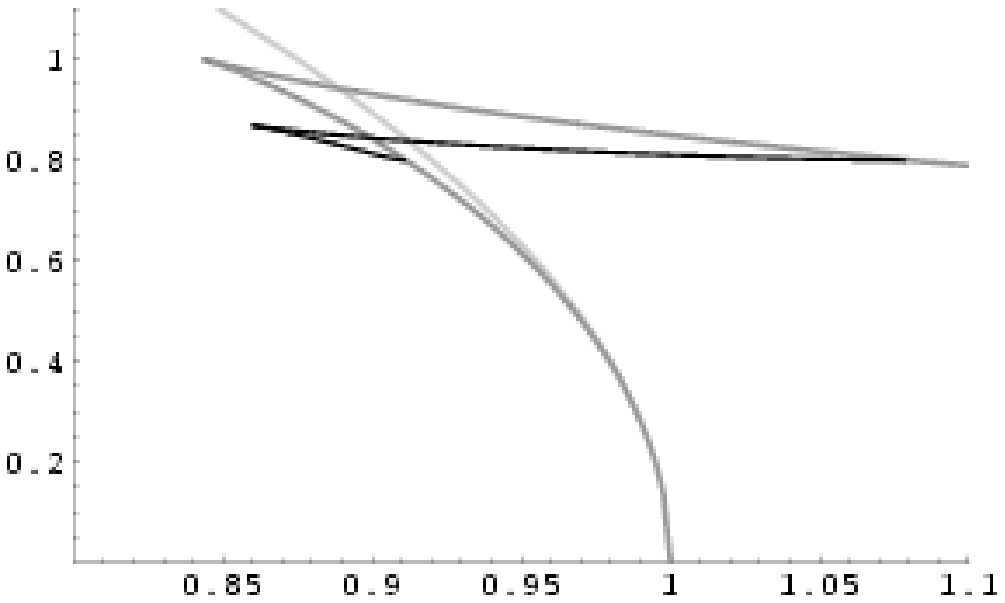}
        \\[15mm]
        \includegraphics[width=7.2cm]{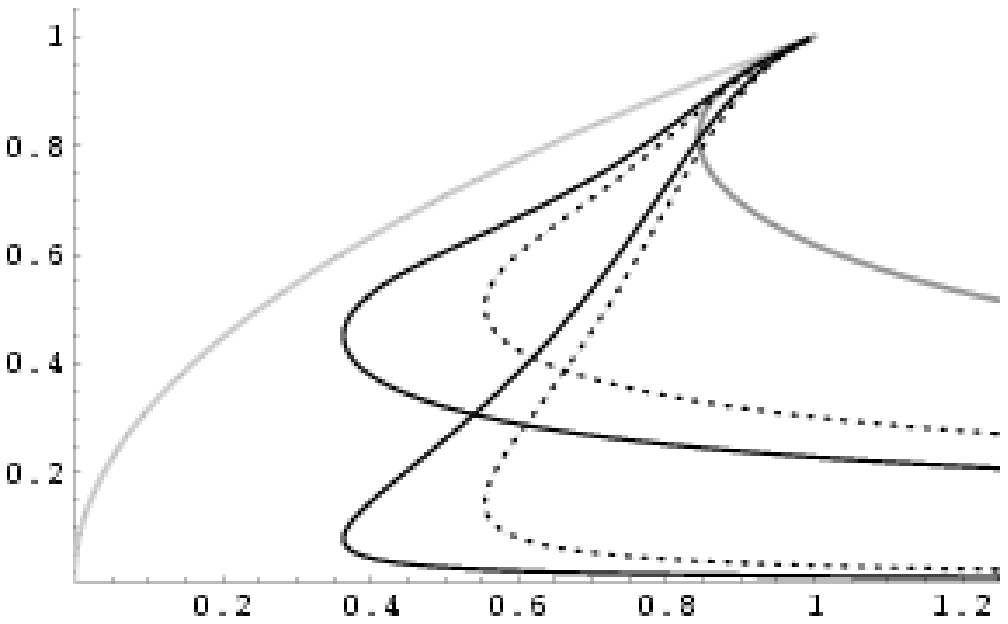}
        & \hspace{6mm}
        \includegraphics[width=7.2cm]{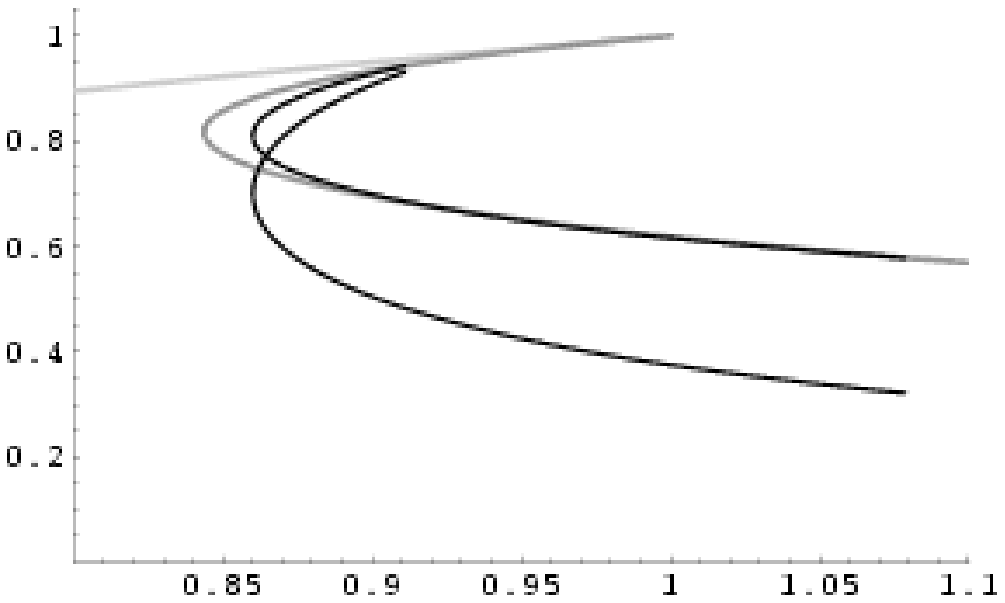}
        \\[23mm]
        \includegraphics[width=7.2cm]{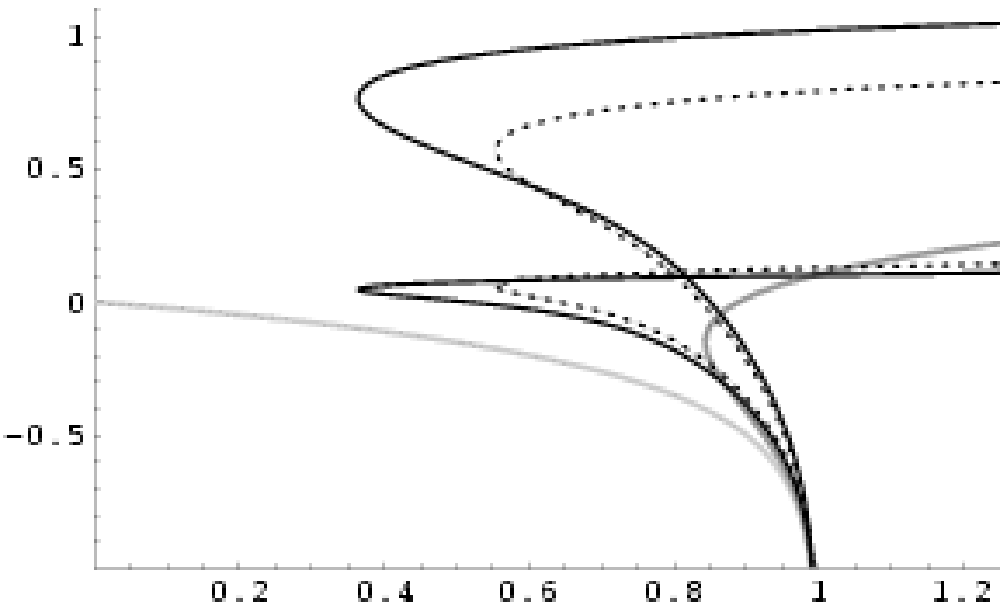}
        &\hspace{6mm}
        \includegraphics[width=7.2cm]{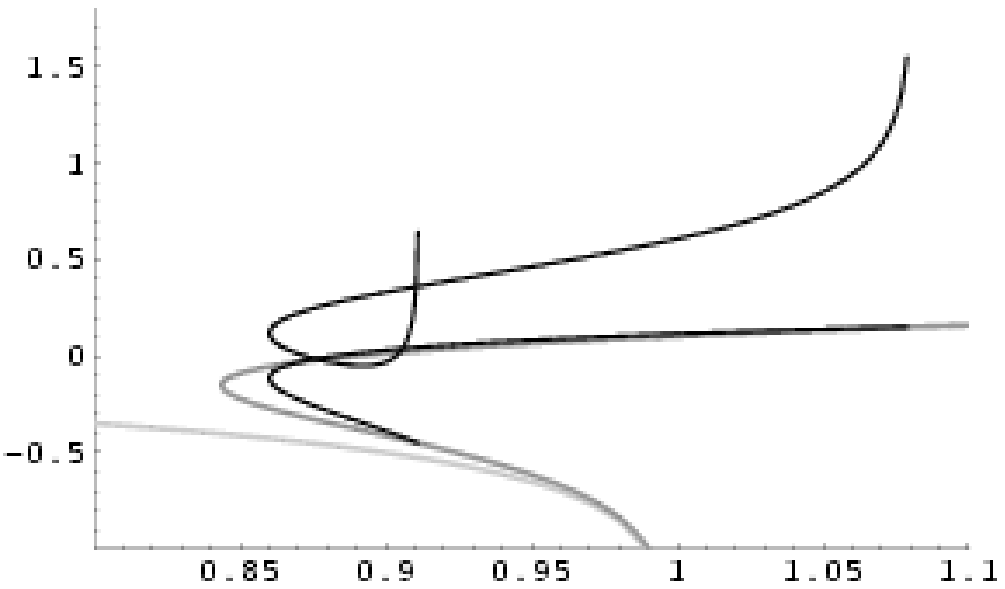}
        \end{tabular}
        }
\begin{picture}(0,0)(0,0)
\put(34,104){\scriptsize$\omega_\textrm{BH}$}
\put(72,99){\scriptsize$\omega_\textrm{BR}$}
\put(102,112){\scriptsize$\omega_\textrm{BH}$}
\put(100,104){\scriptsize$\omega_\textrm{BR}$}
\put(23,19){\scriptsize$\log\tau_\textrm{BH}$}
\put(70,16){\scriptsize$\log\tau_\textrm{BR}$}
\put(69,17){\vector(-2,3){5}}
\put(103,11){\scriptsize$\log\tau_\textrm{BH}$}
\put(102,30){\scriptsize$\log\tau_\textrm{BR}$}
\put(107,29){\vector(1,-4){2.2}}
\end{picture}
\vspace{18mm}
\caption{\small For fixed total mass and some representative values of
  the $a_\textrm{H}^\textrm{BR}$, the various reduced quantities are
  plotted vs. $j^2$. The gray curves correspond to the Myers-Perry black hole
  (light gray) and the black ring (darker gray) respectively.} 
\label{fig:cc2zeroBR}
\end{figure}

As shown in figure \ref{fig:cc2zeroBR2}(a), the reduced area
$a_\rom{H}^\rom{BR}$ of a single black ring takes values
$0<a_\rom{H}^\rom{BR}\le 1$. We can therefore fix the reduced black
ring area at any value between 0 and 1 and then ``grow'' the $S^3$
black hole at the center of the black ring.  
The result is illustrated
for representative values of $a_\rom{H}^\rom{BR}$ in figure
\ref{fig:cc2zeroBR}.

For any value $0<a_\rom{H}^\rom{BR}\le 1$, there exist both a fat and
a thin black ring, and the $S^3$ black hole can be grown from
either. 
This is illustrated most clearly in figure   
\ref{fig:cc2zeroBR}(b), where we have fixed $a_\rom{H}^\rom{BR}=0.8$
and plotted $a_\rom{H}^\rom{total}$
vs.~$j^2$. The standard Myers-Perry black hole ``phase'' is shown in
light gray, 
the black ring ``phase'' in darker gray. The black
saturn configuration with fixed $a_\rom{H}^\rom{BR}=0.8$ (black curve)
starts at the 
thin and fat black ring branches at $a_\rom{H}=0.8$.
Since $J_\rom{Komar}^\rom{BH} = 0$, the $S^3$ black hole contributes no
angular momentum, and hence $j$ decreases as long as the black hole
grows, i.e.~until reaching the cusp of the curve in figure
\ref{fig:cc2zeroBR}(b). 

Figure \ref{fig:cc2zeroBR}(a) shows similarly the growth of an $S^3$
black hole at the center of the ring, but now with
$a_\rom{H}^\rom{BR}$ fixed at smaller values, $a_\rom{H}^\rom{BR}=0.1$
(dotted) and $0.05$ (solid). 
The plot shows the saturn ``phases'' grow from the standard fat black
ring branch; they meet the thin black ring branch at very large values
of $j$ not shown in figure \ref{fig:cc2zeroBR}(a).  
For such small fixed areas of
the black ring, the $S^3$ black hole is allowed to grow very large, and
these saturn ``phases'' dominate the standard black ring branch entropically. 

{}Figures \ref{fig:cc2zeroBR}(c) and \ref{fig:cc2zeroBR}(d) show that
the $S^3$ black hole is rotating, i.e.~it has non-zero angular
velocity $\omega_\rom{BH}$. 
That the $S^3$ black hole rotates despite carrying no intrinsic
angular momentum 
($J_\rom{Komar}^\rom{BH} = 0$) is naturally interpreted as gravitational
frame-dragging: the rotating black ring drags the spacetime
surrounding it and that causes the $S^3$ horizon to rotate. This
interpretation is supported by the fact that the angular velocity
$\omega_\rom{BH}$ follows, 
and is always smaller than, $\omega_\rom{BR}$. 

To gain a better understanding of the physics of the black saturn, we
first focus on the cases of small values of the fixed black ring
area. The relevant plots are figure
\ref{fig:cc2zeroBR}(a) ($a_\rom{H}^\rom{total}$ vs. $j^2$), figure
\ref{fig:cc2zeroBR}(c) (angular velocities vs.~$j^2$), and figure
\ref{fig:cc2zeroBR}(e) (temperatures\footnote{Note that we plot
  temperatures on a logarithmic scale in order to better capture the
  structure of all phases in one plot.} 
vs.~$j^2$) for fixed
$a_\rom{H}^\rom{BR}=0.1$ (dotted) and $0.05$ (solid). 
For these values of the black ring area, the thin black ring is very thin, has large $S^1$
radius and is rotating slowly ($\omega_\rom{BR}$ is small). 
A small $S^3$ black hole at the center of such a thin black ring will hardly
feel the surrounding ring. Indeed, for large $j$, the $S^3$  black
hole has very small angular velocity $\omega_\rom{BH}$ (figure
\ref{fig:cc2zeroBR}(c)), and it has large temperature $\tau_\rom{BH}$
(figure \ref{fig:cc2zeroBR}(e)) which decreases as the black hole
grows. Thus the black hole behaves much like a small-mass Schwarzschild
black hole, and we expect its horizon to be nearly round as long as it
has small area.  

Instead of growing the $S^3$ black hole from the thin black ring
branch, consider starting with the fat black ring with
$a_\rom{H}^\rom{BR}=0.05$ or $0.1$. 
The fat black ring has $j$ near 1, the horizon is flattened out and it
rotates fast. The $S^3$ black hole growing from this configuration
will naturally be highly affected by the
surrounding black ring. 
Consequently, the dragging-effect is much stronger, and indeed figure \ref{fig:cc2zeroBR}(c) shows that the $S^3$ black hole is
rotating fast. Its temperature is very small (figure \ref{fig:cc2zeroBR}(e)), 
so it behaves much
like the highly spinning small area Myers-Perry black hole near
$j=1$. Thus we expect the $S^3$ black hole to flatten out in the plane of the ring
in this regime of black saturn. 

{}Figures \ref{fig:cc2zeroBR}(b), \ref{fig:cc2zeroBR}(d), and
\ref{fig:cc2zeroBR}(f) show the equivalent plots for the black ring area
fixed at a larger value $a_\rom{H}^\rom{BR}=0.8$. In this case, the
distinction between growing the black hole from the thin or fat black
ring branches is less pronounced. The $S^3$ black hole is always
dragged along to that the angular velocity is far from zero, but even
as the black ring becomes fat, the $S^3$ black hole never spins so fast
that it enters the regime of the near-$j=1$ Myers-Perry black
hole as the area goes to zero. This effect can be seen from the
temperature $\tau_\rom{BH}$ which increases as $a_\rom{H}^\rom{BH} \to
0$ --- compare figures 
\ref{fig:cc2zeroBR}(e) and \ref{fig:cc2zeroBR}(f). 

Increasing the fixed value of the black ring area,
$a_\rom{H}^\rom{BR}$, the corresponding black saturn ``phase'' becomes
smaller and smaller, and for fixed $a_\rom{H}^\rom{BR}=1$ we find no
saturn solutions. This is because growing the $S^3$ black hole with
$J_\rom{Komar}^\rom{BH} = 0$ decreases the total angular momentum $j$,
and for the black ring with $j=\sqrt{27/32}$ and
$a_\rom{H}^\rom{BR}=1$, there 
are no black ring solutions with less angular momentum.

\vspace{2mm}

Finally, let us note that it is possible to fix the black ring area to be zero,
$a_\rom{H}^\rom{BR}=0$. The $a_\rom{H}^\rom{BR}=0$ saturn configuration
describes a nakedly singular ring rotating around the $S^3$ black
hole, which is also rotating as it is being dragged along by the ring
singularity. The reduced area of the $S^3$ black hole vs.~$j^2$ for this
configuration is shown as a dotted curve in figure \ref{fig:cc2zeroBH}.


\subsubsection{Fixed area black hole}
\label{s:fixedaBH}

\begin{figure}[t]
\begin{picture}(0,0)(0,0)
\put(42,-2){\footnotesize$a_\textrm{H}$}
\put(134,-60){\footnotesize$j^2$}
%
\put(35,-6){\footnotesize$2\sqrt 2$}
\put(35,-13.5){\footnotesize$2.45$}
\put(39.5,-22.5){\footnotesize$2$}
\put(39.5,-41.5){\footnotesize$1$}
\put(41,-63){\footnotesize$0.039$}
\put(61,-63){\footnotesize$0.29$}
\put(80.5,-63){\footnotesize$0.55$}
\put(102,-63){\footnotesize$\frac{27}{32}$}
\put(106,-63){\footnotesize$0.86$}
\put(116.5,-63){\footnotesize$1$}
\put(123,-63){\footnotesize$1.1$}
%
%
\end{picture}
  \begin{center}
  \hspace{-4.6cm}
  \includegraphics[width=9cm]{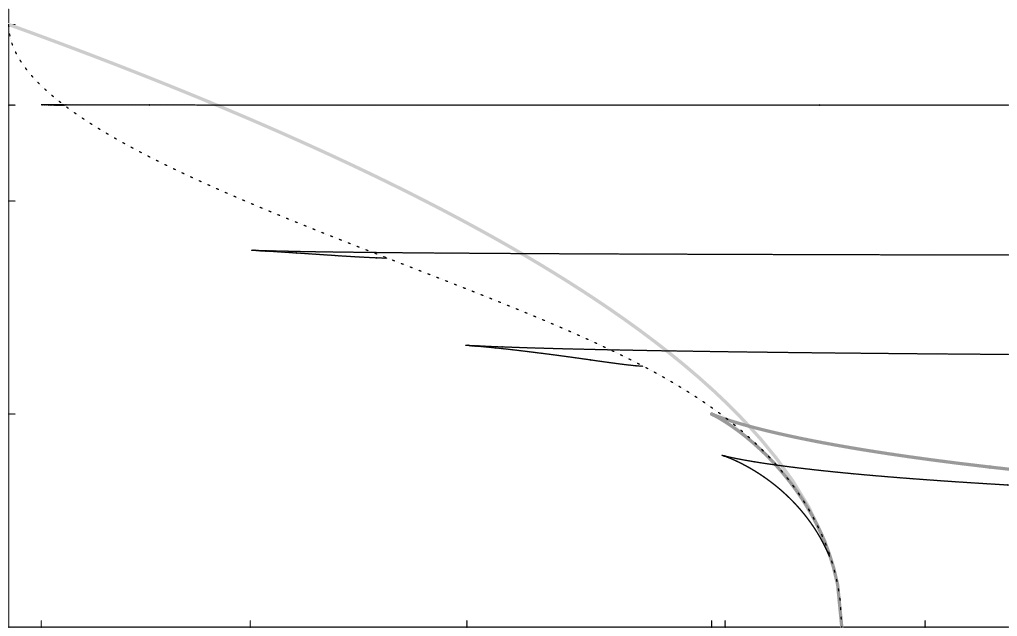}
  \hspace{-8cm}
  \raisebox{0.8cm}{
  \includegraphics[width=2.5cm]{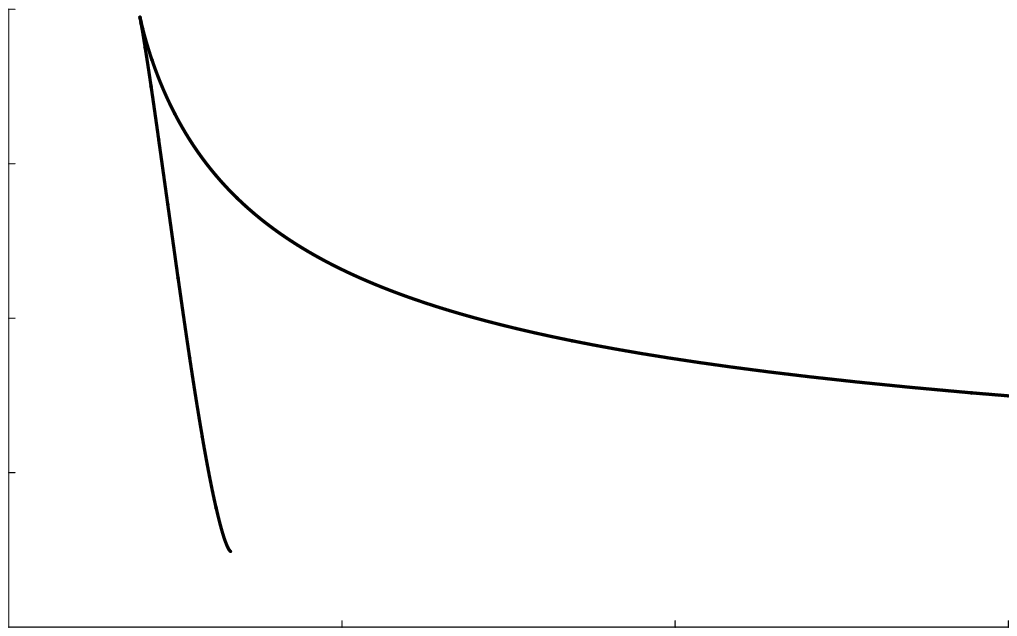}}
  \vspace{4mm}
  \caption{\small Plots of $a_\textrm{H}$ vs.~$j^2$ for different
    representative values of 
    $(a_\textrm{H}^\textrm{BH})^2=6,3,\frac{3}{2},
    \frac{1}{10}$ (black solid curves). The dotted curve corresponds to a
    Myers-Perry black hole surrounded by a nakedly 
    singular ring. Again, the gray curves correspond to the Myers-Perry hole
    (gray) and the black ring (darker gray). The smaller plot zooms in on the small $j$
    part of the $a_\textrm{H}^\textrm{BH}=\sqrt{6}$ curve.}
    \label{fig:cc2zeroBH}
  \end{center}
\begin{picture}(0,0)(0,0)
\put(57,89){\footnotesize$a_\textrm{H}^\textrm{BH}=\sqrt 6$}
\put(56,89){\vector(-2,-3){4}}
\put(96,71){\footnotesize$a_\textrm{H}^\textrm{BH}=\sqrt 3$}
\put(107,63){\footnotesize$a_\textrm{H}^\textrm{BH}=\sqrt{\frac{3}{2}}$}
\put(118,42){\footnotesize$a_\textrm{H}^\textrm{BH}=\sqrt{\frac{1}{10}}$}
\put(61.8,40.5){\scriptsize$0.05$}
\put(71.5,40.5){\scriptsize$0.1$}
\put(49.5,58){\scriptsize$2.453$}
\put(49.5,54){\scriptsize$2.452$}
\put(49.5,50){\scriptsize$2.451$}
\put(51,46){\scriptsize$2.45$}
\end{picture}
\end{figure}

We keep $\bc_2 = 0$ as before, so that $J_\rom{Komar}^\rom{BH} = 0$,
but instead of keeping the black ring area  $a_\rom{H}^\rom{BR}$ fixed
as in the previous subsection we now fix the $S^3$ black hole area
$a_\rom{H}^\rom{BH}$.  
Thus we ``grow'' a black ring around the $S^3$ black hole area of fixed area.
A balanced black ring cannot exist for arbitrarily small angular 
momentum while keeping the configuration in equilibrium, so the black
ring grows from a nakedly singular ring around the Myers-Perry black
hole; this is nothing but the $a_\rom{H}^\rom{BR}=0$ configuration
discussed at the end of the previous section, and shown as the
dotted curve in figure \ref{fig:cc2zeroBH}.  

{}Figure \ref{fig:cc2zeroBH} shows black saturn phases with
fixed black hole area for representative values of
$a_\rom{H}^\rom{BH}$. For
each value of $a_\rom{H}^\rom{BH}$, the corresponding curve has a fat and
a thin black ring phase. Note that the thin ring branches extend to large
values of $j$. 

The large-$j$ tails of the constant $a^\rom{BH}_\textrm{H}$ curves show
that balanced saturn configurations can have very large entropies. It
can be argued \cite{EEF2} that for any fixed value of
$0 < a^\rom{BH}_\textrm{H} < 2\sqrt{2}$, the tails extend to arbitrarily
large $j$. This in turn means that for any $j$ there exist black
saturn configurations with total area $a_\rom{H}^\rom{total}$
arbitrarily close to $2\sqrt{2}$. We refer to \cite{EEF2} for further
details.  

When the $S^3$ black hole area is close to zero, the black saturn
curves approach the phase of the single black ring.  
Since the black hole itself does not carry any intrinsic spin, we can
set its area to zero, $a_\rom{H}^\rom{BH}=0$, and then black saturn
simply reduces to the black ring solution. 

It is worth noting that for large values of $a^\rom{BH}_\textrm{H}$,
the black saturn curves also extend to small values of $j$. For $\bc_2 =
0$, the saturn phases  never reach $j=0$. This is expected because
$j=0$ requires that the black hole and the black ring are
counter-rotating and that is never the case for the $\bc_2=0$. 


\subsubsection{Saturn frame-dragging}
\label{s:test}

Above we have seen that in the presence of the rotating black ring of
black saturn, an $S^3$ black hole with no intrinsic spin
($J_\rom{Komar}^\rom{BH}=0$) can be rotating 
($\omega_\rom{BH} \ne 0$). We have interpreted this 
as a consequence of gravitational frame-dragging. 
We test this interpretation by studying the
geometry of the black saturn configuration (still keeping
$\bc_2=0$). If indeed 
we are seeing frame-dragging, then the effect should be very small
when the black ring is thin and very far from the $S^3$
black hole, and increase as the black ring and the black hole come
closer. We keep $m_\rom{BH}= M^\rom{BH}_\rom{Komar}/M$ fixed and 
let the distance between the black hole and black ring vary.

To characterize the configuration, we first introduce the reduced
inner and outer $S^1$ (horizon) radii of the black ring  
\bea
  r_\rom{inner} = (G M)^{-1/2} \sqrt{G_{\psi\psi}} \Big|_{\rho=0,\,
    \bar{z}=\ka_2} 
  \, , 
  \hspace{1cm}
  r_\rom{outer} = (G M)^{-1/2} \sqrt{G_{\psi\psi}} \Big|_{\rho=0,\,
    \bar{z}=\ka_3} 
  \, .
\eea
It is shown in \cite{EEV} that for a single black ring of fixed mass, 
the inner radius $r_\rom{inner}$ decreases monotonically when going  
from the thin black ring (i.e.~large-$j$) regime to the fat black ring
branch, and that  
$r_\rom{inner} \to 0$ when $j \to 1$ for the fat black rings. However,
when a black hole is present at the center of the ring, as in black
saturn, there is a lower bound on the inner radius of the black 
ring.\footnote{As pointed out in section \ref{s:limits}, there is no
  smooth merger limit for the balanced black saturn system.}

\begin{figure}[t]
\begin{picture}(0,0)(0,0)
\put(16,-3){\scriptsize $\omega_i$}
\put(78,-41){\scriptsize $r_\rom{inner}$}
\put(12,-10){\scriptsize $0.6$}
\put(12,-20.5){\scriptsize $0.4$}
\put(12,-31){\scriptsize $0.2$}
\put(27.5,-44){\scriptsize $3$}
\put(51,-44){\scriptsize $4$}
\put(75,-44){\scriptsize $5$}
%
\put(92,-3){\scriptsize $\omega_i$}
\put(154,-41){\scriptsize $\ell$}
\put(90,-5){\scriptsize $1$}
\put(88,-23.5){\scriptsize $0.5$}
\put(111.2,-44){\scriptsize $1$}
\put(130.7,-44){\scriptsize $2$}
\put(150.2,-44){\scriptsize $3$}
%
%
%
\put(41,-50){\footnotesize Figure \ref{fig:dragging}(a)}
\put(119,-50){\footnotesize Figure \ref{fig:dragging}(b)}
\end{picture}
  \begin{center}
  \includegraphics[width=6cm]{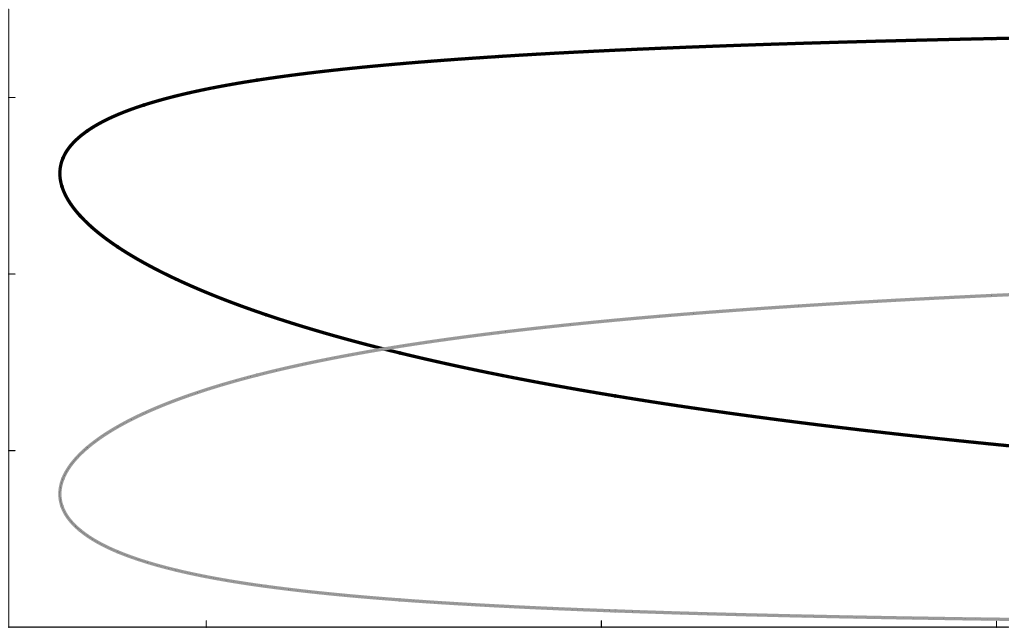}
  \hspace{1.3cm}
  \includegraphics[width=6cm]{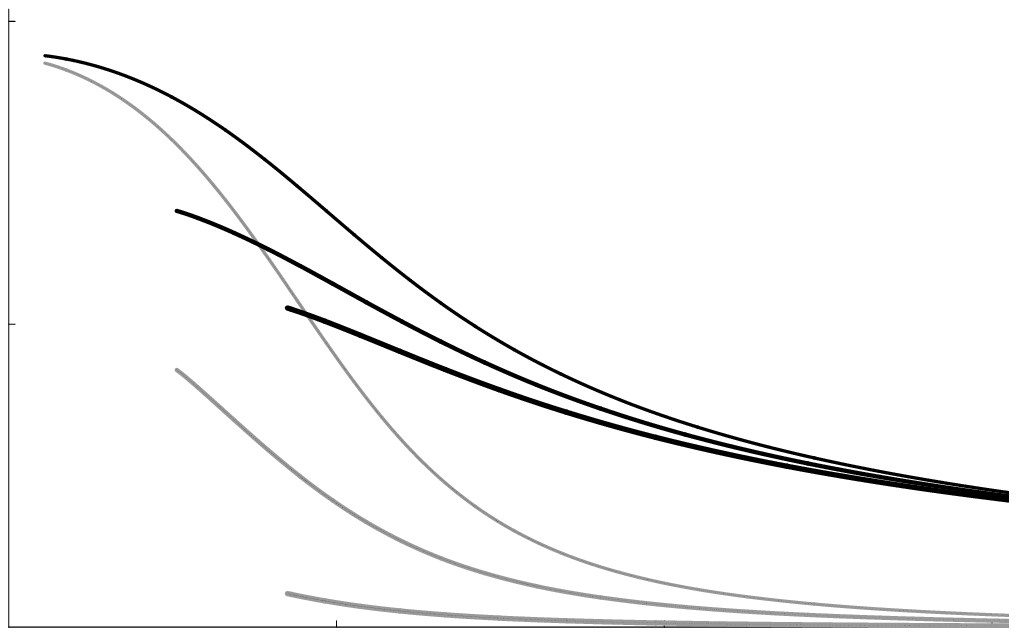}
  \vspace{4mm}
  \bigskip
  \caption{\small Test of frame-dragging: Figure \ref{fig:dragging}(a)
  shows angular velocities $\omega_i$ vs.~the inner radius of the
  black ring $r_\rom{inner}$ for fixed Komar mass
  $m_\rom{BH} = M^\rom{BH}_\rom{Komar}/M=0.5$. 
  Figure \ref{fig:dragging}(b) shows the angular velocities $\omega_i$
  vs.~the proper distance $\ell$ between the black hole and the black
  ring for three different mass distributions: 
  $m_\rom{BH} = M^\rom{BH}_\rom{Komar}/M=0.1$, $0.5$, and  $0.9$.} 
    \label{fig:dragging}
  \end{center}
\begin{picture}(0,0)(0,0)
\put(21,72.5){\scriptsize $\omega_\rom{BR}$}
\put(21,54.5){\scriptsize $\omega_\rom{BH}$}
\put(121,68){\scriptsize $m_\rom{BH}=0.1$}
\put(120,68.5){\vector(-4,0){13}}
\put(120,68.5){\vector(-1,-4){4.2}}
\put(124,64){\scriptsize $m_\rom{BH}=0.5$}
\put(123,64.5){\vector(-4,-1){12.1}}
\put(123,64.5){\vector(-1,-4){5}}
\put(127,60){\scriptsize $m_\rom{BH}=0.9$}
\put(126,60.5){\vector(-4,-1){12.5}}
\put(126,60.5){\vector(-1,-4){5}}
\end{picture}
\end{figure}

{}Figure \ref{fig:dragging}(a) shows the angular velocities of
the $S^3$ black hole and the black ring plotted vs.~$r_\rom{inner}$
for fixed Komar mass $m_\rom{BH}=0.5$. The lower branch of the
$\omega_\rom{BR}$ curve (black) corresponds to the slowly rotating
``thin'' black ring. For large radius, the $S^3$ black hole is not affected
much and $\omega_\rom{BH}$ is correspondingly small (gray curve).
As $r_\rom{inner}$ decreases the black ring
spins faster and so does the $S^3$ black hole.
Clearly there is a lower bound for $r_\rom{inner}$, but
surprisingly, the ring starts growing after reaching this
minimum. It turns out that on the
``upper'' branch of the $\omega_\rom{BR}$ curve, the inner and outer
$S^1$ radii approach each other, so that the ring again becomes thin.
But contrary to the standard thin black rings, the angular velocity
increases as the ring grows. Eventually, as the black ring becomes
thinner, the area $a_\rom{H}^\rom{BR}$ goes to zero leaving just a
nakedly singular black ring around a Myers-Perry black hole (dotted
curve in figure \ref{fig:cc2zeroBH}).  
 
As shown in figure \ref{fig:dragging}(a), the $S^3$ black hole angular
velocity, $\omega_\rom{BH}$, follows that of the black ring. In
particular, $\omega_\rom{BH}$ continues to grow even if the inner
radius of the black ring is growing. This may at first seem to
contradict that the rotation of the $S^3$ black hole is caused by
frame-dragging, since it would seem that the $S^3$ black hole should slow
down as the black ring becomes thinner and its $S^1$ radius grows.
However, since the $S^3$ black hole is itself
rotating, it flattens out in the plane of rotation. To
study this effect we compute the proper distance
between the $S^3$ black hole and the black ring (for fixed mass):
\bea
  \label{theell}
  \ell = (G M)^{-1/2} \int_{\ka_2}^{\ka_1} d\bar{z}\, 
  \sqrt{G_{\bar{z}\bar{z}}} \, .
\eea
As expected, the proper distance $\ell$ increases as the inner radius
of the black ring increases along the lower branch in figure
\ref{fig:dragging}(a).  
But even as the inner radius $r_\rom{inner}$ of the black ring
increases (upper branch), the proper distance $\ell$ continues to
decrease. This confirms that the black hole, as it is spinning faster, 
flattens out into the plane of rotation. Figure \ref{fig:dragging}(b)
shows the angular velocities as functions of the proper
distance $\ell$, for three different mass distributions $m_\rom{BH} =
0.1$, $0.5$, and $0.9$. The angular velocity of the black ring
increases as the proper distance $\ell$ decreases. And 
$\omega_\rom{BH} \to 0$ when $\ell$
becomes large.
This is precisely the behavior one
would expect from frame-dragging. 

Moreover, figure \ref{fig:dragging}(b) shows that the effect of
dragging depends on the relative masses of the black ring and
the $S^3$ black hole: the effect of a thin small-mass black ring on a
large-mass black hole is weak ($m_\rom{BH} = 0.9$), but the effect of
a thick massive black ring on a small-mass black hole is strong
($m_\rom{BH} = 0.1$).

The above analysis gives strong evidence that we are indeed
observing rotational frame-dragging.


\subsection{Black hole with intrinsic spin}
\label{s:BHwspin}

We now take $\bc_2 \ne 0$ and study the more general saturn
 configurations.
When $\bc_2 \ne 0$, the $S^3$ black hole and the black ring have
independent rotation parameters, in particular we can have
$J_\rom{Komar}^\rom{BH} \ne 0$. As a result, the two black objects can
be co- or counter-rotating. We illustrate the physics in two
examples. 


\subsubsection{Counter-rotation and $\Omega_\rom{BH}$=0}
In the previous section, the $S^3$ black hole had no intrinsic
rotation, $J_\rom{Komar}^\rom{BH}=0$, and it was rotating only because
it was dragged along by the black ring. With $\bc_2 \ne 0$ the $S^3$
black hole has its own intrinsic angular momentum $J_\rom{Komar}^\rom{BH}
\ne 0$, and it is possible to let the $S^3$ black hole counter-rotate
in such a way that the intrinsic angular momentum cancels the effect
of the dragging, so that the $S^3$ horizon becomes non-rotating,
$\omega_\rom{BH} = 0$. 

\begin{figure}[t]
  \begin{center}
  \hspace{-.6cm}
  \includegraphics[width=9cm]{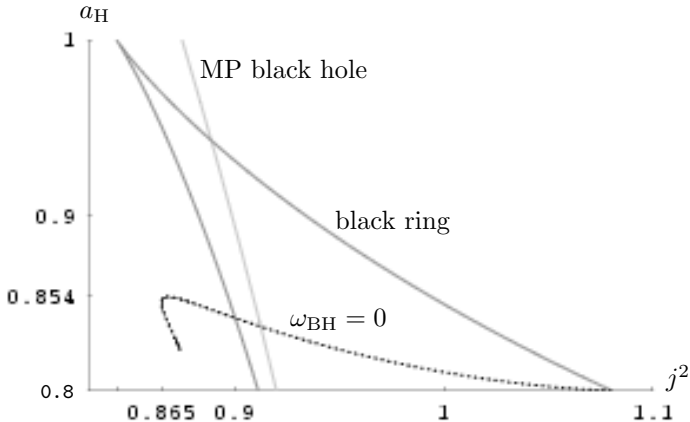}
  \caption{\small  Black saturn with a non-spinning black hole. The
    plot shows $a_\textrm{H}^\rom{total}$ vs. $j^2$ for  
    fixed $a_\rom{H}^\textrm{BR}=0.8$ and $\omega_\textrm{BH}=0$
    (dotted curve).  
    For reference, the dark gray curve is the black ring while the
    lighter gray curve is the Myers-Perry black hole.} 
  \label{fig:OmegaBHzero}
  \end{center}
\begin{picture}(0,0)(0,0)
\put(48,83){\footnotesize$a_\textrm{H}$}
\put(126,34){\footnotesize$j^2$}
\put(64,75){\footnotesize MP black hole}
\put(82,55){\footnotesize black ring}
\put(76,42){\footnotesize $\omega_\textrm{BH}=0$}
\put(42.8,32.5){\scriptsize\texttt{0.8}}
\end{picture}
\end{figure}

As an example of this effect, figure
\ref{fig:OmegaBHzero} shows a curve of black saturn solution with
$\omega_\rom{BH} = 0$ fixed. In addition we have also fixed
$a_\rom{H}^\rom{BR}=0.8$.  
This $\omega_\rom{BH} = 0$ curve starts at the thin black ring branch with
$a_\rom{H}=0.8$ and the black hole grows ``Schwarzschild style'' (zero
angular velocity, high temperature which decreases as the black hole
grows). As the black ring becomes fatter, the black hole
is affected more and more by the ring, and 
at some point its intrinsic counter-rotation can no longer resist the
dragging of the black ring; at this point the $\omega_\rom{BH}=0$
curve ends.  

The possibility of making the $S^3$ horizon non-rotating by turning on
``intrinsic'' angular momentum is reminiscent of the situation for the
4+1d supersymmetric $S^3$ black hole. This black hole also has a
non-rotating horizon, $\Omega = 0$, and it can be shown \cite{GMT}
that this requires angular momentum to be stored in the Maxwell fields
inside the horizon. Similar configurations were also discussed in
\cite{kunz}. 
Of course, for black saturn there are no Maxwell
fields to carry the angular momentum, but the picture of having
contributions to the rotation from ``inside'' the horizon to make
$\omega_\rom{BH}=0$ is common for the two systems.

\subsubsection{Reaching $j=0$}
\label{s:reachjz}

One might have expected the only 
solution with $j=0$ to be the Schwarzschild black hole. However,
taking into account solutions with more than one component of the
horizon, counter-rotation can give $j=0$. For black saturn this is
possible while maintaining balance. 

\begin{figure}[t]
\begin{picture}(0,0)(0,0)
\put(19,-2){\footnotesize$a_\textrm{H}$}
\put(94,-51){\footnotesize$j^2$}
\put(111,-15){\vector(3,-2){12}}
\put(74,-54){\footnotesize $1.19$}
\put(67,-54){\footnotesize $1$}
\put(58.5,-54.5){\scriptsize $\frac{27}{32}$}
\put(10,-5){\footnotesize $2\sqrt{2}$}
\put(10,-16){\footnotesize $2.18$}
\put(15,-35){\footnotesize $1$}
\put(130,-3){\footnotesize $j$}
\put(154,-28){\footnotesize $\omega_\rom{BH}$}
\put(110,-26){\footnotesize $-1$}
\put(147,-26){\footnotesize $1$}
\put(132,-11){\footnotesize $1$}
\put(132,-45.6){\footnotesize $-1$}
\put(132,-31.5){\footnotesize $-0.18$}
\end{picture}
  \begin{center}
  \includegraphics[width=7.5cm]{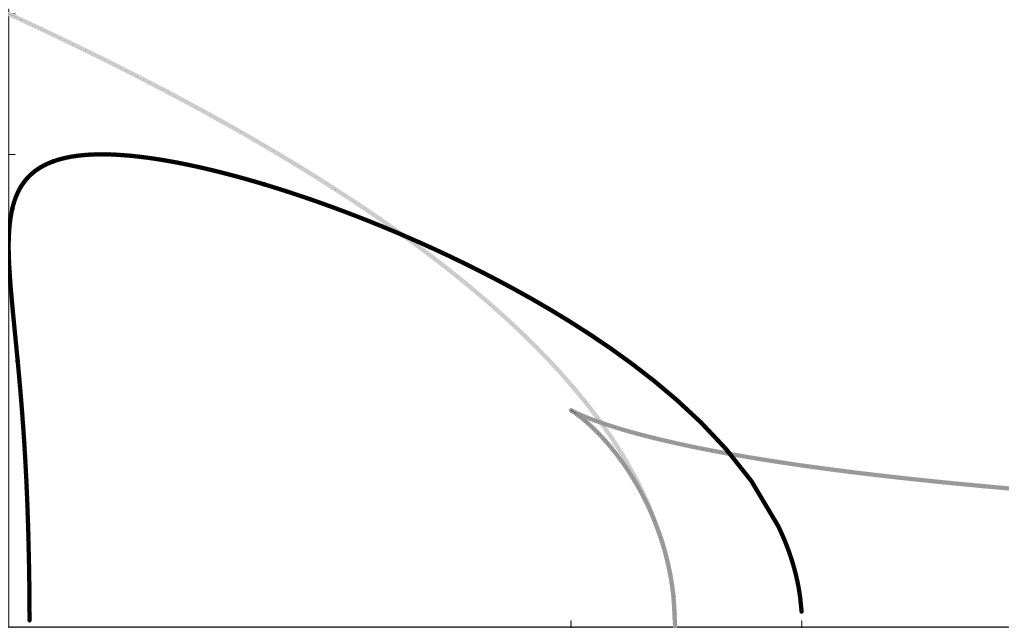}
  \hspace{1.5cm}
  \raisebox{0.3cm}{\includegraphics[width=4cm]{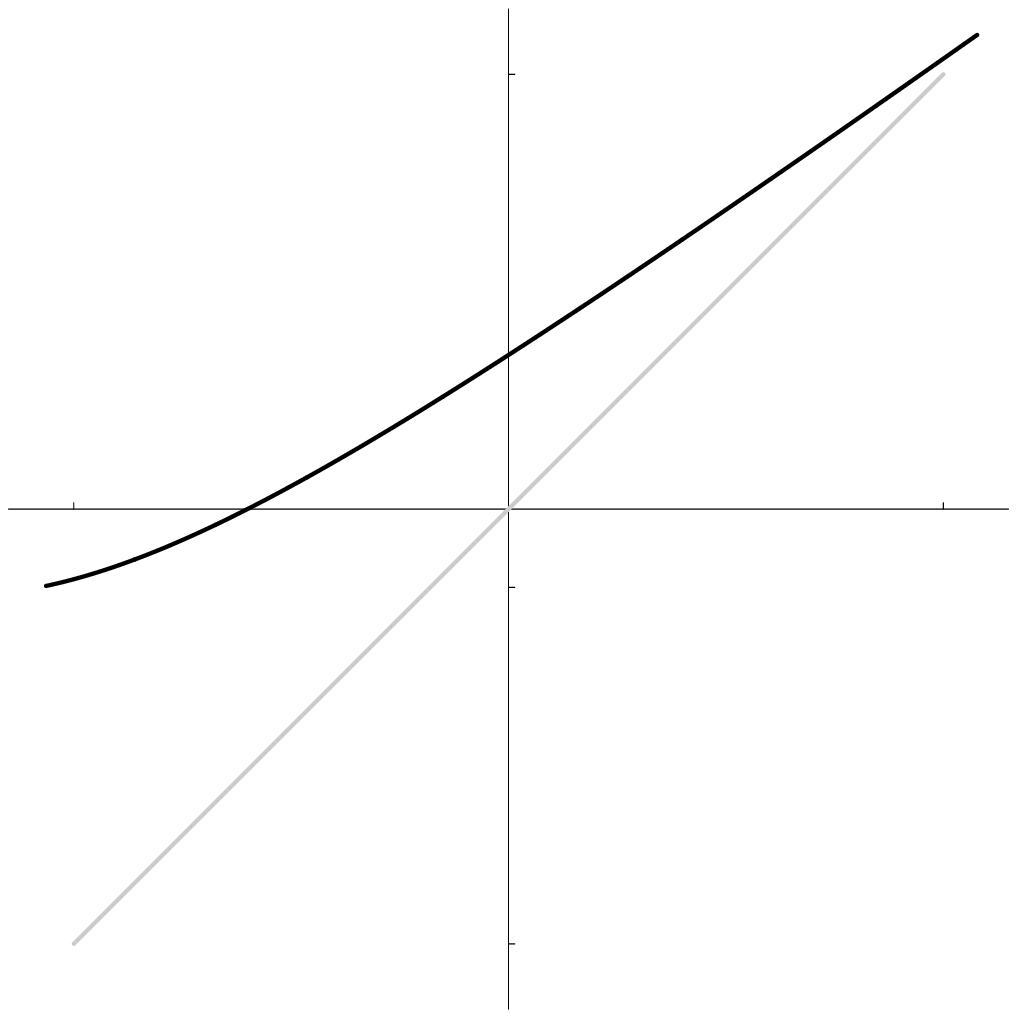}}
\vspace{4mm}
  \caption{\small Left: Plot of $a_\textrm{H}$ vs. $j^2$ for 
    fixed $a_\rom{H}^\textrm{BR}=0.01$ and $\omega_\rom{BR}=0.3$
    (black curve). 
    Included are also the ``phases'' of a single Myers-Perry black hole
    (gray) and the black ring (darker gray). 
    Right: $j$ is plotted vs.~the black hole for black saturn with
    $a_\rom{H}^\textrm{BR}=0.01$ and $\omega_\rom{BR}=0.3$ and for
    comparison with the corresponding curve for a single Myers-Perry
    black hole (light gray). The saturn configuration reaches $j=0$
    and extends to negative $j$.} 
     \label{fig:reachjz}
  \end{center}
\begin{picture}(0,0)(0,0)
\put(32,80){\footnotesize MP black hole}
\put(80,54){\footnotesize black ring}
\put(70,77){\footnotesize $a_\rom{H}^\rom{BR}=0.01$ and $\omega_\rom{BR}=0.3$}
\put(69,75){\vector(-2,-3){7}}
\end{picture}

\end{figure}

{}Figure \ref{fig:reachjz} shows a saturn configuration with
$a_\rom{H}^\rom{BR}=0.01$ and $\omega_\rom{BR}=0.3$ in the phase
diagram $a_\rom{H}^\rom{total}$ vs.~$j^2$. 
To reach $j=0$ requires that the black ring has small area, but
otherwise there is nothing special about the values chosen for 
$a_\rom{H}^\rom{BR}$ and $\omega_\rom{BR}$; they just illustrate the
physics well. For large values $j$, the
black ring and the $S^3$ black hole are co-rotating, as can be seen from
the $j$ vs.~$\omega_\rom{BH}$ plot in figure \ref{fig:reachjz}. As the
angular velocity of the black hole decreases, the total angular
momentum $j$ decreases and the area of the $S^3$ black
hole grows. The area reaches a maximum close to where the black hole
angular velocity vanishes. 
As the $S^3$ black hole counter-rotates, $\omega_\rom{BH} < 0$,
the area decreases. Eventually, the counter-rotation is such that the
total angular momentum at infinity vanishes, $j=0$. 
The black hole can be even more counter-rotating and then $j$ becomes
negative. Note from the $j$ vs.~$\omega_\rom{BH}$ plot in figure
\ref{fig:reachjz} that when the black holes are co-rotating $j$ is 
almost linear in the angular momentum, just as it is for a Myers-Perry
black hole, and the range covered $-1 \lsim \omega_\rom{BH} \lsim 1$,
is nearly the same. 

It is clear from figure \ref{fig:reachjz} that the 4+1d Schwarzschild
black hole and the slowly spinning Myers-Perry black holes are not
unique. We show in section \ref{sec:jzero} that there is a 2-fold
continuous family of black saturn solutions with $j=0$.


\subsection{Non-uniqueness}
\label{s:nonuniq2}

In the previous sections we have examined a number of examples which
--- among other phenomena --- all illustrated non-uniqueness in the
phase diagram $a_\rom{H}$ vs.~$j^2$. It is clear from these examples
that black saturn covers large regions of the phase diagram. We now
explore how large.

\subsubsection{Non-uniqueness in the phase diagram}
\begin{figure}[t!]
\begin{picture}(0,0)(0,0)
\put(137,1){\footnotesize $j$}
\put(62,63){\footnotesize $a_\rom{H}^\rom{total}$}
\put(38,-3){\footnotesize $-1$}
\put(86,-3){\footnotesize $1$}
\put(109.2,-3){\footnotesize $2$}
\put(132,-3){\footnotesize $3$}
\end{picture}
 \centerline{
 \includegraphics[width=10cm]{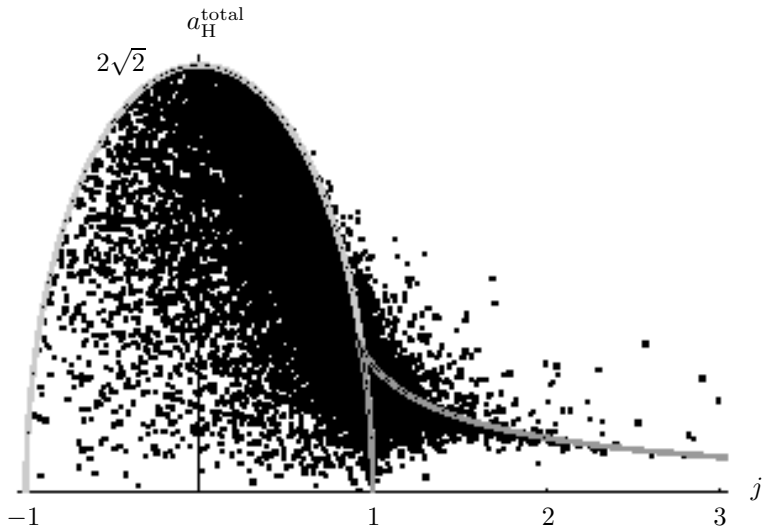}}
\begin{picture}(0,0)(0,0)
  \put(50,62.5){\footnotesize $2\sqrt{2}$}
\end{picture}
\caption{{\small Non-uniqueness in the phase diagram: The plot shows
    the distribution of black saturn for 100.000 randomly chosen black
    saturn configurations.}} 
\label{fig:phasediag}
\end{figure}

To study the region of the phase diagram
covered by black saturn, we choose random sets of points
$(\ka_1,\ka_2,\ka_3)$ satisfying the ordering \reef{order2} and plot 
the corresponding point $(j,a_\rom{H}^\rom{total})$ 
in the phase diagram.\footnote{The BZ parameter $\bc_2$ is fixed in terms of
  $(\ka_1,\ka_2,\ka_3)$ by the balance condition \reef{balance} with
  $\epsilon=+1$.} 
Figure \ref{fig:phasediag} shows the
distribution of 100.000 such points.  

We first note that we find no points with $j<-1$, thus
black saturn takes values of $j \ge -1$. The asymmetry between
positive and negative $j$ is just a choice of rotation direction,
which can be reversed by simply taking $\psi \to -\psi$ in the black
saturn metric.\footnote{It may also be noted that while
  $\omega_\rom{BH}$ takes both positive and negative values, we find
  that $\omega_\rom{BR}$ is always positive.   
The bound  $\omega_\rom{BR}>0$ is intuitively a
consequence of the fact that the black ring needs to rotate in order
to keep the system balanced.}

Next the total area
$a_\rom{H}^\rom{total}$ is always less than the area of the static
Schwarzschild black hole, which has $a_\rom{H}^\rom{Schw} =
2\sqrt{2}$. We believe that there are black saturn configurations with
$a_\rom{H}^\rom{total}$ arbitrarily close to
$a_\rom{H}^\rom{Schw}$. In fact for the dataset shown, we have 
\bea
  \min(a_\rom{H}^\rom{Schw}-a_\rom{H}^\rom{total})=9.5 \cdot 10^{-4} \, .
\eea

The distribution\footnote{The density in the distribution is caused by
  the discrete 
non-uniqueness in regions where both thin and fat black rings exist,
but can also be affected by the particular distribution of
points $(\ka_1,\ka_2,\ka_3)$.}
 of black saturn configurations in the phase diagram
figure \ref{fig:phasediag} indicates that the region bounded by the
Myers-Perry phase (shown in light gray for both positive and negative
$j$) is fully covered by black saturn solutions. But there are also
points outside this region: for $j$ greater than $\sim0.5$, there are
black saturn solutions with total area greater than the Myers-Perry
black hole and the black ring. 

By tuning the distribution, it can be shown \cite{EEF2} that the whole
open strip  
\bea
  0< a_\rom{H}^\rom{total} < 2\sqrt{2} = a_\rom{H}^\rom{Schw}  \, ,
  \hspace{1cm} 
  j\ge 0 \, ,
\eea
is covered with black saturn configurations. For any $j \ge 0$ the high-entropy configurations are black saturn with an almost static $S^3$ black hole (accounting for the high entropy) surrounded by a large thin black ring (carrying the angular momentum). This type of configuration allow us to have black saturns with total area arbitrarily close to the bound set by the static Schwarzschild black hole. Details of this and the structure of the phase diagram are presented in \cite{EEF2}.


\subsubsection{Balanced saturn with zero angular momentum $j$=0}
\label{sec:jzero}

The phase diagram figure \ref{fig:phasediag} strongly indicates that
the $j=0$ black saturn configurations are non-unique. We confirm the
non-uniqueness in this section
by studying the ranges of area, angular velocity and temperature
covered by the balanced $j=0$ saturn solutions.

\begin{figure}[t]
\begin{picture}(0,0)(0,0)
\put(102,-2){\footnotesize$a_\textrm{H}$}
\put(144,-44){\footnotesize $\tau_\rom{BR}$}
\put(8,-46){\footnotesize $\tau_\rom{BH}$}
\put(119,-42){\footnotesize $50\times a_\rom{H}^\rom{BR}$}
\end{picture}
 \begin{center}
 \raisebox{2mm}{\includegraphics[width=0.7cm]{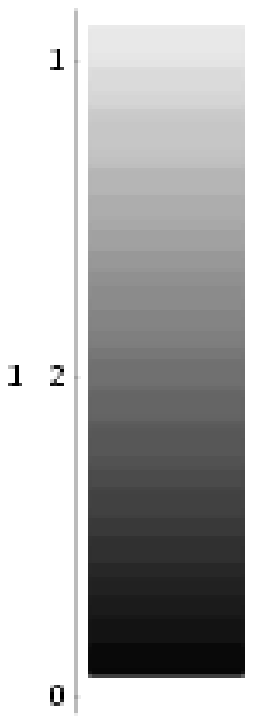}}
 \hspace{7mm}
 \includegraphics[width=11cm]{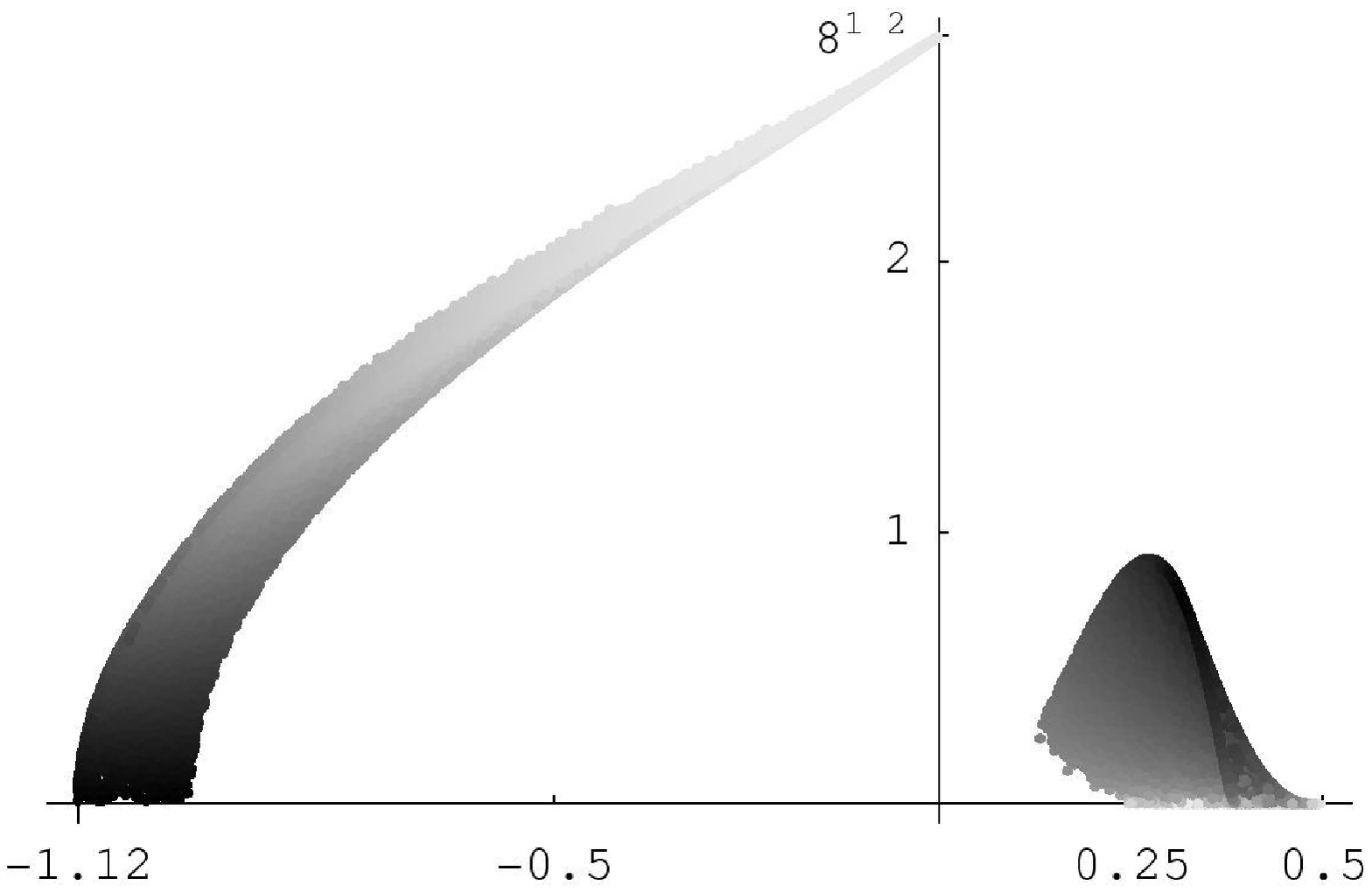}
 \hspace{7mm}
 \raisebox{3mm}{\includegraphics[width=0.7cm]{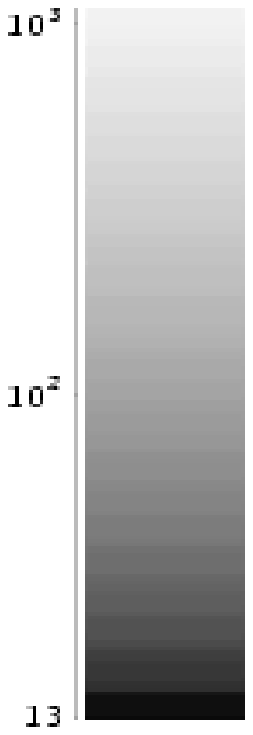}}
\caption{\small Non-uniqueness  with $j=0$: The plot shows the
  areas of the black hole (left) and 
  black ring (right) vs.~their respective angular velocities. Note
  that in order to fit the black hole and black ring areas on the same
  plot we have multiplied the black area by a factor of $50$.
  The black holes are clearly counter-rotating. 
  The points in the plots are colored according to the
  temperature. The black ring is always much hotter than the black
  hole, so different color scales are used for the black hole and
  the black ring.}
\label{fig:jzeroArea}
\end{center}
\begin{picture}(0,0)(0,0)
\put(139,47){\footnotesize $\omega_\rom{BR}$}
\put(28,47){\footnotesize $\omega_\rom{BH}$}
\put(115,93){\footnotesize black ring}
\put(51,93){\footnotesize black hole}
\put(119,71){\footnotesize $50\times a_\rom{H}^\rom{BR}$}
\put(99,107.5){\scriptsize/}
\end{picture}
\end{figure}

{}Figure \ref{fig:jzeroArea} shows the regions of the 
$(\omega_i, a_\rom{H}^i)$ plane covered by the black ring and the 
$S^3$ black hole in saturn configurations with $j=0$. 
Since $\omega_\rom{BH}<0$ and $\omega_\rom{BR}>0$, the two objects are
clearly counter-rotating.
Note that the black ring area $a_\rom{H}^\rom{BR}$ has been multiplied
by a factor of 50 in order for the plot be visible in the same plot as
the $S^3$ black hole. 
The total area $a_\rom{H}^\rom{total}$ never exceeds that of the
4+1d Schwarzschild black hole. 

The points in figure \ref{fig:jzeroArea} are colored according to the 
temperature $\tau_i$ of the corresponding black hole/ring: Light
gray means hot and black means cold. The scales used for the 
black hole and the black ring temperatures are different, as shown in figure
\ref{fig:jzeroArea}. The $S^3$ black hole temperature varies roughly 
between 0 and 3 (roughly like the Myers-Perry black hole which varies
between 0 and 1), while the black ring is much hotter with temperature
varying between $13$ and $\sim 10^3$. This, and the very small area of
the black ring, signals that these are very thin, large radius black
rings.

We further note that there is discrete non-uniqueness in the black
ring sector of $j=0$ black saturn. This can be seen by the ``skirt"
hanging over the righthand-part of the black ring area
vs. $\omega_\rom{BR}$ ``bell''. The rings here have lower temperatures
than the other rings with the same parameters, and it is therefore
natural to interpret this ``skirt" as a fat ring branch. 

\begin{figure}[t]
\begin{picture}(0,0)(0,0)
\put(72,39){\footnotesize$\omega_\rom{BR}$}
\put(165,39){\footnotesize$\omega_\rom{BR}$}
\end{picture}
 \centerline{
 \includegraphics[width=0.8cm]{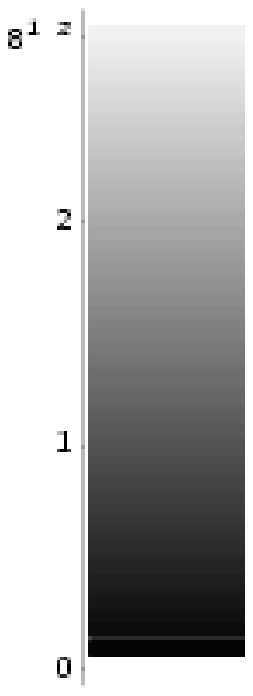}
 \hspace{7mm}
 \includegraphics[width=6cm]{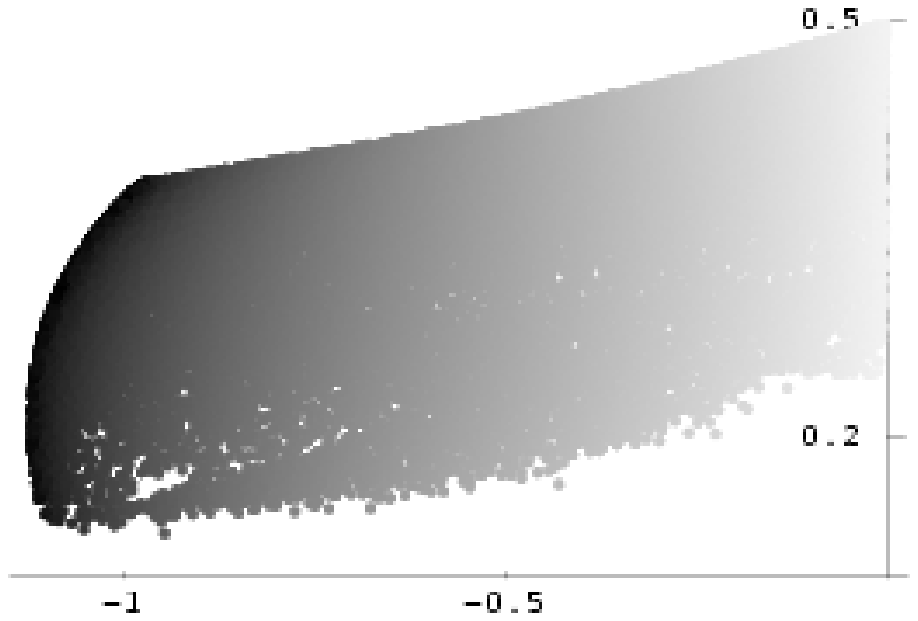}
 \hspace{1cm}
 \raisebox{-3mm}{\includegraphics[width=1cm]{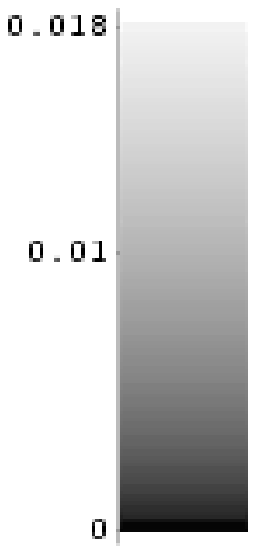}}
 \hspace{7mm}
 \includegraphics[width=6cm]{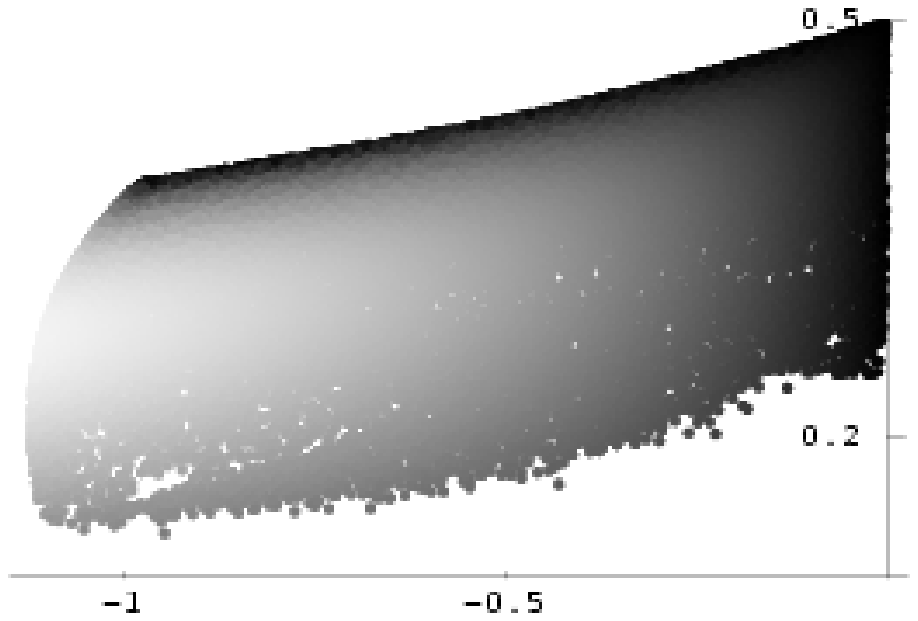}}
\caption{\small Non-uniqueness with $j=0$: The two plots show
  the regions covered in ``angular velocity space'' when the
  angular momentum is fixed to zero, $j=0$. Clearly the ring and the
  black hole 
  are counter-rotating.
  On the left plot, the color shows the area of the black hole, while
  on the right plot it is the area of the black ring. Note again the
  scales are different and the black ring area is much smaller than
  that of the black hole.}
\begin{picture}(0,0)(0,0)
\put(15,36){\footnotesize $\omega_\rom{BH}$}
\put(107,36){\footnotesize $\omega_\rom{BH}$}
\put(1,65){\footnotesize $a_\rom{H}^\rom{BH}$}
\put(93,66){\footnotesize $a_\rom{H}^\rom{BR}$}
\end{picture}
\label{fig:jzeroAngVel}
\end{figure}

The points $(\omega_\rom{BH}, a_\rom{H}^\rom{BH})$ lie in the wedge
shown in figure \ref{fig:jzeroArea}. For each point in this $S^3$
black hole wedge there is one (or two, in case of additional discrete
non-uniqueness) corresponding point(s) in the black ring
``bell''. But it is not clear which $S^3$ black hole goes together
with which black ring(s). That is illustrated better in figure
\ref{fig:jzeroAngVel}, which shows two plots of $\omega_\rom{BR}$
vs.~$\omega_\rom{BH}$. The first is colored 
according to the area of the black hole $a_\rom{H}^\rom{BH}$, while
the second is colored according to the area of the black ring
$a_\rom{H}^\rom{BR}$. Light gray means large area, black 
small area. As shown, different scales are used in the two plots.  

In both figure \ref{fig:jzeroArea} and figure \ref{fig:jzeroAngVel}
certain edges of the plots are rugged, and there are small white
uncovered regions. This is simply due to the finite
number of points generated for each plot, since some regions are
covered less than others (this was also visible in figure
\ref{fig:phasediag}).


\subsubsection{Fixed $j$ plots}

\begin{figure}[thbp]
\begin{picture}(0,0)(0,0)
\put(46,-3){\footnotesize $a_\rom{H}$}
\put(121,-3){\footnotesize $a_\rom{H}$}
%
\put(46,-59){\footnotesize $a_\rom{H}$}
\put(121,-59){\footnotesize $a_\rom{H}$}
%
\put(46,-114){\footnotesize $a_\rom{H}$}
\put(121,-114){\footnotesize $a_\rom{H}$}
\put(32,-51){\footnotesize Figure \ref{fig:combijsq}(a): $j=1.2$ .}
\put(107,-51){\footnotesize Figure \ref{fig:combijsq}(b): $j=0.95$ .}
\put(32,-106){\footnotesize Figure \ref{fig:combijsq}(c): $j=0.93$ .}
\put(107,-106){\footnotesize Figure \ref{fig:combijsq}(d): $j=0.925$ .}
\put(9,-161){\footnotesize Figure \ref{fig:combijsq}(e): $j=0.5$. The $a_\textrm{BR}$ is multiplied } 
\put(9,-165){\footnotesize by a factor of 10.}
\put(83,-161){\footnotesize Figure \ref{fig:combijsq}(f): $j=0.2$. The $a_\textrm{BR}$ is multiplied } 
\put(83,-165){\footnotesize by a factor of 10.}
\end{picture}
\begin{center}
       \mbox{   
        \includegraphics[width=6.5cm]{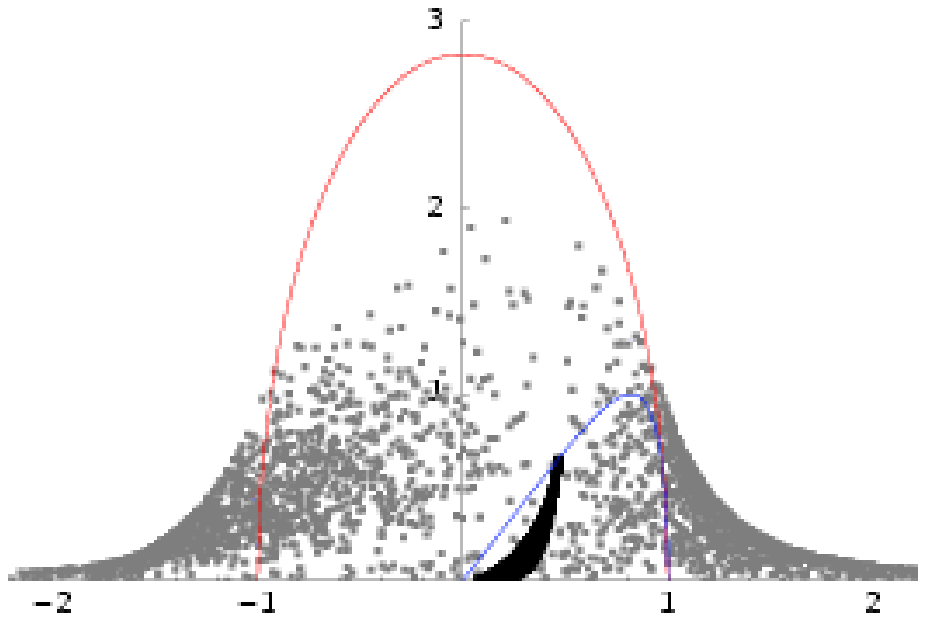}
        \hspace{6mm}
        \includegraphics[width=6.5cm]{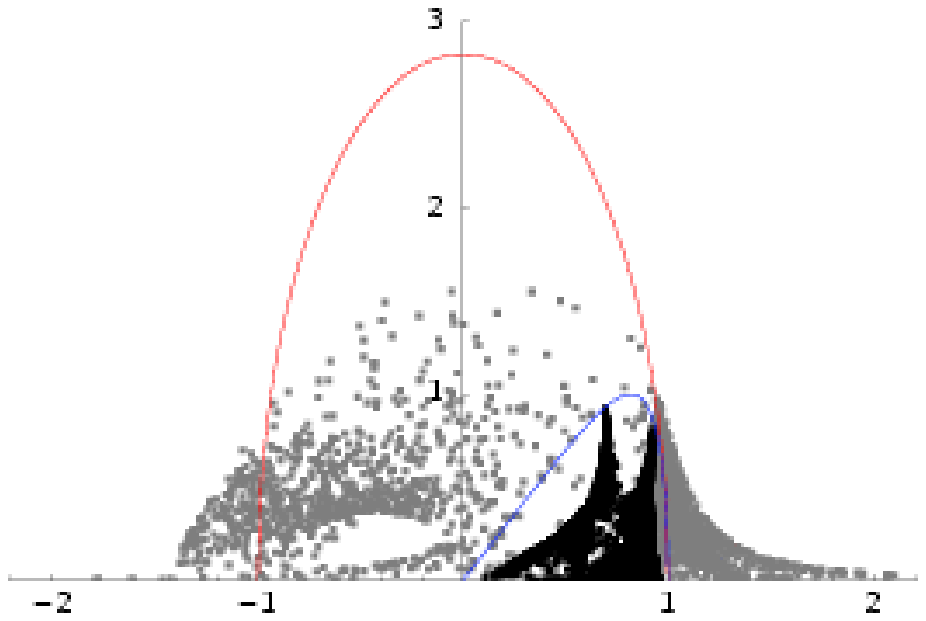}
        }\\[15mm]
        \mbox{
        \includegraphics[width=6.5cm]{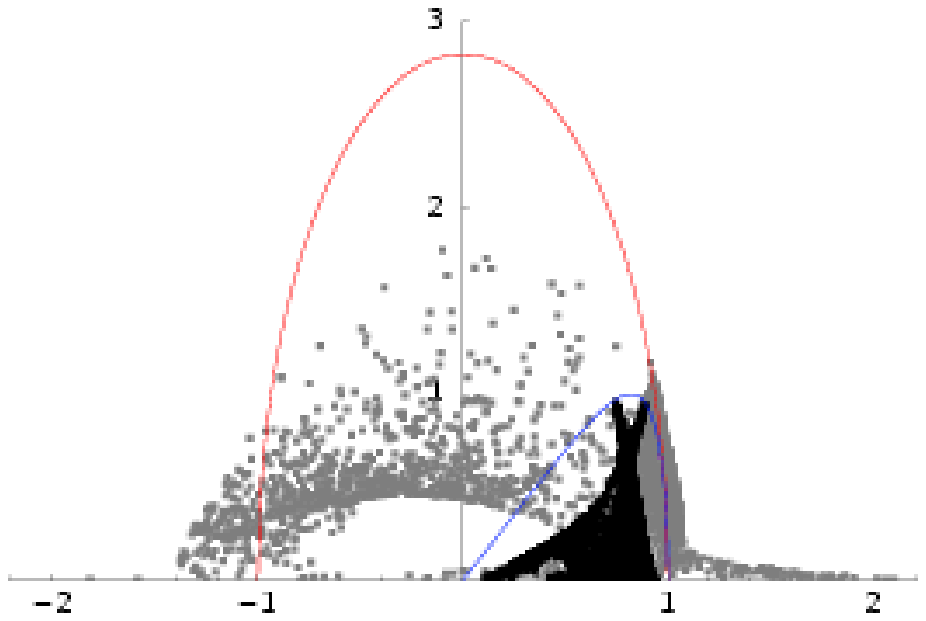}
        \hspace{6mm}
        \includegraphics[width=6.5cm]{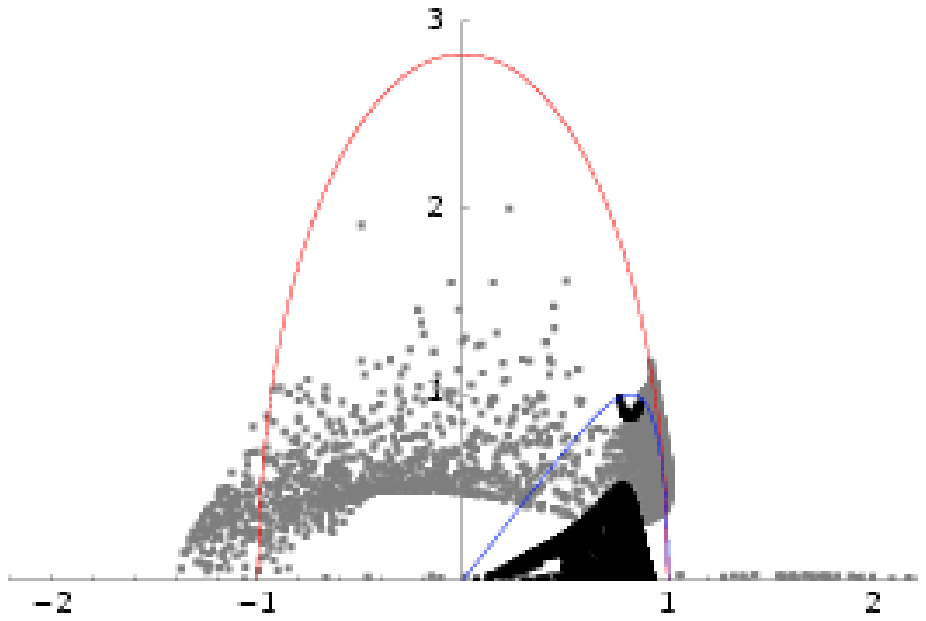}
        }\\[15mm]
        \mbox{
        \includegraphics[width=6.5cm]{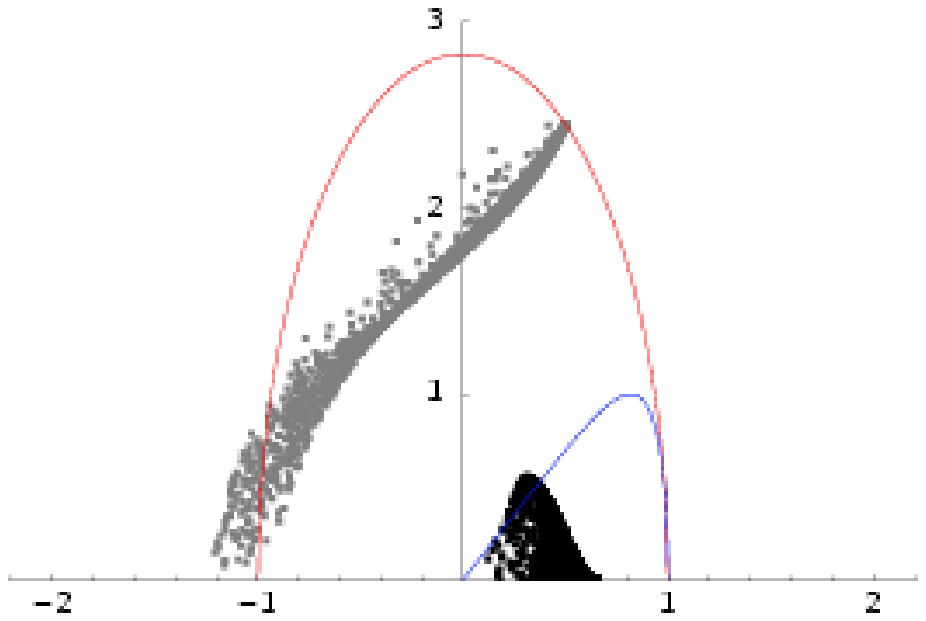}
        \hspace{6mm}
        \includegraphics[width=6.5cm]{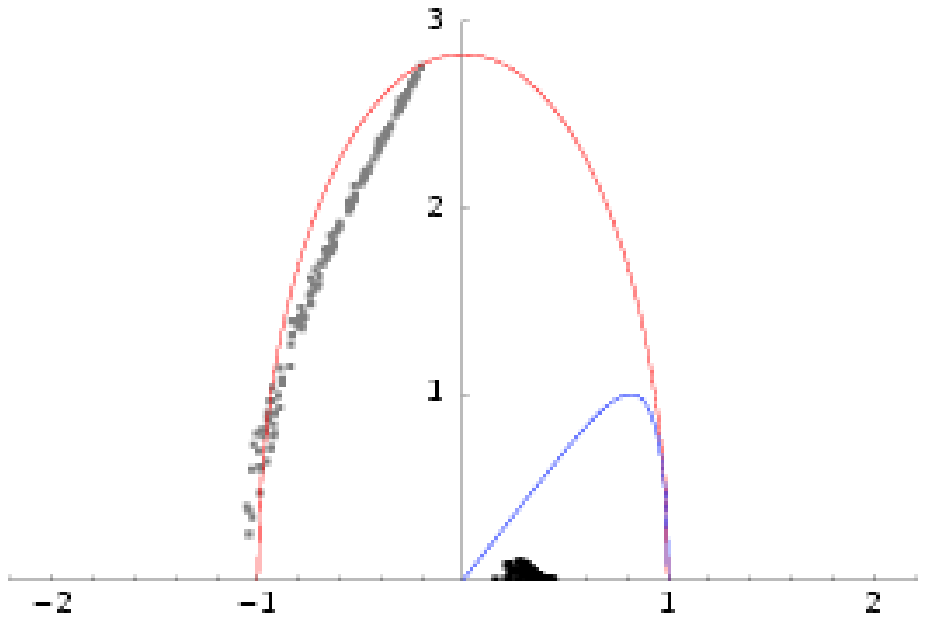}
        }
\vspace{10mm}
\caption{\small For fixed total mass and some representative values of
  $j^2$, the area of the black hole (gray dots) and the area of the
  black ring (black dots) are plotted against their respective angular
  velocities. The superimposed curves correspond to the area of a
  single Myers-Perry black hole against its angular velocity (upper curve), and
  similarly for the black ring. For these curves, the angular momentum $j$ is of course not fixed. Note that for the black hole we included both the positive and negative $\omega_\textrm{BH}$.}
\label{fig:combijsq}
\end{center}
\begin{picture}(0,0)(0,0)
\put(79,159){\footnotesize$\omega_i$}
\put(11,159){\footnotesize$\omega_\rom{BH}$}
\put(85,159){\footnotesize$\omega_\rom{BH}$}
\put(153,159){\footnotesize$\omega_i$}
\put(79,104){\footnotesize$\omega_i$}
\put(11,104){\footnotesize$\omega_\rom{BH}$}
\put(153,104){\footnotesize$\omega_i$}
\put(85,104){\footnotesize$\omega_\rom{BH}$}
\put(78,48){\footnotesize$\omega_i$}
\put(11,48){\footnotesize$\omega_\rom{BH}$}
\put(152,48){\footnotesize$\omega_i$}
\put(84,48){\footnotesize$\omega_\rom{BH}$}
\put(12,192){\footnotesize single MP bh}
\put(34,193){\vector(2,-3){5}}
\put(34,193){\vector(4,-1){21}}
\put(65,178){\footnotesize single black ring}
\put(64,179){\vector(-2,-3){5}}
\end{picture}
\end{figure}

We displayed in the previous section the regions of parameter space 
$(\omega_i, a_\rom{H}^i)$ covered by saturn configurations with
$j=0$. Likewise we can explore non-uniqueness for saturn
configurations with $j$ fixed at other values. Figure
\ref{fig:combijsq} shows $(\omega_i, a_\rom{H}^i)$ plots for
representative values of fixed $j$. 

When $j > \sqrt{27/32} \sim 0.92$, the $S^3$ black hole angular
velocity $\omega_\rom{BH}$ and area $a_\rom{H}^\rom{BH}$ vary over a
large range of values. This is shown in figures
\ref{fig:combijsq}(a)-(d). 
As $j$ becomes smaller than $j =\sqrt{27/32}$, which is the minimum
value of $j$ for the single black ring, the black ring of saturn has
very small area and the range of the $S^3$ black hole parameters are
more constrained, see figures \ref{fig:combijsq}(e)-(f). When the
black ring and $S^3$ black hole are counter-rotating so that $j$ is
negative, the $S^3$ black hole parameters differ only little from the
parameters of the Myers-Perry black hole, and the black ring is very
thin and contributes little to the total area.

\subsection{Solutions with $\epsilon=-1$}
\label{s:negeps}

Up to this point we have examined the physics of black saturn
solutions for which the condition for balance \reef{s:balance} was
imposed with the choice of sign $\epsilon=+1$, and hence $\bc_2 >
-\ka_2^{-1}$.   
Here we briefly discuss the balanced saturn solutions with
$\epsilon=-1$ for which $\bc_2 < -\ka_2^{-1}$. 

As pointed out at the end of section \ref{s:komar}, the solutions with
$\bc_2 > -\ka_2^{-1}$ have positive Komar masses, while for $\bc_2 <
-\ka_2^{-1}$, the Komar mass of the $S^3$ black hole is negative. We
interpret this as an effect of extreme rotational frame-dragging,
which makes $\Omega^\rom{BH} J_\rom{Komar}^\rom{BH}$ so negative that
the Smarr relation \reef{SatSmarr} renders the Komar mass negative.  

In the limit $\bc_2 \to -\ka_2^{-1}$, several of the dimensionful
physical parameters diverge. However, the dimensionless fixed-mass
reduced quantities remain finite. In particular, the dimensionless 
proper distance $\ell$, defined in \reef{theell}, between the black
ring and the $S^3$ goes to zero when $\bc_2 \to -\ka_2^{-1}$. So this
is a singular merger limit which ends in the nakedly singular $\bc_2=
-\ka_2^{-1}$ solution.  

The balanced saturn solutions with $\bc_2 < -\ka_2^{-1}$ occupy only a
small region of the phase diagram $(j,a^\rom{total}_\rom{H})$. They
have $j \simeq -1$ and total area $0<a^\rom{total}_\rom{H} \lsim 1$.
We interpret these solutions as tightly bound gravitational systems;
they probably deserve a closer study than the one provided here.  


\section{Discussion}
\label{s:discussion}

We have presented and analyzed a new exact solution to 4+1-dimensional
vacuum Einstein equations describing Black Saturn: a Myers-Perry black
hole surrounded by a black ring which is balanced by rotation in the
plane of the ring. The system exhibits a number of interesting
properties, such as non-uniqueness and frame-dragging, which were
summarized in the Introduction. 

Most surprising is probably the result
that the 4+1-dimensional Schwarzschild black hole and slowly spinning
Myers-Perry black holes are not unique. Black saturn shows that once
multiple 
black hole horizons are considered (and staticity not assumed for the 
$J=0$ configurations) black holes in 4+1-dimensions have large
degeneracies. This and the structure of the phase diagram for
4+1-dimensional black holes can be found in \cite{EEF2}.  

We expect both black objects in black saturn to have ergoregions
whenever their angular velocities are non-zero. This is always the
case for the black ring, whose ergosurface is expected to have
topology $S^1 \times S^2$ \cite{ER}. The
$S^3$ black hole can be tuned to have zero angular velocity, and it is
natural to expect that the solution, despite having non-vanishing
intrinsic angular momentum, has no ergoregion. Generally, however, we
expect an ergoregion bounded by an $S^3$ ergosurface. The metric in
Weyl axisymmetric coordinates $(\rho, z)$ is sufficiently complicated
that we have not extracted useful equations for the
ergoregions. We hope this will be addressed in future work, and note
that it may be useful to first examine the $\rho=0$ metric in order to
examine the intersections of the ergosurfaces with the plane of the
ring. 
 
It would be desirable to transform the black saturn metric to a 
simpler coordinate system. The supersymmetric concentric black hole -
black ring solutions of \cite{GG} can be written in ring coordinates
$(x,y)$ and it would presumably simplify our solution considerably to
write it in such coordinates. We have presented in appendix
\ref{app:BR} the coordinate transformation from Weyl axisymmetric
coordinates $(\rho,z)$ to ring coordinates $(x,y)$ for the simpler
limit of the black ring without the $S^3$ black hole. We leave it to
future work to convert the full black saturn solution to ring
coordinates. We expect that ring coordinates will make it easier to
study the ergoregions. 
  
Focusing on the plane of the ring, we have numerically checked
examples of co- and counter-rotating configurations, and found no
closed timelike curves (CTCs). While we see no signs of CTCs --- the
horizon areas and temperatures are positive and well-defined in the
full range of parameters --- this should be analyzed in greater detail
than done in this paper. Writing the solution in ring coordinates
$(x,y)$ will likely facilitate such an analysis. 

The 1st law of thermodynamics for black saturn is studied in \cite{EEF2}.
(We refer to refs.~\cite{KCGH} and \cite{Astefanesei:2005ad} for other
works on black ring thermodynamics.) Black saturn is an example of an
equilibrium system of 
two black objects which generally have different temperatures and
different angular velocities. This is therefore only a classical
equilibrium.  
It is shown in \cite{EEF2} that imposing thermodynamic equilibrium, in
the sense of the two objects having equal temperatures and equal
angular velocities, reduces the continuous family of saturns to a
one-parameter family of equilibrium solutions with only discrete
non-uniqueness. The phase diagram of equilibrium solutions is
presented in  \cite{EEF2}.

The saturn system may well be classically unstable.
The black ring of black saturn likely suffers from the same
instabilities as the single black ring \cite{ER}.
Using the Poincar\'e (or ``turning-point'') method, it was argued in
\cite{ALT} that (at least) one mode of instability would appear at the
cusp of the black ring curve in the area vs.~angular momentum phase
diagram. At the cusp, where the thin and fat black ring branches meet,
the black ring has minimum angular 
momentum and maximum entropy (for given mass).
Studying the potential for the radial balance of a black ring, 
evidence was found \cite{EEV} that a thin black ring would be stable
under small 
radial perturbations while a fat black ring would be unstable.
The radial
instability of fat black rings appear exactly at the cusp, and so this
mode is a physical concretization of the mode predicted by the
turning-point method \cite{ALT}.\footnote{Due to the continuous
  non-uniqueness, the implementation of the 
turning-point method for black saturn does not seem possible.
Following \cite{EEV} one can try to compute the radial potential for
the black ring in black saturn, but here one also has to choose to fix
some non-conserved quantities in order to carry out the analysis.}

Under radial perturbations, the analysis of \cite{EEV} indicated that
fat black rings either collapse to $S^3$ black hole (if perturbed
inward) or possibly expand to become a thin black ring (if perturbed
outwards). The latter may not happen in a dynamical process if thin 
black rings suffer from other classical instabilities, such as the
Gregory-Laflamme instability \cite{GL}, not captured by the
turning-point method. Showing that Gregory-Laflamme modes always fit
on (thin) black rings, refs.~\cite{Hovdebo,EEV} argued that thin black
rings very likely suffer from Gregory-Laflamme
instabilities. Likewise, we expect thin black rings of the black
saturn system to be unstable to Gregory-Laflamme instabilities.

While some stability properties of black saturn can be expected to be
inherited from the individual components, the Myers-Perry black hole and the
single black ring, there can also be new instabilities for the black
saturn system, for instance, perturbation of the center-of-mass of the 
$S^3$ black hole away from the center of the ring.

We constructed the black saturn solution using the inverse scattering
method \cite{BZ1,BZ2,BV}. The seed solution and the soliton
transformations invite a number of interesting generalizations of
black saturn: 

\begin{itemize}
\item {\sl Multiple rings of saturn}: It is straightforward to
  generalize the seed solution to include more rod sources that will
  correspond to black ring horizons. One can also add negative density
  rods to facilitate the addition of angular momentum for each black
  ring using (anti)soliton transformations like we did in section
  \ref{s:BZing}. Provided all singularities can be removed as done
  here, and the system balanced, the generated solution will describe
  ``the multiple rings of black saturn''. One interesting property of
  multiple ring solutions is the high degree of continuous
  non-uniqueness as we discussed in the Introduction.

\item {\sl Doubly spinning black saturn}: The 3-soliton
  transformation in section \ref{s:BZing} adds a second
  angular momentum for the $S^3$ black hole when $b_3 \ne 0$. An
  analysis of this solution is required to check that all possible
  singularities can be eliminated. Then it will be interesting to
  study the physics of this doubly spinning saturn system. We expect
  that the second ``intrinsic'' spin $J_\phi^\rom{Komar}$ will only be 
  non-vanishing for the $S^3$ black hole, but that the black 
  ring will also have non-vanishing angular velocity on the $S^2$. 
  This would be interpreted as rotational dragging of the $S^3$ black
  hole on the black ring. 

The 3-soliton construction of section \ref{s:BZing} does not give the
most general doubly spinning black saturn configuration, because the
black ring would not carry independent angular momentum on the
$S^2$. The more general black saturn configuration should be possible
to obtain with the methods recently used to construct the doubly
spinning black ring \cite{Pom2}. 

Doubly spinning black rings likely suffer from superradiant
instabilities \cite{odias}. It would be interesting to see if such an
instability is present also when the black ring is only being dragged
on the $S^2$ by the $S^3$ black hole. 

\item {\sl Dipole black saturn}: Black rings can carry non-conserved
  ``dipole charges'' \cite{RE}. 
Adding dipole charge(s) to the black ring  will give a dipole black
  saturn solution. The techniques \cite{SY} for adding dipole charge
  by combining two or more vacuum solutions should apply here. 

\item {\sl Charged black saturn}: Vacuum solutions can be charged up
  to carry conserved charges --- and for black rings also dipole
  charges. Lifting the solutions to ten dimensions and using boosts
  and dualities it is easy to charge up Myers-Perry black holes and
  black rings to carry, say D1- and D5-charges. The same
  transformations give a D1-D5-charged black saturn configuration
  (although not the most general such solution). 
  For black rings there is a technical difficulty in adding the third
  charge, momentum $P$, as detailed in \cite{EE}. This can be
  overcome by starting with a dipole black ring, and in this way a
  class of non-supersymmetric three-charge black rings have been
  obtained \cite{EEF}.   

Likewise, a D1-D5-P black saturn solution  can be obtained from dipole
black saturn, and this would lead to (a subclass of)
non-supersymmetric generalizations of the supersymmetric
concentric black ring solutions \cite{GG}. It would be interesting if
techniques can be developed to add independent charges to
multi-component black hole systems.

\end{itemize}

As discussed in the Introduction, one motivation for the existence of
black saturn is to think of a thin black ring balanced in the external
potential of the $S^3$ black hole. We have of course seen clear
evidence of the gravitational interactions between the black ring and
the $S^3$ black hole, for instance the rotational dragging  
(see section \ref{s:cc2zero}). So considering the black hole as
providing an external potential should only be seen as a motivation
for the case where the black ring is very thin with large $S^1$ radius
so that the interactions between the objects is negligible.  
 Following the method of \cite{EEV} one can take the system off-shell
 and study the equilibrium of forces on a very thin black ring around
 a small black hole. Presumably this would give a Newtonian balance
 between a string-like tension of the ring and the angular velocity in
 the background gravitational potential of the $S^3$ black hole.  

The balanced black saturn solution presented here has two separate
sectors. These arise from two different ways of imposing the balance
condition, as described in section \ref{s:analysis}. 
 We have focused almost entirely on the sector where the Komar
masses of both the black ring and the $S^3$ black hole are
non-negative. However, the other sector --- for which the $S^3$ black hole
Komar mass is negative --- may also contain interesting physics. We
interpret the possibility of negative Komar mass as a consequence of
extreme rotational dragging experienced by the $S^3$ black hole when
its Komar angular momentum cannot counter the dragging by the black
ring. It would be interesting to understand this strongly interacting
system better.

The 3+1 dimensions double-Kerr solution can be constructed with
methods similar to the ones used in this paper. While it is not
possible for two Kerr-black holes to be balanced by spin-spin
interactions alone, one could ask if it is for two Myers-Perry black
holes in 4+1 dimensions. 
The static solution describing two (or more)
Myers-Perry black holes held apart 
by conical singular membranes is easy to construct using the methods
of \cite{ER2}; this family of solutions was studied in
\cite{TanTeo}. Angular momentum can be added by soliton
transformations similar to the ones used here. It is
not clear if the resulting solution can be made free of singularities
and, even if so, if the black holes can be held apart by the spin-spin
interactions. 

Little is known about what types of black holes are admitted by the
Einstein equations in six and higher-dimensions. The main focus has
been on spacetimes with a single connected black hole horizon, but black
saturn has shown that interesting physics arises in higher-dimensional
multi-black hole systems. It will be interesting to see what exotic
multi-black hole solutions higher-dimensional gravity has to offer.

\subsubsection*{Acknowledgements}
We are grateful to Gary Horowitz, Harvey Reall, Enric Verdaguer and especially
Roberto Emparan for discussions of the inverse scattering method and
saturn physics.  
We thank Chethan Krishnan for pointing out that a clarification was
needed for comparison of the results of the 2- and 3-soliton
transformations. 

PF would like to thank the Theoretical Physics group at
Imperial College London for hospitality 
and also the Center for Theoretical Physics at MIT for warm
hospitality while part of this work was done. 
HE thanks the Aspen Center for Physics for a fruitful stay during the 
summer 2006. HE also thanks the Theoretical Physics group at Uppsala
University and the Niels Bohr 
Institute for hospitality during the final stages of this work. 

HE was supported by a Pappalardo Fellowship in
Physics at MIT and by the US Department of Energy through cooperative
research agreement DE-FG0205ER41360. PF was supported by FI
and BE fellowships from AGAUR (Generalitat de Catalunya), DURSI 2005 SGR
00082, CICYT FPA 2004-04582-C02-02 and EC FP6 program
MRTN-CT-2004-005104.


\appendix

\section{Limits}
In this section we provide details of the Myers-Perry black hole limit
and the black ring limit of the black saturn solution.

\subsection{Myers-Perry black hole}
\label{app:MP}

\begin{figure}[t]
\begin{picture}(0,0)(0,0)
\put(-3,29){$t$}
\put(-3,15){$\phi$}
\put(-3,0){$\psi$}
\put(37,-5){$a_3$}
\put(65,-5){$a_2$}
\end{picture}
\centering{\includegraphics[width=4in]{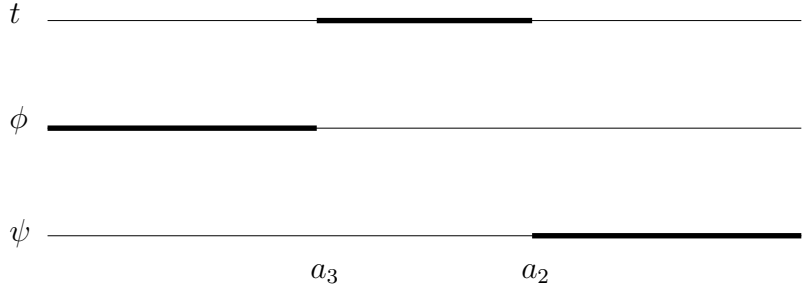}}
\bigskip
\caption{Sources for the Myers-Perry black hole. The timelike rod is
  aligned along $(1,0,\Omega_\psi^{\rom{BH}})$.} 
\label{fig:MPbh} 
\end{figure} 

{}From the rod structure figure \ref{fig:saturnrod}(b) of the full
solution, one can 
see that the Myers-Perry black hole with a single angular momentum is obtained
by eliminating the rod corresponding to the black ring. There is
however an issue of the order of limits. First one can note that from
our general saturn solution, with $c_1$ and $c_2$ arbitrary, the
following two limits result in the same solution:
\bea
  \label{MPlimits}
  &&\rom{Limit}~1:~~~
  a_1\to a_5\, ,~~~\rom{then}~~~a_5\to a_4\, ,. \\
  &&\rom{Limit}~2:~~~
  a_5\to a_4\, ,~~~\rom{then}~~~a_1\to a_4\, ,. 
\eea
As long as $c_1$ and $c_2$ are kept fixed (i.e.~the regularity condition \reef{eqn:c1} is \emph{not} imposed), these two limits are equivalent and give the metric 
\begin{equation}
\begin{aligned}
&G_{tt}&=&-\frac{\mu_3 \left[-c_2^2\mu_2^2\mu_3+(\rho^2+\mu_2 \mu_3)^2\right]}
        {\mu_2\left[c_2^2\mu_3 \rho^2+(\rho^2+\mu_2 \mu_3)^2\right]}\;, \hspace{1cm}
&&G_{t\psi}&=&-\frac{c_2 \mu_3(\rho^2+\mu_2^2)(\rho^2+\mu_3^2)}
        {\mu_2\left[c_2^2\mu_3 \rho^2+(\rho^2+\mu_2 \mu_3)^2\right]} \;,\\
&G_{\psi\psi}&=&~\frac{\mu_2^2(\rho^2+\mu_2\mu_3)^2-c_2^2 \mu_3\rho^2}
        {\mu_2\left[c_2^2\mu_3 \rho^2+(\rho^2+\mu_2 \mu_3)^2\right]}\;, \hspace{1cm}
&&G_{\phi\phi}&=&~\frac{\rho^2}{\mu_3}\;, \\
&e^{2\nu}&=&~\frac{k^2\,\mu_2\left[c_2^2\mu_3 \rho^2+(\rho^2+\mu_2 \mu_3)^2\right]}
        {(\rho^2+\mu_2^2)(\rho^2+\mu_3^2)(\rho^2+\mu_2\mu_3)}\;.
\end{aligned}
\label{eqn:MPweyl}
\end{equation}
To bring the metric given above to an asympotically flat form one has
to perform  change the coordinates according to $t=t'-c_2 \psi'$ and
$\psi=\psi'$. Finally, to show that this solution given in
\eqref{eqn:MPweyl} is indeed the Myers-Perry black hole with a single
angular momentum, one can change to prolate spheroidal coordinates as
done in \cite{Har} and \cite{Pom}. 

Now if we are interested in obtaining the Myers-Perry black hole as a
limit of the 
black saturn configuration, we must remove the black ring in a limit
where the condition  \reef{eqn:c1}  is imposed on $c_1$. Note that in
Limit 1 of  \reef{MPlimits}, $c_1 \to \infty$. This is the reason we
consider Limit 2 in 
\reef{MPlimits}. 

In the parametrization introduced in section \ref{sec:parametrization}
Limit 2 is 
\bea
    \ka_3\to \ka_2\, ,~~~\rom{then}~~~\ka_2\to 0\, .
\eea
By first taking $\ka_3 \to \ka_2$ we eliminate the
divergence in $c_1$ and then the $\ka_2 \to 0$ limit can taken
safely. The resulting metric is \reef{eqn:MPweyl}.

With $\bc_2$ fixed by the balanced condition \reef{balance}, this
limit can only be 
taken for $\epsilon=+1$, and one finds
\bea
  \bc_2 = 1-\frac{1}{2\ka_1} 
\eea
and all physical parameters are then functions of the dimensionless 
parameter $\ka_1$ and the scale $L$, which are related to the standard
Myers-Perry black hole parameters $r_0$ and 
$a$ through
\bea
  r_0^2 =  \frac{L^2}{2\ka_1} \, ,
  ~~~~ 
  a = \frac{L(1-2\ka_1)}{\sqrt{2 \ka_1}} \, .
\eea


\subsection{Black ring limit}
\label{app:BR}
\begin{figure}[t]
\begin{picture}(0,0)(0,0)
\put(-3,36){$t$}
\put(-3,18){$\phi$}
\put(-3,0){$\psi$}
\put(46,-5){$a_5$}
\put(64,-5){$a_4$}
\put(90,-5){$a_3$}
\end{picture}
\centering{\includegraphics[width=5in]{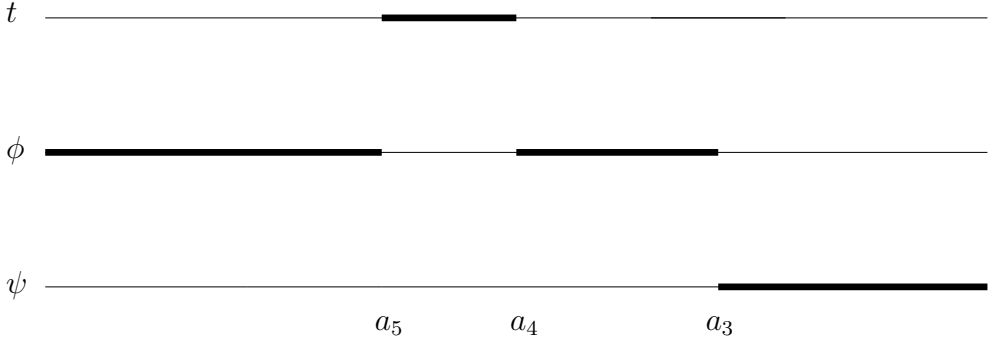}}
\bigskip
\caption{Sources for the black ring. The timelike rod has direction
  $(1,0,\Omega_\psi^{\rom{BR}})$.} 
\label{fig:BR} 
\end{figure}

The $\psi$-spinning black ring is obtained by first setting $c_2=0$,
then taking $a_2=a_3$. We must continue to impose the condition
\reef{eqn:c1} for $c_1$; note that this condition is independent of
$a_2$. We find  
\bea
  \label{psimetricrhoz}
  G_{tt} &=& -\frac{\mu_1 
   \Big[ \mu_5 (\rho^2 + \mu_1 \mu_3)^2 
                (\rho^2 + \mu_1 \mu_4)^2
          -c_1^2 \mu_3 \mu_4(\mu_1-\mu_5)^2 \rho^4
   \Big]}
   {\mu_4 
   \Big[ \mu_5 (\rho^2 + \mu_1 \mu_3)^2 
                (\rho^2 + \mu_1 \mu_4)^2
          +c_1^2 \mu_1^2 \mu_3 \mu_4 (\mu_1-\mu_5)^2 \rho^2
   \Big]} \, , \\[2mm]
  G_{t\psi} &=& -\frac{c_1 
   \mu_3 \mu_5 (\mu_1-\mu_5) (\rho^2 + \mu_1^2) (\rho^2 + \mu_1 \mu_3) 
                (\rho^2 + \mu_1 \mu_4)}
   {
   \Big[ \mu_5 (\rho^2 + \mu_1 \mu_3)^2 
                (\rho^2 + \mu_1 \mu_4)^2
          +c_1^2 \mu_1^2 \mu_3 \mu_4 (\mu_1-\mu_5)^2 \rho^2
   \Big]} \, , \\[2mm]
  G_{\psi\psi} &=& \frac{\mu_3 \mu_5
   \Big[ \mu_5 (\rho^2 + \mu_1 \mu_3)^2 
                (\rho^2 + \mu_1 \mu_4)^2
          -c_1^2 \mu_1^4 \mu_3 \mu_4(\mu_1-\mu_5)^2 
   \Big]}
   {\mu_1 
   \Big[ \mu_5 (\rho^2 + \mu_1 \mu_3)^2 
                (\rho^2 + \mu_1 \mu_4)^2
          +c_1^2 \mu_1^2 \mu_3 \mu_4 (\mu_1-\mu_5)^2 \rho^2
   \Big]} \, , \\[2mm]
  e^{2\nu} &=& k^2
   \frac{\mu_3 (\rho^2 + \mu_1 \mu_5)
               (\rho^2 + \mu_3 \mu_4)
               (\rho^2 + \mu_4 \mu_5)\Big[
         \mu_5 (\rho^2 + \mu_1 \mu_3)^2 
                (\rho^2 + \mu_1 \mu_4)^2
          +c_1^2 \mu_1^2 \mu_3 \mu_4 (\mu_1-\mu_5)^2 \rho^2\Big]}
  {4 \mu_1 (\rho^2 + \mu_1 \mu_3)
           (\rho^2 + \mu_1 \mu_4)
           (\rho^2 + \mu_3 \mu_5)^2
           (\rho^2+\mu_1^2)
           (\rho^2+\mu_3^2)
           (\rho^2+\mu_4^2)
           (\rho^2+\mu_5^2)}\nonumber\\
\eea
Note that for $c_1=0$ we obtain the metric for the static black ring. The corresponding rod structure is depicted in figure \ref{fig:BR}.

To verify that this solution really describes the $\psi$-ring, we
rewrite the metric in ring coordinates $(x,y)$, i.e.
\bea
  \label{psimetricxy}
  ds^2 &=& -\frac{F(y)}{F(x)} \left( dt+C_\lambda  \,  R\, \frac{1+y}{F(y)} d\psi\right)^2 
  \nonumber \\[1mm]
       && + \frac{R^2}{(x-y)^2} F(x)
       \left[
         -\frac{G(y)}{F(y)} d\psi^2
         -\frac{dy^2}{G(y)}
         +\frac{dx^2}{G(x)}
         +\frac{G(x)}{F(x)} d\phi^2
       \right] \, ,
\eea 
where 
\bea
  G(\xi) = (1-\xi^2)(1+\nu \xi) \, , 
  ~~~~
  F(\xi) = (1+\lambda \xi) \, ,
  ~~~~
  C_\lambda = \sqrt{\lambda (\lambda-\nu)\frac{1+\lambda}{1-\lambda}} \, .
\eea
The coordinate transformation from Weyl coordinates $(\rho, z)$ to
  ring coordinates $(x,y)$ is
\bea
  \rho = \frac{R^2 \sqrt{-G(x)G(y)}}{(x-y)^2} \, ,
  \hspace{7mm}
  z = \frac{R^2(1-x y)[2+\nu (x+y)]}{2(x-y)^2} \, .
\eea
Note that
\bea
  d\rho^2 + dz^2 = K(x,y) \left[ -\frac{dy^2}{G(y)} + \frac{dx^2}{G(x)} \right] \, ,
\eea
with
\bea
  K(x,y) = -\frac{R^4}{4(x-y)^3} [x+y+\nu(1+xy)]
  [2+\nu(1+x+y-x y)] [2+\nu(-1+x+y+x y)] \, .~\nonumber \\
\eea
The rod endpoints are related to the parameters $\nu$ and $\lambda$ as
\bea
  a_1 = R^2  \alpha\, ,
  ~~~
  a_5 = - \frac{R^2}{2} \nu \, ,
  ~~~
  a_4 =  \frac{R^2}{2} \nu \, ,
  ~~~
  a_3 =  \frac{R^2}{2} \, .
\eea
Here $\alpha < -\nu/2$ is a constant which will be determined below.
With this choice, $\rho^2 + (z-a_i)^2$ is a perfect square for
  $i=3,4,5$ (but not for $i=1$ for choice of $\alpha< -\nu/2$) so we
  have simple expressions for $\mu_i = R_i - (z-a_i) =  \sqrt{\rho^2 +
  (z-a_i)^2}  - (z-a_i)$: 
\bea
  \mu_5 &=& - \frac{R^2 (1-x)(1+y)(1+\nu y)}{(x-y)^2} \, , \\
  \mu_4 &=& - \frac{R^2 (1-x)(1+y)(1+\nu x)}{(x-y)^2} \, , \\
  \mu_3 &=& - \frac{R^2 (1-y^2)(1+\nu x)}{(x-y)^2} \, .
\eea
The expression for $\mu_1$, however,  involves an explicit squareroot, 
$R_1= \sqrt{\rho^2 + (z-a_1)^2}$. We write $\mu_1=R_1-(z-a_1)$, but
  keep $R_1$ unevaluated. Then collect powers of $R_1$ and simplify
  the expressions for each metric component when only even powers of
  $R_1$ are replaced by their expression in terms of $x,y$. Then we
  end up in general for each metric component with expressions of the
  form 
\bea
  g_{\mu\nu} \sim \frac{p_0 + p_1 R_1}{q_0 + q_1 R_1} \, ,
\eea
where $p_{0,1}$ and $q_{0,1}$ are functions of $x,y$.
Now it turns out, as can explicitly be verified, that for all cases,
  $p_0/q_0 = p_1/q_1$ so that $g_{\mu\nu} = p_0/q_0$. Thus we have
  eliminated the squareroot $R_1$ from the expressions, and indeed
  $p_0/q_0$ is a simple function of $x,y$.  
For example, we find
\bea
  G_{tt} = -\frac{2-2\alpha (1+y) -\nu (1-y)}{2-2\alpha (1+x) -\nu (1-x)} \, .
\eea
We bring $G_{tt}$ to the standard from given in \reef{psimetricxy} by choosing
\bea
 \label{thea1}
 \alpha = \frac{\nu (1+ \lambda)-2\lambda}{2(1-\lambda)} \, .
\eea
The condition that $\alpha \le -\nu/2$ is then simply that $\nu \le \lambda$, while $-\infty<\alpha$ gives $\lambda <1$. We have therefore recovered the bound
\bea
  0 < \nu \le \lambda < 1 
\eea
on the black ring parameters $\lambda$ and $\nu$.

With this choice \reef{thea1} for $\alpha$ we have
\bea
  e^{2\nu} = \frac{k^2 (1-\nu)^2}{1-\lambda} \frac{R^2}{(x-y)^2} \frac{F(x)}{K(x,y)} \, ,
\eea 
so that choosing the integration constant $k$ as 
\bea
  k^2=\frac{(1-\lambda)}{(1-\nu)^2}\, 
\eea
we recover the $x,y$-part of the metric \reef{psimetricxy}. With these
choices for $\alpha$ (the position of the ``fake'' rod endpoint) and
$k$, the full metric \reef{psimetricrhoz} becomes \reef{psimetricxy}. 

Note that in the black ring limit with $\bc_2 = 0$, the periods
\reef{Deltas} are 
$\Delta \psi=\Delta \phi= 2\pi k$, which with $k$ given above agrees
precisely with the result for the black ring \cite{RE}. Likewise all
physical parameters of the neutral black ring are reproduced from this
limit of the black saturn solution.



\end{document}